\documentclass[a4paper, traditabstract]{aa} 

\usepackage[english]{babel} 
\usepackage[utf8]{inputenc} 
\usepackage[T1]{fontenc}
\usepackage[mathscr]{eucal}
\usepackage{graphicx}
\usepackage{amsmath}
\usepackage{epstopdf}
\usepackage{epsf,color}
\usepackage[breaklinks, colorlinks, citecolor=blue]{hyperref} 
\usepackage{natbib}
\usepackage{multirow}
\usepackage{longtable,lscape, supertabular}
\usepackage{txfonts}
\usepackage{amsfonts}
\usepackage{amssymb}
\usepackage{newlfont}
\usepackage{textcomp}
\usepackage{times}
\usepackage{units}
\usepackage{mathrsfs}
\usepackage{xcolor}

\begin{document}

\title{Impact of polarised extragalactic sources on the measurement of CMB B-mode anisotropies}
\author{
G.~Lagache\inst{1} \and
  M.~B\'ethermin\inst{1} \and
           L.~Montier\inst{2}\and
            P.~Serra\inst{3} \and 
           M.~Tucci\inst{4}  
           }
\institute{
Aix Marseille Univ, CNRS, CNES, LAM, Marseille, France \email{guilaine.lagache@lam.fr}\goodbreak \and
CNRS,  IRAP,  9  Av. colonel  Roche,  BP  44346,  F-31028 Toulouse cedex 4, France \goodbreak \and
 Jet Propulsion Laboratory, California Institute of Technology, Pasadena, California 91109, USA \goodbreak \and
D\'epartement de Physique Th\'eorique and Center for  Astroparticle Physics, Universit\'e de Gen\`eve,~~24 quai Ansermet, CH--1211 Gen\`eve 4, Switzerland \goodbreak
}
\date{}

\abstract{
One  of  the  main  goals  of  cosmology  is  to  search for  the  imprint  of  primordial gravitational waves in the polarisation filed of the cosmic microwave background to probe  inflation  theories. One of the obstacles in detecting the primordial signal is that the cosmic microwave background B-mode polarisation must be extracted from among astrophysical contaminations. Most efforts have focus on limiting Galactic foreground residuals,  but extragalactic foregrounds cannot be ignored at the large scale ($\ell \lesssim 150$), where the primordial B-modes are brightest.  We present a complete analysis of extragalactic foreground contamination
that is due to polarised emission of radio and dusty star-forming galaxies. We update or use current models that are validated using the most recent measurements of source number counts, shot noise, and cosmic infrared background power spectra. 
We predict the flux limit (confusion noise) for future cosmic microwave background (CMB) space-based or balloon-borne experiments (IDS, PIPER, SPIDER, LiteBIRD, and PICO), as well as ground-based experiments (C-BASS, NEXT-BASS, QUIJOTE, AdvACTPOL, BICEP3+Keck, BICEPArray, CLASS, Simons Observatory, SPT3G, and S4). The telescope aperture size (and frequency) is the main characteristic that affects the level of confusion noise. Using the flux limits and assuming mean polarisation fractions independent of flux and frequency for radio and dusty galaxies, we computed the B-mode power spectra of the three extragalactic foregrounds (radio source shot noise, dusty galaxy shot noise, and clustering). We discuss their relative levels and compare their amplitudes to that of the primordial tensor modes parametrised by the tensor-to-scalar ratio $r$.  At the reionisation bump ($\ell$=5), contamination by extragalactic foregrounds is negligible. While the contamination is much lower
than the targeted sensitivity on $r$ for large-aperture telescopes at the recombination peak ($\ell$=80), it is at a comparable level for some of the medium- ($\sim$1.5m) and small-aperture telescope ($\le$0.6m) experiments. For example, the contamination is at the level of the 68\% confidence level uncertainty on the primordial $r$ for the LiteBIRD and PICO space-based experiments. These results were obtained in the absence of multi-frequency component separation (i.e. considering each frequency independently). We stress that extragalactic foreground contaminations have to be included in the input sky models of component separation methods dedicated to the recovery of the CMB primordial B-mode power spectrum. Finally, we also provide some useful unit conversion factors and give some predictions for the SPICA B-BOP experiment, which is dedicated to Galactic and extragalactic polarisation studies. We show that SPICA B-BOP will be limited at 200 and 350\,$\mu$m by confusion from extragalactic sources for long integrations in polarisation, but very short integrations in intensity.
}

\authorrunning{Lagache et al.}
\titlerunning{}

\maketitle

\section{Introduction}
\label{sect:intro}
The $\Lambda$CDM model is the standard model of cosmology. It is the simplest parametrisation of the Hot Big Bang model, with two principal ingredients: $\Lambda$ refers to a cosmological constant (i.e. the energy density of the vacuum), and CDM stands for cold dark matter, that is, dark matter particles that move slowly. Because it is very successful in predicting a wide variety of observations related to the cosmic microwave background (CMB), the large-scale structure, and gravitational lensing, the $\Lambda$CDM model has reached the status of a paradigm. In this paradigm, an era of early exponential expansion of the Universe, dubbed cosmic inflation, has been proposed to explain why the Universe as revealed by the CMB radiation is almost exactly Euclidean and so nearly uniform in all directions. While the basic $\Lambda$CDM model fits all the data (with parameters known at the percent level), the physics of inflation is still unknown. Thus, one of the central goals of modern cosmology is to determine the nature of inflation. One generic prediction is the existence of a background of gravitational waves, which produces a distinct, curl-like signature in the polarisation of the CMB. This is referred to as primordial B-mode polarisation (which is due to tensor perturbations).
The detection of this primordial B-mode polarisation would provide clear proof that inflation did occur in the early Universe.
Unfortunately, cosmic inflation does not provide a unique prediction for the amplitude of the primordial tensor modes
parametrised by the tensor-to-scalar ratio $r$. We are in a situation where there is no natural range for $r$, in particular, there is no relevant lower bound. The natural goal is to be able to measure $r$ beyond doubt for the Higgs inflation (which is an inflation scenario where the inflaton  field is the Higgs boson), that is, r$\ge\sim$ 2$\times$10$^{-3}$ at 5$\sigma$. If this does not lead to a detection, it will discard the whole class of large-field models. If the inflaton field was nothing but the Higgs field, this would have tremendous consequences for physics. Thus a precise measurement of (or upper bound on) $r$ is essential to constrain inflation physics. The current 95\% CL upper limit on $r$ as measured by \textit{Planck}\footnote{\textit{Planck} (\url{http://www.esa.int/Planck}) is a project of the European Space Agency (ESA) with instruments provided by two scientific consortia funded by ESA member states and led by Principal Investigators from France and Italy, telescope reflectors provided through a collaboration between ESA and a scientific consortium led and funded by Denmark, and additional contributions from NASA (USA).} combined with  ground-based CMB experiments is $r<0.056$ \citep{planck2018_cosmo} at a pivot scale of k= 0.002/Mpc. The search for the primordial B-mode is an outstanding challenge that has motivated a number of experiments designed to measure the anisotropies of the CMB in polarisation with an ever-increasing precision. 

B-modes are also generated by gravitational lensing of E-mode polarisation, providing a unique window into the physics of the evolved Universe and invaluable insights into late-time physics, such as the effect of dark energy and the damping of structure formation by massive neutrinos. 
These lensing B-modes are a nuisance for the primordial B-modes. Several approaches have been studied for the CMB B-mode delensing using large-scale structure surveys \citep[galaxies or the cosmic infrared background (CIB), e.g.][]{Smith2012, Sherwin2015, Manzotti2017}, or assuming that the lensing potential can be estimated internally from CMB data \citep[e.g.][]{Carron2017, Sehgal2017}.

In addition to instrumental challenges, future experiments targeting $r\sim$10$^{-3}$ will have to solve the critical problem of component separation. In addition to lensing, polarised Galactic foreground contamination dominates the amplitude of the large-scale CMB B-modes by several orders of magnitude. The capabilities of future experiments to remove the contamination due to polarised Galactic emissions have been investigated for example by \cite{Errard2016}, \cite{Remazeilles2016} and \citet{Philcox2018}.  We investigate the polarisation fluctuations caused by extragalactic contaminants: radio galaxies and dusty star-forming galaxies (DSFG).  While polarised compact extragalactic sources are expected to be a negligible foreground for CMB B-modes near the reionisation peak ($\ell<$10), they are expected to be the dominant foreground for $r= 10^{-3}$ when delensing has been applied to the data, from the recombination peak to higher multipoles, $\ell>$50 \citep{Curto2013}.

Extragalactic radio sources are typically assumed to be Poisson distributed in the sky. The clustering of radio sources is strongly diluted by the broad distribution in redshifts of objects that contribute at any flux density. The contribution of clustering to the angular power spectrum is therefore small and can be neglected if sources are not subtracted down to very faint flux limits,
$S\ll10$\,mJy \citep{gon05}. 

For DSFG, we have to consider polarisation fluctuations not only for the Poisson distribution of point sources, but also for the clustering, that is, the CIB anisotropies \citep[e.g.][]{knox2001,negrello2004,viero2013, planckXXX}. The CIB power spectrum can be represented as the sum of two contributions that are usually called the one-halo and two-halo terms. The one-halo represents the correlation of galaxies in the same dark matter halo (pairs of galaxies inside the same halo); the two-halo, capturing the galaxy correlations in different dark-matter haloes, describes the large-scale clustering.
While we expect some polarisation fluctuations from the one-halo (which is close to Poisson fluctuations), polarised two-halo fluctuations are expected to be null, provided there is no correlation of the polarisation of galaxies within distinct halos. We could have a contribution from the large-scale clustering because of galaxy spin alignments in the filamentary dark-matter structure \citep[e.g.][and references therein]{Codis2018, Piras2018}. However, as recently shown by \cite{Feng2019}, this contribution is $>$100 and $\gtrsim1000$ times lower than the shot noise of DSFG at $\ell$=100 and $\ell$=1000, respectively. Thus we consider that it has a negligible effect because it is extremely weak. 

We compute the expected level of polarised fluctuations from the shot noise of radio galaxies and DSFG and from the CIB one-halo using current or updated models for a large set of future CMB space-based or balloon-borne experiments (IDS, PIPER, SPIDER, LiteBIRD, and PICO)\footnote{The meaning of all abbreviations is: inflation and dust surveyor, primordial inflation polarisation explorer, lite (light) satellite for the studies of B-mode polarisation and inflation from cosmic background radiation detection, and probe of inflation and cosmic origins for IDS, PIPER, LiteBIRD and PICO, respectively.}, as well as ground-based experiments (C-BASS, NEXT-BASS, QUIJOTE, AdvACTPOL, BICEP3+Keck, BICEPArray, CLASS, SO, SPT3G, and S4)\footnote{The meaning of all abbreviations is: C-band all-sky survey,  next band all-sky-survey, Q-U-I joint Tenerife, advanced Atacama cosmology telescope polarimeter, background imaging of cosmic extragalactic polarisation, cosmology large angular scale surveyor, Simons observatory, south pole telescope and stage-4 for C-BASS, NEXT-BASS, QUIJOTE, AdvACTPOL, BICEP, CLASS, SO, SPT, and S4, respectively.}. Our predictions use a point-source detection limit that is self-consistently computed for each experiment (taking the sensitivities into account and determining confusion noises using our number count models). We also include some predictions for SPICA B-POP. An accurate computation of the flux detection limit is mandatory to predict the shot noise of radio sources because changing the flux cut by 30\% affects the shot noise by 30\%, while it is less important for DSFG: a small variation in the flux cut leads to only a small variation in shot-noise power \citep{Planck_CIB_2011}.
 
Our work extends previous studies that concentrated either on a single experiment \citep[e.g.][]{dezotti2015}, a restricted frequency area \citep[e.g.][]{Bonavera17b, Curto2013}, a given galaxy population (e.g. radio galaxies; \citealt{Puglisi2018}), or on high multipoles (e.g. \citealt{Gupta2019} for $\ell\gtrsim$2000; e.g. \citealt{Datta2019} for CMB EE). We are the first to use our radio and DSFG models in combination with the CIB and CMB contamination and instrument noise to iteratively predict the confusion noise that is due to extragalactic sources for all experiments and then derive the level of polarised fluctuations. 

The paper is organised as follows.  We present the evolutionary models for radio sources and DSFG and discuss their polarised emission in Sects.\,\ref{radio_sources} and \ref{sec:DSFG_model}. In Sect.\,\ref{sec:polar_formalism} we give the formalism for computing polarised shot noise from galaxy number counts in intensity. We then describe our halo model of CIB anisotropies that is used to compute the polarisation power spectra that arise from the clustering of DSFG (Sect.\,\ref{extended_halo_model}). We use these models to compute the flux limit (caused by the fluctuations of the background sky brightness below which sources cannot be detected individually, i.e. the confusion noise) for a large number of future CMB experiments and for SPICA B-POP (Sect.\,\ref{sec:conf_noise}). The flux limits allow us to compute the expected level of radio and dusty galaxy polarised shot noises, which we discuss (together with the polarised one-halo) in Sect.\,\ref{compar_polar_Cell}, and which we compare to the CMB primordial B-mode power spectrum in Sect.\,\ref{compar_CMB_Cell} for all experiments. We conclude in Sect.\,\ref{conclu}.

\section{Radio sources}
\label{radio_sources}
In this section, we present the evolutionary model we are choosing to describe the number counts of radio galaxies (Sect\,\ref{NC_radio}), and its update (Sect\,\ref{NC_radio_up}). We then discuss the polarised emission of radio galaxies (Sect\,\ref{radio_polar}). Finally, we compute the shot noise using our model and compare it with observations from CMB experiments (Sect\,\ref{SN_radio}).

\subsection{\label{NC_radio} Number counts at cm to mm wavelengths}

Number counts of extragalactic radio sources are well determined at
radio frequencies $\nu\la10$\,GHz down to flux densities of
$S\la1$\,mJy (and even $S\la0.03$\,mJy at 1.4\,GHz) based on data from deep and large area surveys 
 \citep[e.g.][]{Bondi2008, dez10, Bonavera2011, massardi2011, Miller2013, Smolcic2017, Puglisi2018, Huynh2020}.
At higher frequencies, that is, from tens of GHz to millimetre (mm) wavelengths, observational
data on radio sources are mainly provided by CMB experiments  \citep[e.g.][]{planck_PCCS2, Datta2019,Gralla2020,Everett2020}. Space
missions such as WMAP and \textit{Planck}, which cover the full sky, were
able to detect only bright sources, with flux densities higher than a
few hundred mJy at best. On the other hand, the better angular resolution of ground-based experiments allows them to reach deeper in flux density,
but on smaller areas of the sky. The uncertainties on number counts are
therefore still large, especially in the frequency range where the CMB
dominates, that is, between 70 and 300\,GHz.

Evolutionary models for extragalactic radio sources \citep[e.g.][]{tof98,dez05,mas10} are able to provide a good fit to data on luminosity functions and multi-frequency source counts from $\sim100$\,MHz to
$\ga5$\,GHz. They adopt a schematic description of radio source
populations, divided into steep- and flat-spectrum (or blazars)
sources, according to the spectral index of the power-law spectrum,
$S(\nu)\propto \nu^\alpha$, at GHz frequencies that is lower or higher than
$-0.5$. A simple power law is also used to extrapolate spectra to
high frequencies, $\nu\gg5\,$GHz. However, especially for blazars,
real source spectra are generally more complex than a power law, which
can hold only for limited frequency ranges. As a consequence, these models tend to
over-predict the number counts of radio sources at $\nu\ga100\,$GHz,
as measured by the Atacama Cosmology Telescope (ACT) at 148\,GHz
\citep{mar11}, for instance, or by \textit{Planck} in all the High Frequency Instrument
(HFI) channels \citep{planck_xiii_2011,planck_pip_vii}. The main reason
for this disagreement is the spectral steepening observed in \textit{Planck} radio source catalogues above $\sim70$\,GHz \citep{planck_xiii_2011,planck_xv_2011,planck_pip_xlv} 
that was previously suggested by other data sets \citep{gon08,sad08}.

A first attempt of taking this steepening in blazar spectra into
account was made by \citet{tuc11}. They described the spectral
behaviour of blazars at cm--mm wavelengths statistically
by considering the main physical mechanisms responsible for the
emission. In agreement with classical models of the synchrotron
emission in the inner jets of blazars \citep{bla79,kon81,mar85}, the
spectral high-frequency steepening was interpreted as caused, at least
partially, by the transition from the optically thick to the
optically thin regime. The frequency $\nu_M$ at which the spectral
break occurs depends on the relevant physical parameters of AGNs:
the redshift, the Doppler factor ($\delta$), and the linear dimension of
the region (approximated as homogeneous and spherical) that is mainly
responsible for the emission at the break frequency. In particular,
\citet{tuc11} showed that the break frequency can be written in an
approximated form as
\begin{equation}
\nu_M\approx C(\alpha_{fl},\alpha_{st},S_{\nu_0})
\frac{D_L}{r_M\,\sqrt{(1+z)^3\delta}}\,,
\end{equation}
where $D_L$ is the luminosity distance of the sources, and $C$ is a
function of the spectral indices before and after the break
frequency ($\alpha_{fl}$ and $\alpha_{st}$ respectively) and of the
flux density $S_{\nu_0}$ at a reference frequency (typically 5\,GHz;
see their Appendix\,B). Finally, the parameter $r_{M}$ is the distance
from the AGN core of the jet region that dominates the emission at the
frequency $\nu_M$ (for a conical jet model, this parameter can
  be easily related to the dimension of the emitting jet region). It
defines the dimension and thus the compactness of the emitting region
at that frequency. This is the most critical parameter for determining
$\nu_M$ because the uncertainty on its actual value is large.

Based on 5\,GHz number counts and on information of spectral
properties of radio sources at GHz frequencies, the \citet{tuc11}
model provided predictions of number counts at cm/mm wavelengths by
extrapolating flux densities of radio sources from low (1--5\,GHz) to
high frequencies. The model considered three populations of radio
sources (steep-, inverted-, and flat-spectrum sources), and a
different high-frequency spectral behaviour for each of them. Here we
focus on blazars, which are the dominant class at $\nu\ga70$\,GHz.
The most successful model studied in the paper (referred to as ``C2Ex'')
assumes different distributions of the break frequency for BL\,Lac
objects and flat-spectrum radio quasars (FSRQs). According to this, most FSRQs should bend their otherwise flat spectra between 10 and
100\,GHz, whereas in BL\,Lac, spectral breaks are expected typically at $\nu\ga100$\,GHz (implying that the observed synchrotron
radiation comes from more compact emitting regions than
FSRQs). This dichotomy has indeed been found in the \textit{Planck} radio
catalogues \citep{planck_xiii_2011,planck_pip_xlv}. This model
provides a very good fit to all the data of bright ($S\ga100$\,mJy)
radio sources for number counts and spectral index distributions up to
$\sim$500-- 600\,GHz \citep{planck_xiii_2011,planck_pip_vii}.

\begin{figure*}
\centering
\includegraphics[width=6.5cm]{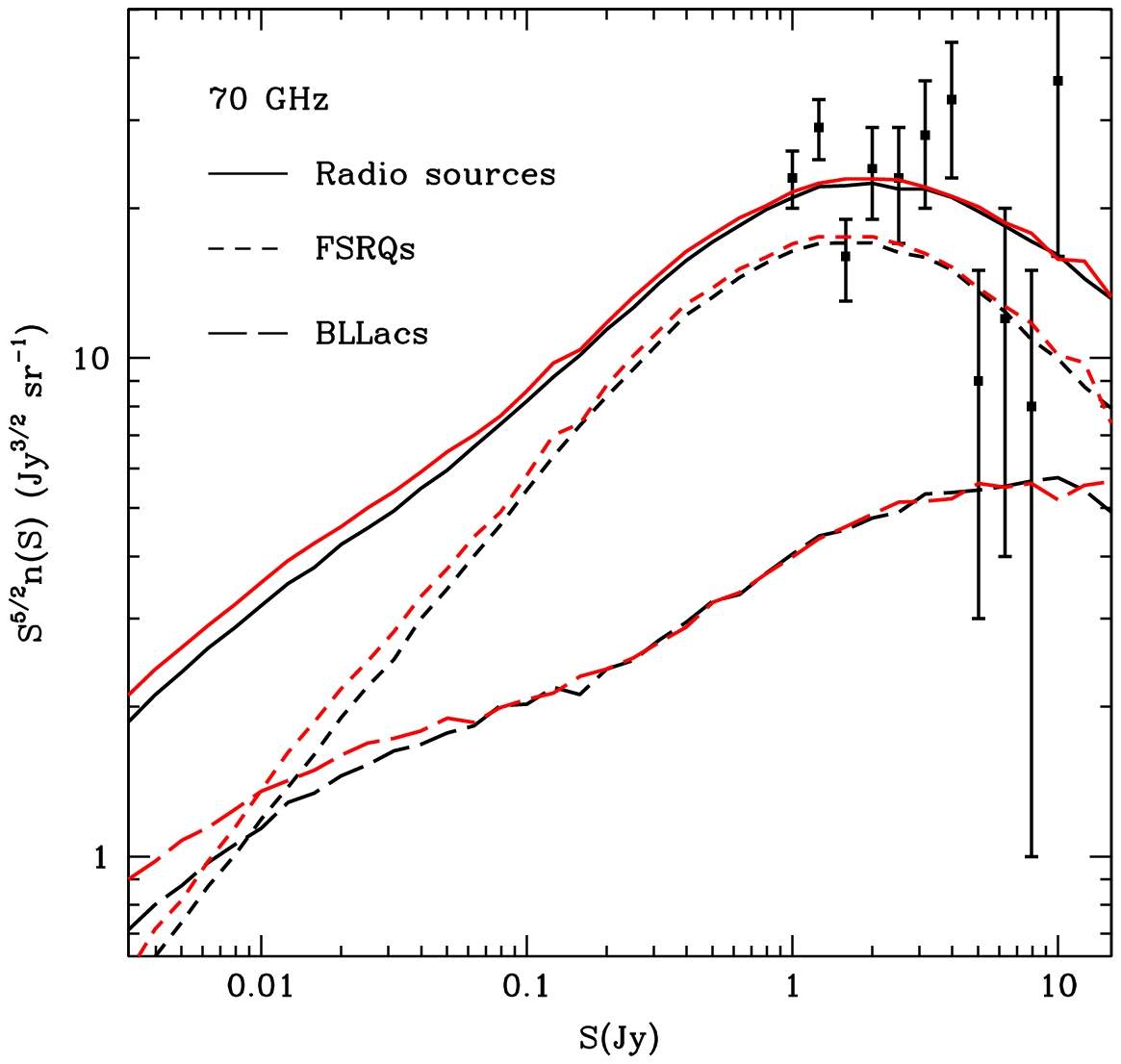}
\includegraphics[width=6.5cm]{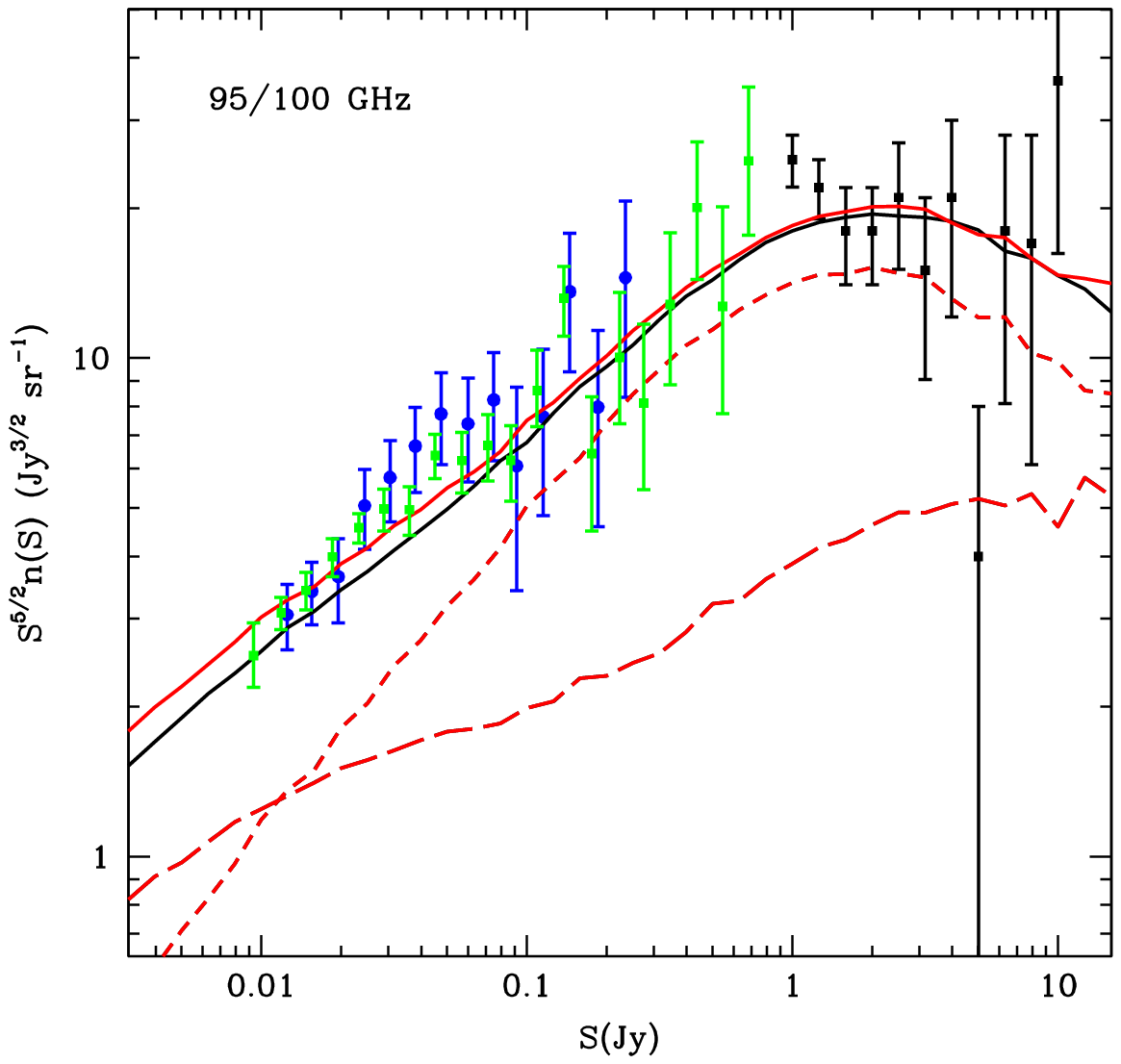}
\includegraphics[width=6.5cm]{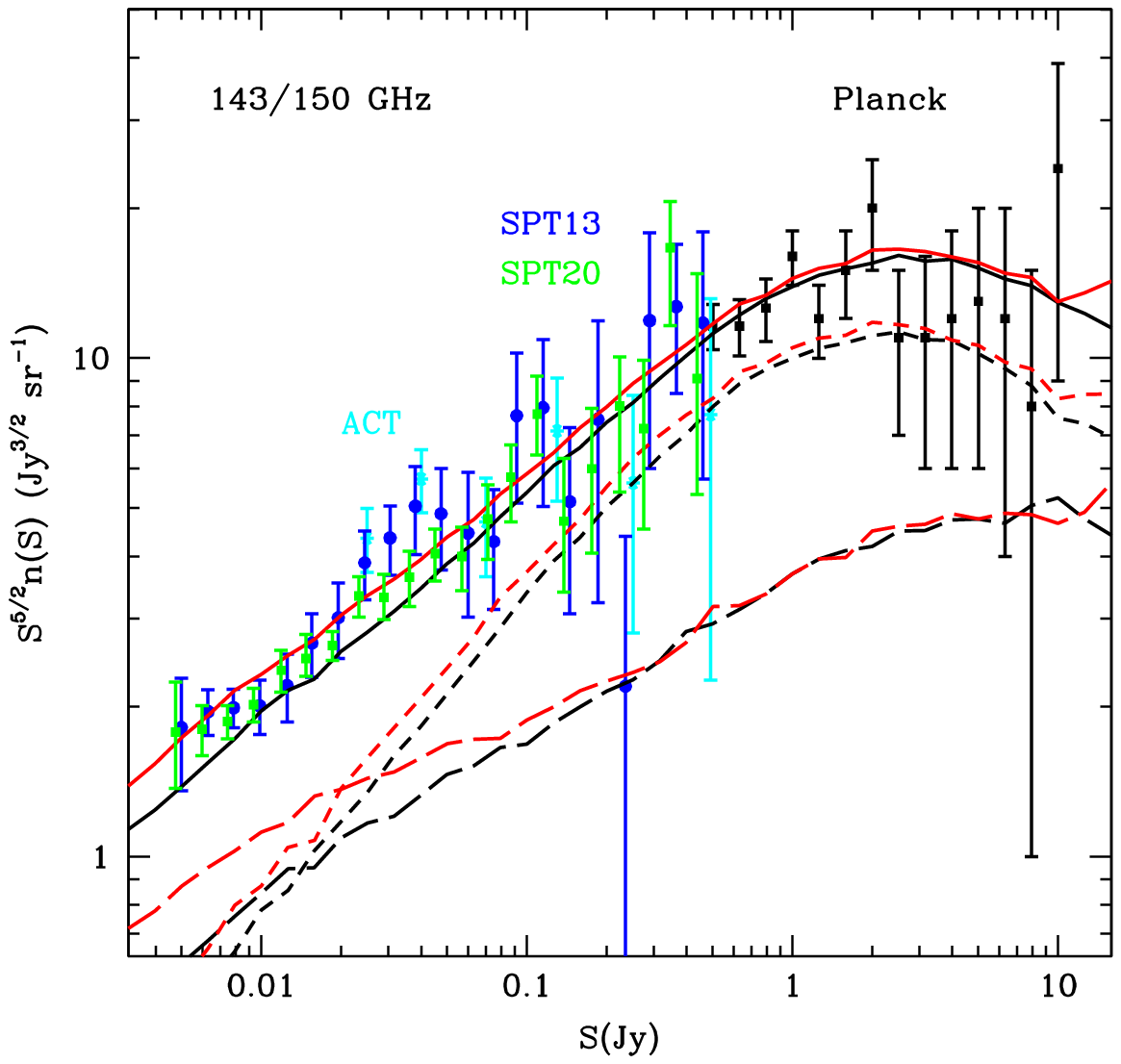}
\includegraphics[width=6.5cm]{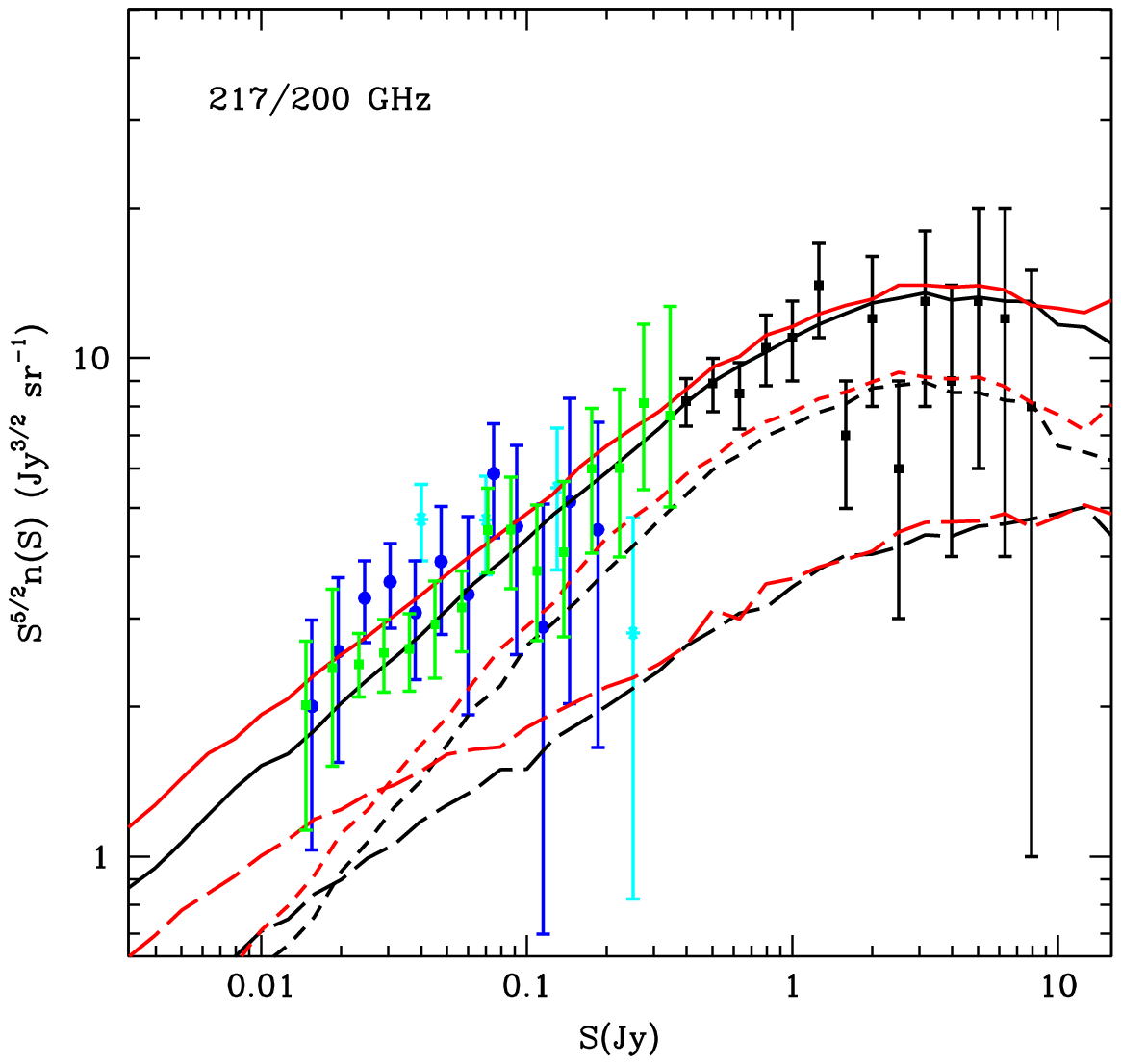}
\caption{Normalized differential number counts ($S^{5/2}n(S)$) from the \citet{tuc11} model (black lines) and observations (\textit{Planck}, black points; ACT, cyan points; SPT, blue and green points) between 70 and 220\,GHz. Red lines represent the model, but with the updated parameter values as described in the text. 
The very recent measurements from ACT at 148\,GHz \citep[cyan dots from][]{Datta2019} and from SPT-SZ \citep[green dots from][]{Everett2020} are not used in the fit, but are shown for comparison.
  }
\label{fig:ns_radio}
\end{figure*}

A partial agreement is also found when other surveys, deeper in flux than \textit{Planck}, are considered.  In Fig.\,\ref{fig:ns_radio} we compare the number counts from the model with observational data at frequencies between 70 and 220\,GHz. Beyond \textit{Planck}, data are from ACT \citep[150, 218\,GHz;][]{mar14, Datta2019}  and SPT and SPT \citep[95, 150, 220\,GHz;][]{moc13}. The model tends to underestimate SPT/ACT counts in the flux density range [20,60]\,mJy. Very recently, however, \cite{Everett2020} presented the number counts from the full 2500 square degrees of the SPT-SZ survey; they extended previous SPT results (see the green points in Fig.\,\ref{fig:ns_radio}). These new data agree better with the C2Ex model estimates at 220\,GHz.
 
\subsection{\label{NC_radio_up}  Updated model for number counts}
\label{mod:rad}
The recent data from ACT and SPT experiments give us the opportunity
to better constrain the model parameters for blazars. We described
above that the break frequency depends on a set of physical parameters
related to AGNs. \citet{tuc11} imposed most of them on the
basis of observational constraints (as the redshift distribution of
the different radio source populations; the Doppler factor; spectral indices) and on the basis of typical assumptions for AGN model
(equipartition condition, narrow conical jets, etc.). The only free
parameter used in the model is the distance $r_M$ to the AGN core of
the emitting jet region at the break frequency. In the best model of
\citet{tuc11}, $r_M$ is taken to be log-uniformly distributed in the
range of [0.3,\,10]\,pc for FSRQs and in the range of [0.01,\,0.3]\,pc for
BL\,Lacs.

We now determine the best estimate of the $r_M$ range by fitting number
counts from \textit{Planck}, ACT \citep{mar14} and SPT \citep{moc13} between 70 and 220\,GHz. This is
done only for BL\,Lacs, while for FSRQs we maintain the same range of
$r_M$ values as before. We verified  that a change in the
$r_M$ interval for this class of objects does not improve the fit of the number counts at sub-Jy level significantly (i.e. for
ACT/SPT data). This is not surprising because FSRQs provide the
dominant contribution to number counts of bright sources, with
$S\gg100$\,mJy (see Fig.\,\ref{fig:ns_radio}). At these flux levels,
the strong constraints come from \textit{Planck} measurements, which are
already well described by the model. On the other hand,
at fainter fluxes, the relevance of BL\,Lacs increases, and we expect them to become the dominant population at a few
dozen mJy. This is exactly the range of fluxes in which the model slightly
underestimates the observed number counts. By increasing the contribution
of BL\,Lacs, we should remove or reduce the discrepancy between model
and SPT/ACT data without affecting the predictions for the very bright
sources.

Jointly with $r_M$, we considered the spectral index $\alpha_{st}$ of
blazars after the break frequency (i.e. in the optically thin regime)
as an additional free parameter in the fit. \citet{tuc11} assumed this to be distributed as a Gaussian around $\langle\alpha_{st}\rangle=-0.8$ with a
dispersion of 0.2, in agreement with the canonical values for the
optically thin synchrotron spectral index. No differences between the
two classes of blazars were considered. However,
\citet{planck_xiii_2011,planck_pip_xlv} found that the average
spectral index of blazars after the spectral break is somewhat flatter
than $-0.8$.

The results of the fit give more compact radio-emission regions in BL\,Lacs than previous values, with $0.0025\le r_M\le0.05$\,pc, that is, about a factor 5 smaller than before. In addition, the
average high-frequency spectral index is flatter,
$\langle\alpha_{st}\rangle=-0.7$, consistent with the trend observed
in \textit{Planck} data.

Number counts predicted by the updated model differ mainly at low- to intermediate-flux densities, $S<0.1\,$mJy, and
provide an improved fit to observational data at 95 and 150\,GHz (see
Fig.\,\ref{fig:ns_radio}). The reduced $\chi^2$ is now very close to 1. 
SPT data at 95\,GHz are still slightly higher, between 20 and 60\,mJy, but the discrepancy is reduced
and is not significant. The change in
the average value of $\langle\alpha_{st}\rangle$ also produces a small
increase in the number counts of FSRQs at $\nu>100$\,GHz.
Number counts from the updated model are provided at https://people.lam.fr/lagache.guilaine/Products.

\subsection{Statistical properties of polarised emission \label{radio_polar}}
 
Polarisation in radio sources is typically observed to be a few
percent of the total intensity at cm or mm wavelengths (e.g. \citep{mur10, bat11, sajina2011, mas13,Galluzzi2019}, and only very few objects show a fractional
polarisation, $\Pi=P/S$, as high as $\sim$10\%. Steep-spectrum radio
sources are on average more polarised than flat-spectrum sources at
$\nu\la20$\,GHz \citep{tucci04,kle03}. Their fractional polarisation
strongly depends on the frequency, from $\sim2.5$\% at 1.4\,GHz
to $\sim5.5$\% at 10.5\,GHz \citep{kle03}. At low frequencies,
flat-spectrum sources are instead characterised by an almost constant
and low degree of polarisation ($\sim\,2.5$\%).

Extensive studies of high-frequency polarisation properties have been
conducted by \citet{tuc12} and \citet{mas13} using the Australia Telescope
20\,GHz (AT20G) survey \citep{mur10}. This is a quite deep
survey in intensity (with a completeness level of 91\% at
$S\ge100$\,mJy and 79\% at $S\ge50$\,mJy in regions south of
declination $-15^{\circ}$) with a high detection rate in
polarisation. Moreover, simultaneous measurements at 5 and 8\,GHz are
also available for a consistent fraction of objects. These analyses
found that the distribution of the polarisation degree (in blazars) is
well described by a log--normal function \citep[see also][]{bat11}
with an average fractional polarisation of $\sim3$\%. No clear correlation between the
fractional polarisation and the flux density was observed, with a
slight dependence on the frequency of the polarisation degree.

At frequencies $\nu>20$\,GHz, polarisation measurements of very bright
sources ($S\ga1$\,Jy) seem to indicate an increase in fractional
polarisation with frequency. Using the VLA for polarisation
measurements of a complete sample of the WMAP catalogue, \citet{bat11}
found that $\langle \Pi^{\mathrm{rad}} \rangle=2.9,\,3.0,$ and 3.5\% at 8.4,
22, and 43\,GHz, respectively, and a fractional polarisation that is typically higher
at 86\,GHz than at 43\,GHz. This was confirmed by measurements at
86\,GHz from \citet{agu10}, obtained with the IRAM 30\,m Telescope.
They found that for sources with detected polarisation at 15\,GHz, the fractional polarisation at
86\,GHz is higher than at 15\,GHz by a mean factor of $\sim2$. 
However, these results were not confirmed using new data and/or improved data analysis procedures \citep{Hales2014, Bonavera17a, Galluzzi2017, Puglisi2018, Trombetti2018, Datta2019, Gupta2019}.  No significant trends of the polarisation degree with flux density or with frequency are found at the frequencies of interest for CMB B-mode search. Latest measurements of fractional polarisation at $\nu>50$\,GHz vary from $\sim$1.5 to 3.5\% and are obtained either using log-normal fits to the distribution of observed polarisation fractions, or using stacking or statistical approaches. To compute the radio source contamination in polarisation to the CMB B-mode (Sect.\,\ref{contrib_Bmode}), we assumed a constant $\langle \Pi^{\mathrm{rad}} \rangle$=2.8\%, in agreement with the recent \textit{Planck} \citep[e.g.][]{Puglisi2018}, SPT \citep{Gupta2019} and ACT \citep{Datta2019} measurements, and radio source follow-ups from 90 to 220\,GHz. 

\subsection{Shot-noise predictions\label{SN_radio} }
In this section, we compare the shot-noise level from residual radio sources found in observational data with values expected from our
reference model, to confirm the validity of the model. As the radio shot noise level is highly sensitive to the flux limit, we also provide some useful empirical relations that allow us to compute the shot-noise level as a function of the flux limit.

\subsubsection{Shot-noise levels in current CMB experiments}
\label{SN_radio}

We report the residual shot-noise level in ACT and SPT data estimated by \cite{dun13} and \cite{geo14}, and compare them with predictions from the
\citet{tuc11} model before and after our update in Table\,\ref{tab_rad:1}. The agreement is quite good for both cases, although the shot-noise level of the updated model is closer to the observational estimates.

\begin{table*}
\centering
\begin{tabular}{ccccccc}
\hline
& \multicolumn{2}{c}{ACT} & &\multicolumn{3}{c}{SPT} \\
\hline
$\nu$\,[GHz] & 148 & 218 & & 95 & 150 & 220 \\
$S_{cut}$\,[mJy] & 15 & 15 & & & 6.4 & \\
\hline
Dunkley+13  & 3.2 $\pm$ 0.4  & 1.4 $\pm$ 0.2 &  & 7.2 $\pm$ 0.8  & 1.4 
$\pm$ 0.2  & 0.7 $\pm$ 0.1 \\
George+14 & & & & 7.81 $\pm$ 0.75  & 1.06 $\pm$ 0.17 & \\
\hline
Tucci+11 &  2.6 & 1.4 & & 5.9 & 1.0 & 0.48 \\
Updated & 3.2 & 1.7 & & 6.6 & 1.3 & 0.67 \\
\hline
\end{tabular}
\caption{Shot-noise power of residual radio sources, $D_{\ell}=\ell(\ell+1)C_{\ell}/2\pi$ [$\mu$K$_{CMB}^2$], at $\ell=3000$, estimated in ACT and SPT data, and predicted by models.}
\label{tab_rad:1}
\end{table*}

In Table\,\ref{tab_rad:2} we report auto- and
cross-power spectra (shot noise only) due to residual radio sources in \textit{Planck}
data according to the updated model. We also compute the error of
these predictions due to an uncertainty in the flux cut of 20 and
30\%\,.  Moreover, we give a tentative estimate of the error
associated with the uncertainty on the model that is computed as the
difference between results from the old and the updated model. The
uncertainties we find are probably quite conservative, but they are
nevertheless smaller than the errors due to a 20\% uncertainty
in $S_{cut}$ at frequencies where radio sources are dominant
(i.e. $\nu\le217$\,GHz).

The consistency between the measured Poisson amplitude in the \textit{Planck} auto- and cross-power spectra at 100, 143, and 217\,GHz
with the updated model discussed here has previously been investigated in \citet[][see their Table\,20]{planck_xi_2015}. The agreement is
good, except at 100\,GHz, where the predicted amplitude is significantly lower than the observed value. However, this discrepancy was
attributed by the authors to a residual unmodelled systematic effect in the data rather than to a foreground modelling error. Moreover, the Poisson power
at 100\,GHz is found to be smaller in \citet{Planck_ll_2018}, which agrees better with the model prediction (7.8\,Jy$^2$/sr for our model with a flux cut of 340\,mJy compared to 10.5\,Jy$^2$/sr for \textit{Planck,}  but with an unknown flux cut).

\begin{table}
\centering
\begin{tabular}{ccccccc}
\hline
$\nu_1$ & $\nu_2$ & $S_{cut}$ & $C_{\ell}$ &
 \multicolumn{2}{c}{$\sigma[S_{cut}]$} &
  $\sigma$[model] \\
& & [Jy] & [Jy$^2$sr$^{-1}$] & 20\% & 30\% & \\
\hline
 30 &  30 & 0.43 & 18.36 & 3.30 & 4.97 & 0.45 \\
 30 &  44 & & 15.48 & 2.87 & 4.29 & 0.50 \\
 30 &  70 & & 12.30 & 2.32 & 3.50 & 0.57 \\
 30 & 100 & &  9.58 & 1.80 & 2.70 & 0.62 \\
 30 & 143 & &  7.28 & 1.34 & 2.05 & 0.58 \\
 30 & 217 & &  5.65 & 1.05 & 1.57 & 0.57 \\
 30 & 353 & &  5.44 & 1.06 & 1.58 & 0.79 \\
 30 & 545 & &  4.67 & 0.91 & 1.37 & 0.87 \\
 30 & 857 & &  4.04 & 0.78 & 1.18 & 0.97 \\
 44 &  44 & 0.76 & 25.11 & 4.43 & 6.63 & 0.77 \\
 44 &  70 & & 15.34 & 2.70 & 4.24 & 0.65 \\
 44 & 100 & & 10.60 & 1.97 & 2.88 & 0.55 \\
 44 & 143 & &  7.68 & 1.36 & 2.12 & 0.45 \\
 44 & 217 & &  6.11 & 1.12 & 1.65 & 0.45 \\
 44 & 353 & &  8.06 & 1.49 & 2.21 & 0.99 \\
 44 & 545 & &  7.42 & 1.39 & 2.10 & 1.16 \\
 44 & 857 & &  6.60 & 1.24 & 1.88 & 1.36 \\
 70 &  70 & 0.50 & 13.46 & 2.53 & 3.75 & 0.63 \\
 70 & 100 & &  8.71 & 1.66 & 2.43 & 0.56 \\
 70 & 143 & &  6.32 & 1.14 & 1.79 & 0.47 \\
 70 & 217 & &  5.04 & 0.92 & 1.39 & 0.44 \\
 70 & 353 & &  5.98 & 1.13 & 1.69 & 0.86 \\
 70 & 545 & &  5.23 & 0.99 & 1.50 & 0.98 \\
 70 & 857 & &  4.59 & 0.87 & 1.30 & 1.11 \\
100 & 100 & 0.34 &  7.76 & 1.47 & 2.21 & 0.51 \\
100 & 143 & &  5.36 & 0.98 & 1.52 & 0.48 \\
100 & 217 & &  4.26 & 0.78 & 1.18 & 0.47 \\
100 & 353 & &  4.36 & 0.82 & 1.23 & 0.73 \\
100 & 545 & &  3.75 & 0.70 & 1.06 & 0.81 \\
100 & 857 & &  3.25 & 0.61 & 0.91 & 0.88 \\
143 & 143 & 0.25 &  4.83 & 0.92 & 1.36 & 0.46 \\
143 & 217 & &  3.60 & 0.68 & 1.00 & 0.46 \\
143 & 353 & &  3.31 & 0.62 & 0.92 & 0.61 \\
143 & 545 & &  2.82 & 0.52 & 0.78 & 0.66 \\
143 & 857 & &  2.43 & 0.45 & 0.67 & 0.70 \\
217 & 217 & 0.20 &  3.22 & 0.61 & 0.90 & 0.44 \\
217 & 353 & &  2.70 & 0.50 & 0.75 & 0.55 \\
217 & 545 & &  2.31 & 0.42 & 0.63 & 0.59 \\
217 & 857 & &  1.99 & 0.36 & 0.55 & 0.62 \\
353 & 353 & 0.40 &  4.86 & 0.87 & 1.30 & 0.75 \\
353 & 545 & &  4.27 & 0.75 & 1.13 & 0.96 \\
353 & 857 & &  3.69 & 0.65 & 0.98 & 1.04 \\
545 & 545 & 0.60 &  5.79 & 1.00 & 1.49 & 1.07 \\
545 & 857 & &  5.16 & 0.89 & 1.33 & 1.36 \\
857 & 857 & 1.0 &  7.38 & 1.21 & 1.80 & 1.59 \\
\hline
\end{tabular}
\caption{\label{SN_radio_Planck} Auto- and cross-power spectra due to residual radio
    sources for \textit{Planck} according to the updated model for the flux cuts reported
    in the Table. Flux cut values correspond to those used to compute some conservative point-source masks inside the \textit{Planck} collaboration for consistency analysis.}
\label{tab_rad:2}
\end{table}

\subsubsection{Shot-noise level as a function of flux limits}

It can be useful to know the dependence of the shot-noise level from
residual radio sources on the flux cut $S_{lim}$. We considered
the \textit{Planck} frequencies, and a range of flux limits between 1\,mJy and
1\,Jy, that is, more or less the range covered by CMB experiments.

We start with auto-power spectra. We know that differential number
counts for radio sources scale approximately as $n(S)\propto S^{-2}$,
and power spectra as $C_{\ell}\propto S_{lim}$. Therefore it is
convenient to consider the quantity
$\mathcal{D}_{SN}=C_{\ell}/S_{lim}$. At a given frequency, we fit
$\mathcal{D}_{SN}\equiv\mathcal{D}_{SN}(S_{lim})$ as a double power
law:
\begin{equation}
\mathcal{D}_{SN}(S_{lim})=\frac{2A}{\bigg(\frac{S_{lim}}{S_0}\bigg)^{\alpha} +
\bigg(\frac{S_{lim}}{S_0}\bigg)^{\beta}}\,.
\label{e1}
\end{equation}
$\mathcal{D}_{SN}(S_{lim})$ from the updated model and the best fits
given by Eq.\,\ref{e1} are shown in Fig.\,\ref{f2}. The parameters of
the fits are provided in Table\,\ref{t3}.

Cross-power spectra depend on the flux cuts at the two considered frequencies. In order to describe $C^{\nu_1,\,\nu_2}_{\ell}\equiv
C^{\nu_1,\,\nu_2}_{\ell}(S_{lim}^{\nu_1},\,S_{lim}^{\nu_2}),$ we chose
to use a sixth-degree polynomial function. After computing cross-power
spectra in an uniform grid of $\log(S_{lim}/{\mathrm Jy})$ between $-3$
and 0, 
we determined the polynomial fit using the IDL routine
SFIT. For arbitrary flux limits (but always between 1\,mJy and 1\,Jy)
at frequencies $\nu_1$ and $\nu_2$, cross-power spectra can be
estimated by means of
%
%

\begin{equation}
\log\bigg[C^{\nu_1,\,\nu_2}_{\ell}(S_{lim}^{\nu_1},\,S_{lim}^{\nu_2})\bigg]=
\sum_{i,j=0}^6\,K_{i,\,j}\,\bigg[{\log(S^{\nu_1}_{lim})+3 \over
  0.2}\bigg]^j
\,\bigg[{\log(S^{\nu_2}_{lim})+3 \over 0.2}\bigg]^i\,,
\label{e2}
\end{equation}

where $K_{i,\,j}$ are the coefficients of the fit{\footnote{$K_{i,\,j}$ are provided at https://people.lam.fr/lagache.guilaine/Products.}.
We verified that the fit has a typical error of 2-3\%, with maximum errors of about 10--15\% (usually at the
borders of the grid). Figure\,\ref{f2} also shows examples of cross-power spectra and the corresponding fits when
$S_{lim}^{\nu_1}$ is fixed. 

\begin{table}
\centering
\begin{tabular}{ccccc}
\hline
$\nu$ & $\log(A)$ & $\log(S_0)$ & $\alpha$ & $\beta$ \\
\hline
  30  &  1.715 & -2.610 &  0.1658 &  -0.509 \\
  44  &  1.558 & -3.000  & 0.1223 &  -0.656 \\
  70  & 1.406  &-3.231 &  0.0967  & -0.754 \\
 100 &   1.290 & -3.307 &  0.0829 &  -0.966 \\
 143 &   1.240 & -3.293 &  0.0948 &  -0.769 \\
 217 &   1.204 & -3.173 &  0.1152 &  -0.479 \\
 353  &  1.118 & -3.035 &  0.1222 &  -0.410 \\
 545  &  1.094 & -1.639 &  0.2154 &  -0.198 \\
 857  &  0.991 & -1.012 &  0.2999 &  -0.161 \\
\hline
\end{tabular}
\caption{Best-fit parameters of Eq.\,\ref{e1} as a function of frequency. \label{t3}}
\end{table}

\begin{figure*}
\centering
\includegraphics[width=6cm]{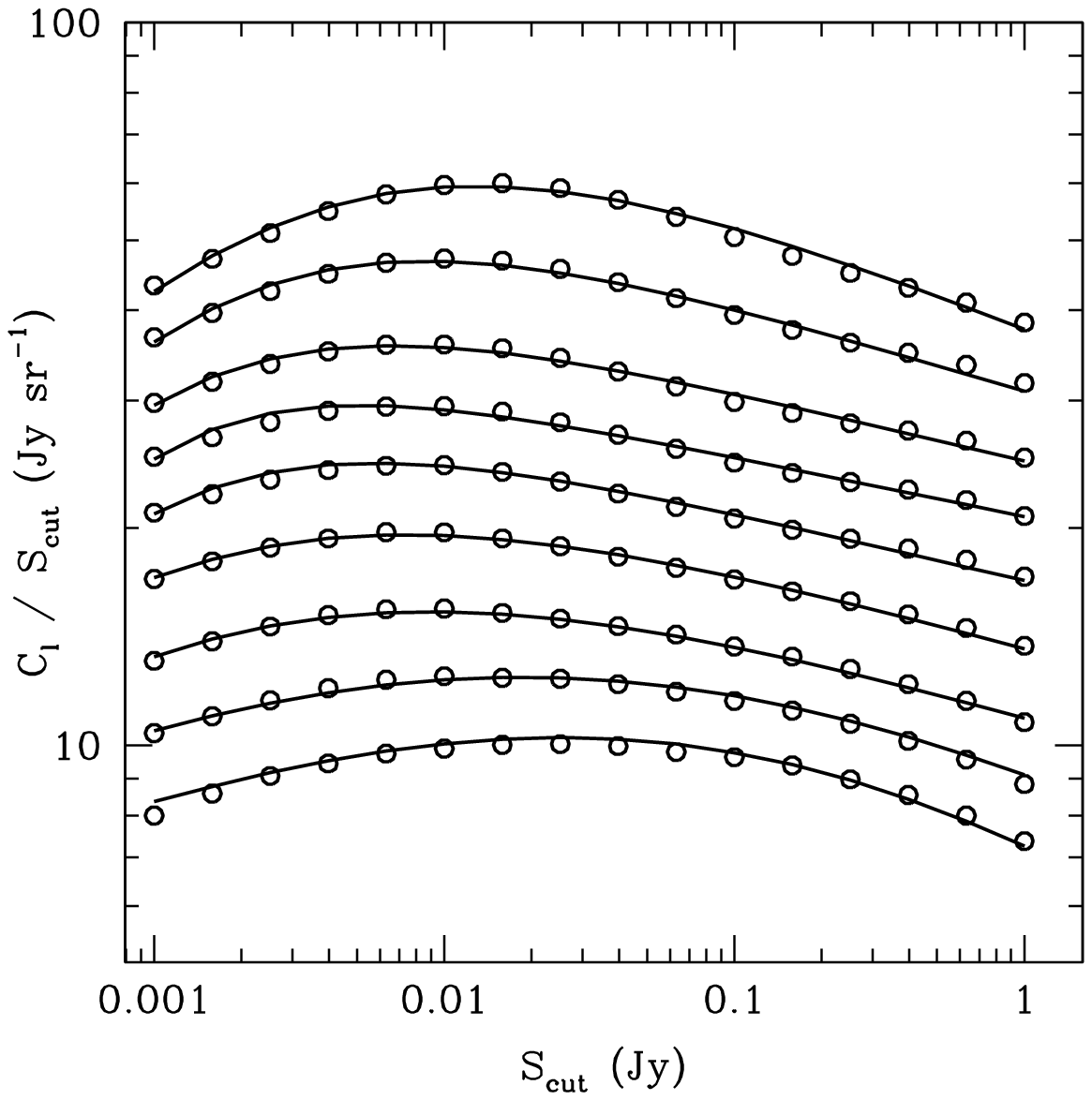}
\includegraphics[width=6cm]{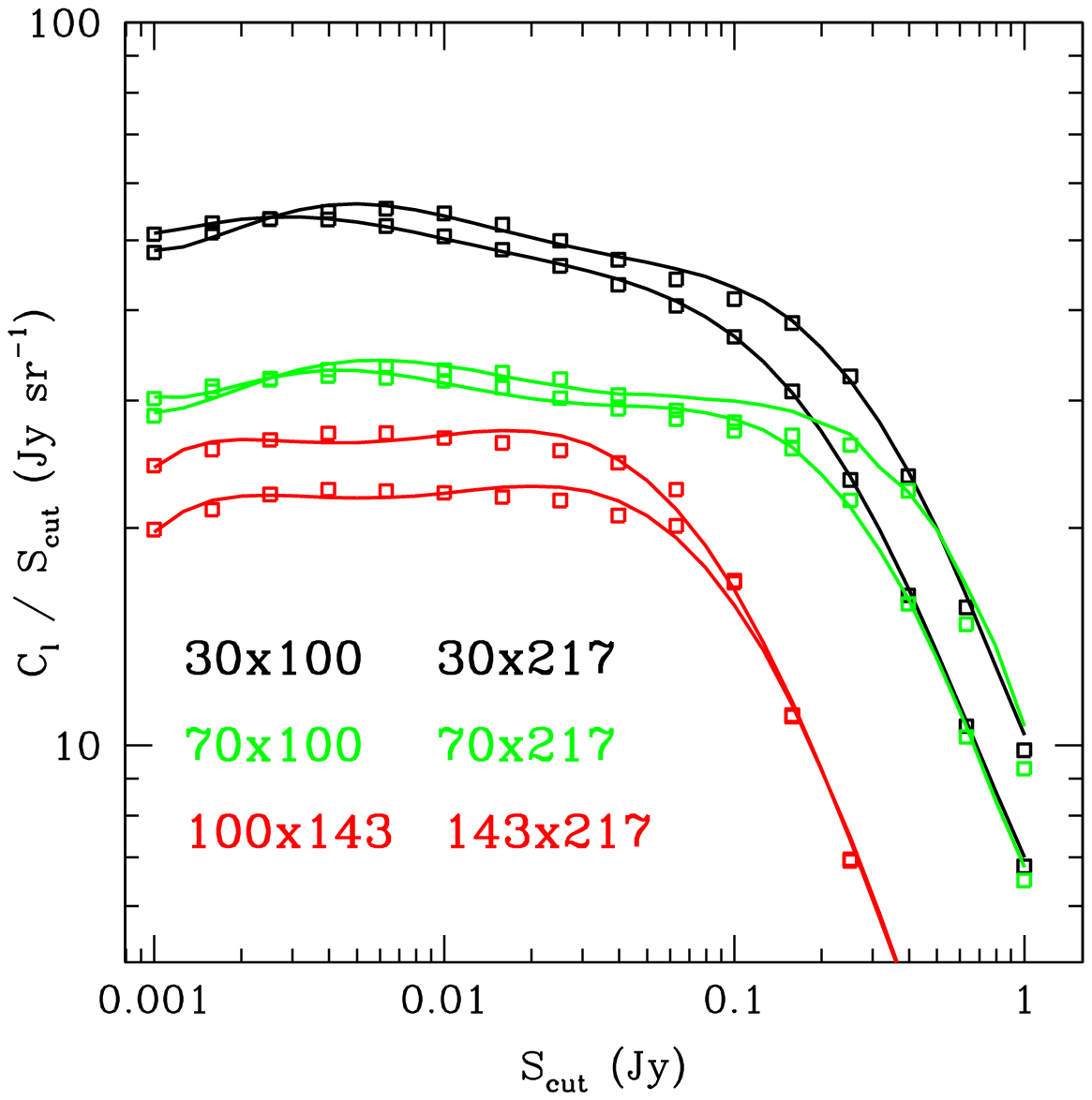}
\caption{({\it Left panel}) Power spectra (divided by the flux limit) of
  residual radio sources as a function of the flux limit from 30 to
  857\,GHz (from top to bottom). Points are from the updated T11 model; solid lines are fits
  using Eq.\,\ref{e1} with parameters given in Table\,\ref{t3}. ({\it Right panel})
  Cross-power spectra at the frequencies indicated in the figure as a
  function of the flux limit $S^{\nu_2}_{lim}$
  ($S^{\nu_1}_{lim}$ is equal to 0.4\,Jy for $\nu_1=30,\,70\,$GHz and
  0.1\,Jy for $\nu_1=100,\,143\,$GHz). Solid lines are obtained from
  Eq.\,\ref{e2}.}
\label{f2}
\end{figure*}
  
\section{Dusty star-forming galaxies \label{sec:DSFG_model}}
Similarly to the previous section, we present here the evolutionary model we chose to describe the number counts of DSFGs (Sect\,\ref{NC_DSFG}). We then discuss their polarised emission (Sect\,\ref{DSFG_polar}). Finally, we compute the shot noise using our model and compare it with recent observations (Sect\,\ref{SN_DSFG}).

\subsection{Model for the number counts \label{NC_DSFG}}
\label{mod:ir}
Since their discoveries in the 1990s, DSFGs have revolutionized the field of galaxy formation and evolution \citep[e.g.][]{casey2014}. 
The continuous advent of new experiments (either space-based -- \textit{ISO}, \textit{Spitzer},  \textit{Herschel},  \textit{Planck} -- or ground-based, e.g. SCUBA/JCMT, Laboca/APEX, IRAM, and ALMA) makes the study of high-z dusty galaxies one of the most important areas of extragalactic astronomy. Accompanying the new measurements, many empirical or semi-analytical models have been developed in the past 20 years  \citep[e.g.][]{lagache2003, bethermin2011, gruppioni2011, lapi2011, cai2013, casey2018, popping2020}. We chose to use the model of \cite{bethermin2012_model} here because it provides one of the best fits to the number counts from the mid-IR to radio wavelengths, including counts per redshift slice in the SPIRE bands. Moreover, it gives a reasonable CIB redshift-distribution, which is important for computing cross-power spectra \citep{bethermin2013}. Finally,  as it has been developed in-house, it can be run for numerous wavelengths and different bandpasses, which is mandatory for our analysis.

The model is based on the main assumption that star-forming galaxies have two modes of star formation: main sequence (MS) and starburst (SB). Main-sequence galaxies are secularly evolving galaxies with a tight
correlation between stellar mass ($M_{\star}$) and star formation rate (SFR) at a given redshift.
The evolution of MS and SB galaxies is based on the \cite{sargent2012} formalism, which jointly used  the mass function of
star-forming galaxies, the redshift evolution of the sSFR  (specific star formation rate, sSFR = SFR/$M_{\star}$),
and its distribution at fixed $M_{\star}$, with a separate contribution from MS and SB galaxies to reproduce IR luminosity functions.
The model uses redshift-dependent templates for the spectral energy distributions (SED) of MS and SB, based on fits of \cite{draineli2007} models to \textit{Herschel} observations of distant galaxies as presented in \cite{magdis2012}. Finally, as strongly lensed sources contribute $\sim$20\%  to (sub-)mm counts around 100\,mJy, magnification caused by strong lensing ($\mu>2$) is also included in the model (see  \cite{bethermin2012_model} for more details).

We show in Fig.\,\ref{counts_B12} the comparison of the model with some measured far-IR/sub-millimetre counts.  We also show the counts from \cite{bethermin2017}, obtained using an updated version of the two star-formation mode galaxy evolution model of \cite{bethermin2012_model}, combined with abundance matching  to populate a dark matter light cone and thus simulate the clustering.  \cite{bethermin2017} produced 2 deg$^2$ simulated maps (called SIDES) and extracted the sources as done in the observations. They convincingly showed that the limited angular resolution of single-dish instruments has a strong effect on far IR and sub-millimetre continuum observations. In particular, at 350 and 500\,$\mu$m, they reported that the number counts measured by Herschel between 5 and 50\,mJy are biased towards high values by a factor $\sim$2. When  these resolution effects are taken into account, they reproduce a large set of observables very well, such as number counts and their evolution with redshift and CIB power spectra. This demonstrates that any model should thus underestimate the measured single-dish number counts from $\sim$100 to 1000 $\mu$m in a given range of fluxes (see Fig.\,4 and 5 in \citealt{bethermin2017}). This is indeed the case for \citealt{bethermin2012_model} (Fig.\,\ref{counts_B12}), which agrees very well with the intrinsic SIDES model (and not with the observed SIDES counts). We also show in Fig.\,\ref{counts_B12} the recent counts obtained from the \textit{ALMA} ALPINE program \citep{bethermin2020} at 850\,$\mu$m, which are not affected by blending due to limited angular resolution, and agree well with the model.
 For bright fluxes ($\gtrsim$1\,Jy), the redshift grid of the model is too coarse to estimate the Euclidian plateau properly. We therefore directly computed the value of the plateau using Eq.\,6 of  \cite{planck_counts_2013}. Although it is mostly systematically $\sim 1\sigma$ lower, the model agrees to first order with the Euclidian plateau measured by \textit{Planck} \citep{planck_counts_2013}. For the purpose of this paper, number counts at such bright fluxes are not relevant, as their contribution to shot noise and confusion noise is negligible.  For example, at 272\,GHz (1.1\,mm), the confusion noise has converged for a flux cut of $\sim$10 mJy (i.e. the confusion noise for sources with flux $<$10\,mJy is nearly equal to that of sources with flux $<$10\,Jy). Therefore we are very confident in our use of the \cite{bethermin2012_model} model to compute the shot-noise levels from DSFG. We clearly validate the use of our model to compute the confusion noises in Sect.\,\ref{CF_validation}. The \cite{bethermin2017} model could not be used for this purpose as it does not give any analytical predictions and the volume of the dark-matter simulation is too small to derive accurate predictions for the large-volume surveys discussed here.

\begin{figure*}
\centering
\includegraphics[width=14cm]{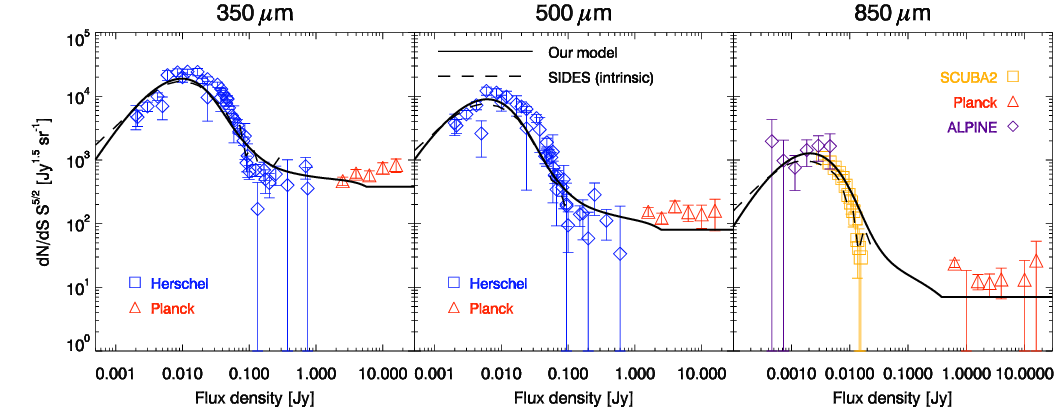}
\caption{Galaxy number counts at 350, 500, and 850\,$\mu$m. The model we used \citep{bethermin2012_model} is shown with the continuous line. It agrees very well with the most recent \cite{bethermin2017}  model (SIDES, long-dashed line).  Measurements are from \textit{Herschel} at 350 and 500 \,$\mu$m \citep{Olivier2010, Clements2010, Glenn2010, Bethermin2012_counts}, \textit{SCUBA2,} and \textit{ALMA} at 850\,$\mu$m \citep{geach2017,bethermin2020}, and \textit{Planck} at very bright fluxes \citep{planck_counts_2013}.  The models are below the \textit{Herschel} measurements at intermediate fluxes because these measurements are biased high due to the relatively low angular resolution combined with galaxy clustering (as demonstrated in \citealt{bethermin2017}).}
\label{counts_B12}
\end{figure*}

\subsection{Polarised emission \label{DSFG_polar}} 
Little is known about the polarisation emission of dusty galaxies. Dust enshrouding star-forming galaxies absorbs UV radiation from stars,  and re-emits light at longer wavelengths, which is responsible for the  far-IR SED of CIB galaxies. Thermal emission from interstellar dust in CIB galaxies, as in our Galaxy, is polarised because the dust grains are aligned with interstellar magnetic fields. The degree of polarisation is not very well known; 
it is likely to be low because the complex structure of galactic magnetic fields with reversals along the line of sight and the disordered alignment of dust grains reduce the global polarised flux when it is integrated over the whole galaxy.\\

Very few measurements exist for individual galaxies. Measurements at 850~$\mu$m of M82 by \cite{greaves2002} gave a global net polarisation degree of only 0.4\%, while Arp 220 measurements at 850~$\mu$m  by \cite{seiffert2007} gave a 99\% confidence upper limit of 1.54\%.
Using the stacking technique with \textit{Planck} data on a sample of $\sim$4700 DSFG, \cite{Bonavera17b} estimated the average fractional polarisation at 143, 217, and 353\,GHz.  They obtained a mean fractional polarisation $\langle \Pi \rangle$ of  3.10$\pm$0.75 and 3.65$\pm$0.66\%\ at 217 and 353\,GHz, respectively, after correcting for noise bias. The uncorrected value of  $\langle \Pi \rangle$ at 217\,GHz is  1.15$\pm$0.74\%, implying that the detection is significant at the 1.55$\sigma$ level. At 353\,GHz, the detection significance increases from  2.8$\sigma$ to 5.5$\sigma$ before and after the correction.  \cite{Trombetti2018} revisited these estimates, exploiting the intensity distribution analysis of the \textit{Planck} polarisation maps. They did not detect any polarisation signal from DSFG at a similarly high significance as \cite{Bonavera17b}. They derived a 90\% confidence upper limit on the median fractional polarisation at 353\,GHz of $\langle \Pi \rangle \lesssim$2.2\%. The upper limit at the same confidence level is looser at 217\,GHz, $\langle \Pi \rangle \lesssim$3.9\%, where dusty galaxies are substantially fainter. These upper limits are consistent with the median values reported
in \cite{Bonavera17b}, which are 1.3$\pm$0.7 and 2.0$\pm$0.8\%\ at 217 and 353\,GHz, respectively.
Recently, \cite{Gupta2019} identified 55 sources as DSFG in their SPT sample, and no polarisation signal was detected for these sources. Their 95\% confidence level upper limits are quite high and consistent with earlier results.
Finally, \cite{dezotti2018} made an estimate for spiral galaxies seen edge-on based on the average value of the Stokes Q parameter measured using the Planck dust polarisation maps of the Milky Way. They estimated a mean  polarisation degree averaged over all possible inclination angles of 1.4\%.
These low values of fractional polarisation are understood as due to the complex structure of galactic magnetic fields and to the disordered alignment of dust grains. 
To study the contamination from polarised emission of DSFG to the CMB B-modes (Sect.\,\ref{contrib_Bmode}), we accordingly adopted $\langle\Pi^{\mathrm{IR}}\rangle$=1.4\%.

\subsection{Shot-noise predictions \label{SN_DSFG}}
\citealt{bethermin2017} (see also \citealt{Negrello2005, valiante2016}) showed that counts obtained from single-dish antenna observations in the far-IR to mm are biased high  
because of source multiplicity and clustering in the large beams (10 to 30 arcsec). This may cause strong discrepancies between shot noises measured from the integral of the observed number counts and shot noises measured from CIB power spectra. For Herschel/SPIRE,   another complexity is introduced into the comparison: the beam profile and aperture efficiency vary across the passband and return a relative spectral response function (RSRF) that is different for point sources and extended emission. To compare model predictions to shot-noise measurements from CIB power spectra, we therefore also ran the model with the RSRF for extended source. Comparisons between model and observations are given in Tables\,\ref{sn_SPIRE} and \ref{HFI_SN} for \textit{Herschel}/SPIRE and \textit{Planck}/HFI, respectively. The shot-noise levels from observations are obtained either by fitting the CIB power spectra using the halo model \citep{viero2013, planckXXX} or by fitting the total power spectra using a parametric model and assuming a power law for the CIB \citep{mak2017}. In the first case, there is a strong degeneracy between the one-halo term and the shot noise, especially at the \textit{Planck} angular resolution. 

It is very difficult to derive any conclusion from Tables\,\ref{sn_SPIRE} and \ref{HFI_SN} because i) some measured values are incompatible (i.e. when the shot noise derived with a higher flux limit is lower than that derived with a lower flux limit). This is the case for \textit{Planck} at 545 and 353\,GHz and for \textit{Herschel} at the three wavelengths. ii) the model is not systematically higher or lower than the measurements. In the frequencies of interest ($\nu \lesssim$500\,GHz), observations and model predictions agree by 20\%, which we assume to be the uncertainty in our prediction. We stress that in contrast to the radio, a small variation in the flux limit S$_{lim}$ leads to only a small variation in shot-noise power. For example, changing S$_{lim}$ by 30\% leads to a variation of the shot-noise level seen by \textit{Planck} by less than 1\% at 217\,GHz \citep{Planck_CIB_2011}\\

\begin{table*}
\caption{\label{sn_SPIRE} \textit{Herschel}/SPIRE shot-noise levels as measured from CIB anisotropies and predicted using the integral of the number counts as modelled by \cite{bethermin2012_model}. Values for the shot noise are given in the photometric convention $\nu$I$_{\nu}$=cst, obtained using either the point source or the extended emission RSRF (see text for more details). Flux limits are coming from CIB power spectra analyses and are much higher than SPIRE sensitivity.}
\begin{center}
\begin{tabular}{c||cccc||cccc}
\hline
Wavelength & Flux limit$^1$  & Measured$^1$  & Predicted & Predicted & Flux limit$^2$  & Measured$^2$  & Predicted & Predicted \\
 & & & point source & extended & & &point source & extended  \\
\tiny{} [$\mu$m] & [mJy] &  [Jy$^2$ sr$^{-1}$] &  [Jy$^2$ sr$^{-1}$] &  [Jy$^2$ sr$^{-1}$]  & [mJy] &  [Jy$^2$ sr$^{-1}$] &  [Jy$^2$ sr$^{-1}$] &  [Jy$^2$ sr$^{-1}$] \\ \hline
250 & 300 & 8.2$\times$10$^3$ & 9983 & 9485 & 600 & $<$7063 & 11033 & 10455 \\ \hline
350 & 300 & 5.8$\times$10$^3$ & 5631 &  5122 & 600 & 4571  & 5929 & 5386 \\ \hline
500 & 300 & 2.3$\times$10$^3$ & 2193 & 1745 & 600 & 1518 & 2262 & 1799 \\ 
\hline
\end{tabular}
\end{center}
$^1$ From \cite{viero2013}.\\
$^2$ From \cite{serra2016}. 
\end{table*}


\begin{table*}
\begin{center}
\caption{\label{HFI_SN} Observed and predicted \textit{Planck}/HFI shot-noise levels. Values for the shot noise are given in the photometric convention $\nu$I$_{\nu}$=cst.}
\begin{tabular}{l||lll||lll}
\hline
Frequency & Flux limit$^1$ & Measured$^1$ & Predicted & Flux limit$^2$ & Measured$^2$& Predicted \\
\tiny{}[GHz] &  [mJy] & [Jy$^2$ sr$^{-1}$] &  [Jy$^2$ sr$^{-1}$] &  [mJy] &  [Jy$^2$ sr$^{-1}$] &  [Jy$^2$ sr$^{-1}$] \\ \hline
857 & 710 & 4966 & 5594 & 1000 & 5929 & 5761 \\ \hline
545 & 350 & 1859 & 1664 & 600 & 1539 & 1700  \\ \hline
353 & 315 & 315 &  275 & 400 & 226 & 277 \\ \hline
217 & 225 & 23  & 21 & -  & - & - \\ 
\hline
\end{tabular}
\end{center}
$^1$ From \cite{planckXXX}, shot noise from their Table\,9, corrected to $\nu$I$_{\nu}$=constant and corrected from the calibration difference between PR1 and PR2 data releases (at 545 and 857\,GHz). At 217\,GHz, the contribution from radio sources has also been removed.\\
$^2$ From \cite{mak2017}.
\end{table*}
  \section{Polarised shot noise from point sources: formalism \label{sec:polar_formalism}}

We explain below why we expect a polarisation term if galaxies have random orientations. We define the complex linear polarisation of a source with flux $S$,
 \begin{equation}
P_s = S \Pi  \exp(2i\psi)
,\end{equation}
where $\Pi$ is the fractional polarisation, and $\psi$ is the polarisation angle.

If the polarisation angles of different sources are uncorrelated, then
 \begin{equation}
<P_s > = 0 \,,
\end{equation}
but the variance is non-zero \citep{dezotti1999},
 \begin{equation}
 \sigma^2_P = \frac{1}{\pi} \int_0^\pi |\,P_s - <P_s>\,|^2 d\psi = S^2 \Pi^2 \,.
\end{equation}
 
We derive the shot-noise fluctuations of polarised point sources following \cite{tucci04}. For Poisson-distributed sources, the temperature power spectrum follows
\begin{equation}
C_\ell^{TT} = \int_0^{S_{limit}} {S^2 \frac{dN}{dS} dS} \,.
\end{equation}

We can consider a similar expression for the polarisation power spectrum,
\begin{equation}
C_\ell^{P} = \int_0^{P_{limit}} P^2 \frac{dN}{dP} dP \,,
\end{equation}
where $P=\sqrt{Q^2 + U^2}$ and $C_\ell^{P}  = C_\ell^{Q} + C_\ell^{U} = C_\ell^{EE} + C_\ell^{BB}$.\\

Because the emission will contribute equally to EE and BB on average, we can consider
\begin{equation}
C_\ell^{EE} = C_\ell^{BB} = \frac{1}{2}C_l^P \,.
\end{equation}

The power spectrum due to sources with a given fractional polarisation is
 \begin{equation}
C_\ell^P(\Pi) = \int_0^{\Pi S_{lim}}  P^2 \frac{dN}{dP} dP =\, \Pi^2  \int_0^{S_{lim}} {S^2 \frac{dN}{dS} dS} \,,
\end{equation}
assuming that $\Pi$ does not vary with S.
When the distribution of fractional polarisation for all sources is considered, the power spectrum becomes
 \begin{equation}
C_\ell^P = \int_0^1 {\mathcal{P}(\Pi) C_\ell^P(\Pi) d\Pi} = <\Pi^2> C_l^{TT} \,,
\label{eq_polar}
\end{equation}
where is $\mathcal{P}(\Pi)$ is the probability density function of fractional polarisation.\\

This formulation is very convenient, as $C_\ell^P$ is defined as a function of a flux cut derived in total intensity. Thus it assumes that sources are masked from polarisation maps using total intensity data. This is the case with current CMB experiments and will probably also be most likely the case with future CMB data with the use of higher angular resolution and sensitivity surveys to remove the source contamination. With this formulation, we can also consider different source populations with different fractional polarisations. \\

The probability density function can be constrained from the observed distributions of fractional polarisations. However,  because of the lack of constraints at CMB frequencies ($\sim$90-200\,GHz) for radio and dusty galaxies, we considered a fix polarisation fraction for each population (see Sect.\ref{contrib_Bmode}). 

\section{Clustering of dusty star-forming galaxies}
\label{extended_halo_model}
To compute polarisation power spectra due  to the clustering of CIB galaxies, we used the halo model, which provides a 
phenomenological description of the galaxy  clustering at all relevant angular scales \citep{cooray2002}. 
Assuming that all galaxies are located in virialised dark matter halos, the CIB clustering power spectrum is expressed as 
the sum of two components: a one-halo term, accounting for correlations
between galaxies in the same halo, and a two-halo term,  due to correlations between galaxies belonging to separated dark matter halos. 
The first term, together with the shot-noise power spectrum, dominates the small-scale clustering, 
and the second is prominent at large angular scales. Thus, 
the total CIB angular power spectrum at frequencies $\nu$ and $\nu^{\prime}$ can be written as
\begin{equation}
\label{eqn:ang_power_total}
C^{\nu\nu^{\prime}}_{\mathrm{tot}}(l) \equiv C^{\nu\nu^{\prime}}_{\mathrm{clust}}(l) + C^{\nu\nu^{\prime}}_{\mathrm{SN}} = C^{\nu\nu^{\prime}}_{\mathrm{1h}}(l) + C^{\nu\nu^{\prime}}_{\mathrm{2h}}(l) + C^{\nu\nu^{\prime}}_{\mathrm{SN}} \,.
\end{equation}
In the following section, after briefly introducing the model and its main 
parameters (we refer to \cite{shang2012,viero2013,planckXXX} 
for a detailed discussion), we show that the amplitudes 
of CIB polarisation power spectra are a small fraction of 
the one-halo term of the clustering spectra at most, and we derive upper limits 
on these amplitudes by fitting the model to 
current measurements of CIB angular power spectra 
from \textit{Herschel}/SPIRE \citep{viero2013}.

\subsection{Halo model with luminosity dependence}
In the Limber approximation \citep{limber1954}, the CIB clustering power spectrum at 
frequencies $\mathrm{\nu}$ and $\mathrm{\nu^{\prime}}$ is
\begin{eqnarray}
\label{eqn:clustering}
C^{\nu\nu^{\prime}}_{\mathrm{clust}}(l) &=& \int\frac{dz}{\chi^2}\frac{d\chi}{dz}a^2(z)\bar{j}(\nu,z)
\bar{j}(\nu^{\prime},z)P^{\nu\nu^{\prime}}(k=l/\chi,z),
\end{eqnarray}
where the term $\chi(z)$ denotes the comoving distance at redshift z, and 
$a(z)$ is the scale factor. 
The total emissivity from all CIB galaxies $\bar{j}_{\nu}(z)$ is computed from the 
luminosity function $dn/dL$ as                             
\begin{eqnarray}
\label{eqn:j0}
\bar{j}_{\nu}(z)&=&\int dL \frac{dn}{dL}(L,z)
\frac{L_{(1+z)\nu}}{4\pi},\end{eqnarray}
where the galaxy luminosity $L_{\nu(1+z)}$ 
is linked to the observed flux $S_{\nu}$ as 
\begin{eqnarray}
L_{\nu(1+z)} &=& \frac{4\pi\chi^2(z)S_{\nu}}{(1+z)}\,.
\end{eqnarray}
Finally, the term $P^{\nu\nu^{\prime}}(k,z)$ is the 3D
power spectrum of the emission coefficient, expressed as
\begin{eqnarray}
\label{eqn:3dpower}
\langle\delta\,j(\vec{k},\nu)\delta\,
j(\vec{k}^{\prime},\nu^{\prime})\rangle &=&
(2\pi)^3\bar{j}_{\nu}\bar{j}_{\nu^{\prime}}
P^{\nu\nu^{\prime}}_j\delta^3(\vec{k}-\vec{k^{\prime}}).
\end{eqnarray}
This term includes the two-halo and one-halo term. Expressing the luminosity of central and satellite galaxies as 
$L_{{cen},\nu(1+z)}(M_{H},z)$ and  $L_{{sat},\nu(1+z)}(m_{SH},z)$ (where $M_{H}$ and $m_{SH}$ denote the halo
and sub-halo masses, respectively), Eq.\,\ref{eq_jnu} can be written 
as the sum of the contributions from central and satellite galaxies as
\begin{eqnarray}
\label{eqn:j1}
\bar{j}_{\nu}(z)&=& \int dM \frac{dN}{dM}(z)\frac{1}{4\pi}
 \Big\{\frac{}{}N_{\mathrm cen}L_{{\mathrm cen},(1+z)\nu}(M_{\mathrm H},z)\\
\nonumber
& &+\int dm_{\mathrm SH} \frac{dn}{dm}(m_{\mathrm SH},z)
 L_{{\mathrm sat},(1+z)\nu}(m_{\mathrm SH},z)\Big\}\,.
\end{eqnarray}
Here $dN/dm$ and $dn/dm$ 
denote the halo and sub-halo mass function from \cite{tinker2008} and  \cite{tinker2010}, respectively, and 
$N_{\mathrm{cen}}$ is the number of central galaxies inside a halo, which 
was assumed to be equal to zero if the mass of the host halo is lower 
than $M_{\mathrm{min}} = 10^{11}$M$_{\odot}$ \citep{shang2012} and one otherwise.\\
Introducing $f^{\mathrm{cen}}_{\nu}$ and $f_{\nu}^{\mathrm{sat}}$ 
as the number of central and satellite galaxies weighted by their 
luminosity as
\begin{eqnarray}
f_{\nu}^{\mathrm cen}(M,z) = N_{\mathrm cen}
 \frac{L_{{\mathrm cen},(1+z)\nu}(M_{\mathrm H},z)}{4\pi},
\label{eqn:fcen}
\end{eqnarray}
\begin{eqnarray}
f_{\nu}^{\mathrm sat}(M,z) &=& \int_{M_{\mathrm min}}^{M}dm
 \frac{dn_{}}{dm}(m_{\mathrm SH},z|M) \\\nonumber
&& \times \frac{L_{{\mathrm sat},(1+z)\nu}(m_{\mathrm SH},z)}{4\pi},
\label{eqn:fsat}
\end{eqnarray}
the 3D CIB power spectrum at the observed frequencies
$\nu,\nu^{\prime}$ in Eq.~\ref{eqn:3dpower} can be expressed 
as the sum of one-halo term and two-halo term as
\begin{eqnarray}
\label{eqn:pj1h}
P^{}_{{\mathrm 1h},\nu\nu^{\prime}}(k,z)&=&
 \frac{1}{\bar{j}_{\nu}\bar{j}_{\nu^\prime}}\int_{M_{\mathrm min}}^{\infty}dM
 \frac{dN}{dM}\\\nonumber
&&\times\, \left\{f_{\nu}^{\mathrm cen}(M,z)f_{\nu^\prime}^{\mathrm sat}(M,z)u(k,M,z)
\right. \\\nonumber
&&\quad +f_{\nu^{\prime}}^{\mathrm cen}(M,z)f_{\nu}^{\mathrm sat}(M,z)
 u(k,M,z)\\\nonumber
&&\left.\quad +f_{\nu}^{\mathrm sat}(M,z)f_{\nu^{\prime}}^{\mathrm sat}(M,z)
 u(k,M,z)^2\right\},\\
\label{eqn:pj2h}
P^{}_{{\mathrm 2h},\nu\nu^{\prime}}(k,z)&=&
 \frac{1}{\bar{j}_{\nu}\bar{j}_{\nu^\prime}}D_{\nu}(k,z)
 D_{\nu^{\prime}}(k,z)P_{\mathrm lin}(k,z),
\end{eqnarray}
where
\begin{eqnarray}
D_{\nu}(k,z)&=&\int_{M_{\mathrm min}}^{\infty}dM\frac{dN}{dM}b(M,z)u(k,M,z)\\
\nonumber
&&\times\, \left\{f_{\nu}^{\mathrm cen}(M,z)+f_{\nu}^{\mathrm sat}(M,z)\right\}.
\label{eqn:pjf}
\end{eqnarray}
The term $u(k,M,z)$ is the Fourier transform of the halo 
density profile \citep{navarro1997} with a concentration parameter 
from \cite{duffy2010}, and $b(M,z)$ denotes the halo bias 
\citep{tinker2010}. The linear dark matter 
power spectrum $P_{lin}(k)$ in Eq.~\ref{eqn:pj2h} is computed using 
{\tt CAMB} (\url{http://camb.info/}).\\
The parametrisation of the term $L_{(1+z)\nu}(M,z)$ is the key ingredient 
of the model. Following \cite{shang2012}, we assumed a simple 
parametric function to describe the link between galaxy 
luminosity and its host dark matter halo, where the dependence of the 
galaxy luminosity on frequency, redshift, and halo mass is factorised in 
three terms as
\begin{eqnarray}
L_{(1+z)\nu}(M,z)= L_0 \Phi(z) \Sigma(M) \Theta[(1+z)\nu].
\label{eqn:lfunc}
\end{eqnarray}
The free normalisation parameter $L_0$ is constrained by the data and has no physical meaning.
The galaxy SED is modelled as \citep[see][and reference therein]{blain2002}
\begin{eqnarray}
\Theta (\nu,z) \propto
\left\{\begin{array}{ccc}
\nu^{\beta}B_{\nu}\,(T_{\mathrm d})&
 \nu<\nu_0\, ;\\
\nu^{-2}&  \nu\ge \nu_0\,;
\end{array}\right.
\label{eqn:thetanu}
\end{eqnarray}
where the Planck function $B_{\nu}$ has an emissivity 
index $\beta=1.5$, \citep{planck_dust2014,serra2016}. 
The power-law functional form at frequencies
$\mathrm{\nu\geq\nu_0}$ has previously been used in a number of similar analyses 
\citep{hall2010,viero2013,shang2012,planckXXX}, and it agrees better with observations than the exponential Wien tail.
The free parameter $\mathrm{T_d}$ is the mean temperature 
of the dust in CIB galaxies, averaged over the considered redshift range.  
We assumed a redshift-dependent, global normalisation of the $L$--$M$
relation of the form
\begin{eqnarray}
\Phi (z)= \left(1+z\right)^{\delta},
\label{eqn:phiz}
\end{eqnarray}
and we considered a log-normal function to describe the luminosity-mass 
relation as
\begin{eqnarray}
\Sigma(M) = M \frac{1}{(2\pi\sigma^2_{L/M})^{0.5}}\mathrm{exp}\Big[-\frac{(\mathrm{log}_{10}M - \mathrm{log}_{10}M_{\mathrm{eff}})^2}{2\sigma_{L/M}^2}\Big].
\end{eqnarray}
The term $\sigma_{L/M}$ (fixed to $\sigma_{L/M}=0.5$, as in 
\citealt{shang2012,viero2013,planckXXX,serra2016}) accounts for the range of halo
masses that contribute most to the IR luminosity. The 
parameter $M_{eff}$ describes a 
narrow range of halo masses around M$_{eff}\sim10^{12}$M$_{\odot}$ associated with a peak in 
the star-formation efficiency that is caused by various mechanisms that suppress star formation in
high and low halo masses \citep{benson2003,silk2003,bertone2005,croton2006,dekel2006,bethermin2012,behroozi2013}.

\subsection{Results}
We constrained the main parameters of our halo model 
using six measurements of CIB angular auto- and cross-power spectra at $250$, $350$, and $500$ $\mu$m 
from \textit{Herschel}/SPIRE \citep{viero2013} in the multipole range 
$200<l<23000$, and assumed the extended flux limit case. To further constrain the model, we also computed the star formation 
rate density in the range $\mathrm{0<z<6}$, and we fit to the compilation of star formation rate density measurements from 
\cite{madau2014}. 

We performed a Monte Carlo Markov chain (MCMC) analysis of the 
parameter space using a modification of the publicly available 
code {\tt CosmoMC} \citep{lewis2002}, and varied the following 
set of four halo model parameters:
\begin{eqnarray}
\mathscr{P} \equiv \{M_{\mathrm{eff}}, T_d, \delta, L_\mathrm{0} \},
\end{eqnarray}
together with six free parameters $\mathrm{A_{i=1,...6}}$ for 
the amplitudes of the shot-noise power spectra. 
We obtained a good fit to the data, with a total $\chi^2$ of $104.9$ for $97$ degrees of freedom. 
Mean values and marginalised limits for all free parameters used in the fit and comparison between
\textit{Herschel}/SPIRE measurements of the CIB power spectra with our best estimates of the one-halo, two-halo, and shot-noise, 
are shown in \cite{serra2016}. Shot noises derived from this model are very close to those found for the \cite{bethermin2017} simulations. This gives us confidence about the level of the one-halo term.

\begin{table*}
\caption{CMB space-based and balloon-borne experiments. From left to right: Experiment name, frequency, angular resolution, sky fraction, and instrument noise ($\sigma^P_{inst}$, in polarisation). The standard deviations ($\sigma$) in mJy give the contributions of instrument noise, radio and dusty (IR) galaxies, CIB clustering, and CMB, to the total noise ($\sigma_{tot}$) when a point-source flux is measured (in intensity). They are corrected for the flux lost by the aperture photometry procedure.  S$_{lim}$ is the point-source flux limit  (computed from $\sigma_{tot}$ using Eq.\,\ref{eq:Slim})}. SN$_{radio}$ and SN$_{IR}$ are the radio and dusty galaxy shot noises, respectively, corresponding to a flux cut equal to S$_{lim}$.
\centering
\tiny
\begin{tabular}{l|llll|llllll|lll}
\hline
Experiment & Freq. & FWHM & f$_{sky}$ & $\sigma^P_{inst}$ & $\sigma_{inst}$ & $\sigma_{rad}$ & $\sigma_{IR}$ & $\sigma_{clust}$ & $\sigma_{CMB}$ & $\sigma_{tot}$ & S$_{lim}$ & SN$_{radio}$ & SN$_{IR}$ \\
 & GHz  & arcmin & \% & $\mu$K$_{CMB}$.arcmin &  mJy & mJy & mJy & mJy & mJy & mJy & mJy & Jy$^2$/sr & Jy$^2$/sr \\
\hline
PLANCK &   30 &  32.30 & 100 &  210.00 &  8.21 &  28.18 &   1.53 &      1.56 &    104.20 &    108.30 &    541.40 &     22.75 &      0.07 \\
&     44 &     27.90 &    100 &    240.00 &     17.00 &     26.78 &      1.40 &      1.34 &    148.70 &    152.10 &    760.50 &     25.21 &      0.07 \\
&    70 &     13.10 &    100 &    300.00 &     23.35 &      8.03 &      0.66 &      0.81 &     61.26 &     66.06 &    330.30 &      9.02 &      0.06 \\
&    100 &      9.70 &    100 &    117.60 &     12.15 &      4.99 &      0.85 &      1.25 &     53.94 &     55.54 &    277.70 &      6.42 &      0.19 \\
&    143 &      7.20 &    100 &     70.20 &      8.63 &      2.96 &      1.68 &      2.21 &     40.35 &     41.46 &    207.30 &      4.04 &      1.29 \\
&   217 &      4.90 &    100 &    105.00 &     11.15 &      1.34 &      3.87 &      4.05 &     16.83 &     20.99 &    105.00 &      1.79 &     14.95 \\
&    353 &      4.90 &    100 &    438.60 &     28.69 &      1.58 &     14.53 &     17.33 &     10.46 &     38.03 &    190.20 &      2.47 &    209.40 \\
\hline
IDS &    150 &      7.20 &      3 &      5.50 &      0.71 &      2.95 &      1.93 &      2.56 &     42.01 &     42.24 &    211.20 &      4.03 &      1.72 \\
&    180 &      6.00 &      3 &      5.50 &      0.68 &      1.98 &      2.74 &      3.26 &     29.01 &     29.39 &    147.00 &      2.62 &      5.02 \\
&    220 &      4.40 &      3 &      9.50 &      0.91 &      0.96 &      3.60 &      3.49 &     12.06 &     13.13 &     65.64 &      1.14 &     16.01 \\
&    250 &      3.60 &      3 &     11.00 &      0.83 &      0.62 &      4.21 &      3.65 &      6.11 &      8.33 &     41.65 &      0.70 &     32.74 \\
&    280 &      4.90 &      3 &     16.00 &      1.51 &      1.21 &      7.90 &      8.85 &     14.94 &     19.17 &     95.87 &      1.47 &     62.41 \\
&    320 &      3.90 &      3 &     24.00 &      1.51 &      0.81 &      8.90 &      8.73 &      6.11 &     13.98 &     69.92 &      1.03 &    124.80 \\
&    360 &      3.20 &      3 &     41.00 &      1.67 &      0.63 &      9.76 &      8.61 &      2.50 &     13.37 &     66.87 &      0.94 &    223.10 \\
\hline
PIPER &    200 &     21.00 &     85 &     31.40 &     14.13 &     25.77 &     14.12 &     41.04 &    634.30 &    636.40 &   3182 &     36.03 &     10.82 \\
&    270 &     15.00 &     85 &     45.90 &     13.69 &     12.48 &     22.80 &     59.77 &    277.70 &    285.60 &   1428 &     16.50 &     55.07 \\
&    350 &     14.00 &     85 &    162.00 &     30.90 &      9.55 &     41.51 &    112.50 &    163.30 &    205.10 &   1026 &     11.11 &    210.00 \\
&    600 &     14.00 &     85 &   2659.2 &     53.56 &     11.12 &    132.00 &    388.60 &     17.24 &    414.40 &   2072 &     15.08 &   2125 \\
\hline
SPIDER &  94 &     42.00 &      7 &     11.00 &      4.51 &     82.96 &      5.34 &     13.25 &   1717 &   1719 &   8593 &     93.16 &      0.39 \\
&    150 &     30.00 &      7 &     14.00 &      7.52 &     44.93 &      8.65 &     27.86 &   1265 &   1266 &   6329 &     64.84 &      2.41 \\
\hline
LiteBIRD &  40 &     69.30 &    100 &     35.10 &      5.14 &    193.40 &      8.01 &      7.05 &    938.00 &    957.80 &   4789 &     99.25 &      0.17 \\
&     50 &     56.80 &    100 &     21.10 &      3.86 &    151.50 &      6.42 &      6.51 &   1122 &   1133 &   5663 &     98.33 &      0.18 \\
&     60 &     49.00 &    100 &     18.20 &      4.05 &    111.40 &      4.96 &      6.34 &   1167 &   1172 &   5862 &     92.26 &      0.18 \\
&     68 &     41.60 &    100 &     11.30 &      2.67 &     75.03 &      3.58 &      5.61 &    973.60 &    976.50 &   4883 &     78.92 &      0.18 \\
&     78 &     36.90 &    100 &      9.70 &      2.56 &     58.44 &      3.12 &      5.94 &    893.80 &    895.80 &   4479 &     71.37 &      0.20 \\
&     89 &     33.00 &    100 &      8.40 &      2.49 &     48.10 &      3.03 &      6.74 &    817.70 &    819.10 &   4096 &     63.61 &      0.25 \\
&    100 &     30.20 &    100 &      5.80 &      1.89 &     42.59 &      3.30 &      8.07 &    771.50 &    772.80 &   3864 &     57.73 &      0.35 \\
&    119 &     26.30 &    100 &      4.20 &      1.51 &     36.04 &      4.23 &     11.12 &    689.70 &    690.70 &   3454 &     49.22 &      0.68 \\
&    140 &     23.70 &    100 &      4.40 &      1.75 &     31.64 &      5.83 &     15.93 &    652.10 &    653.10 &   3266 &     43.97 &      1.50 \\
&    166 &     25.50 &    100 &      4.80 &      2.38 &     35.63 &     10.00 &     30.18 &    909.00 &    910.20 &   4551 &     50.11 &      3.94 \\
&    195 &     23.20 &    100 &      5.80 &      2.86 &     30.48 &     14.50 &     43.88 &    792.50 &    794.40 &   3972 &     42.24 &      9.56 \\
&    235 &     21.30 &    100 &      5.70 &      2.59 &     25.27 &     22.54 &     69.29 &    655.90 &    660.40 &   3302 &     33.72 &     26.83 \\
&    280 &     13.90 &    100 &      7.30 &      1.94 &     10.61 &     23.22 &     58.45 &    226.90 &    235.70 &   1178 &     13.91 &     66.67 \\
&    337 &     12.20 &    100 &      8.60 &      1.53 &      7.39 &     32.68 &     79.31 &    128.80 &    154.90 &    774.70 &      8.79 &    172.10 \\
&    402 &     10.80 &    100 &     15.80 &      1.61 &      5.80 &     44.02 &    101.80 &     62.44 &    127.40 &    637.00 &      6.93 &    399.00 \\
\hline
PICO &     21 &     38.40 &    100 &     19.10 &      0.44 &     35.70 &      2.00 &      2.44 &     83.31 &     90.69 &    453.40 &     23.56 &      0.07 \\
&     25 &     32.00 &    100 &     13.50 &      0.37 &     24.99 &      1.46 &      1.57 &     71.28 &     75.57 &    377.80 &     18.18 &      0.06 \\
&     30 &     28.30 &    100 &      8.30 &      0.28 &     21.97 &      1.28 &      1.23 &     73.55 &     76.79 &    383.90 &     16.71 &      0.06 \\
&     36 &     23.60 &    100 &      5.90 &      0.24 &     17.68 &      1.07 &      0.94 &     67.08 &     69.38 &    346.90 &     13.83 &      0.05 \\
&     43 &     22.20 &    100 &      5.70 &      0.30 &     17.40 &      1.04 &      0.93 &     81.80 &     83.64 &    418.20 &     14.84 &      0.05 \\
&     52 &     18.40 &    100 &      4.00 &      0.26 &     13.24 &      0.85 &      0.84 &     76.41 &     77.56 &    387.80 &     12.31 &      0.05 \\
&     62 &     12.80 &    100 &      4.40 &      0.27 &      6.97 &      0.56 &      0.60 &     47.11 &     47.63 &    238.20 &      7.10 &      0.05 \\
&     75 &     10.70 &    100 &      3.50 &      0.25 &      5.35 &      0.55 &      0.67 &     43.47 &     43.80 &    219.00 &      6.01 &      0.06 \\
&     90 &      9.50 &     100 &      2.10 &      0.18 &      4.53 &      0.65 &      0.89 &     44.03 &     44.28 &    221.40 &      5.45 &      0.11 \\
&    108 &      7.90 &    100 &      1.70 &      0.16 &      3.26 &      0.83 &      1.09 &     36.22 &     36.39 &    181.90 &      4.08 &      0.26 \\
&    129 &      7.40 &    100 &      1.50 &      0.17 &      3.01 &      1.27 &      1.68 &     38.50 &     38.67 &    193.40 &      3.97 &      0.71 \\
&    155 &      6.20 &    100 &      1.30 &      0.15 &      2.10 &      1.81 &      2.16 &     28.64 &     28.85 &    144.30 &      2.75 &      2.06 \\
&    186 &      4.30 &    100 &      3.50 &      0.32 &      0.91 &      2.15 &      1.98 &     10.83 &     11.26 &     56.29 &      1.08 &      5.97 \\
&    223 &      3.60 &    100 &      4.30 &      0.34 &      0.61 &      3.04 &      2.56 &      6.33 &      7.51 &     37.53 &      0.68 &     17.07 \\
&    268 &      4.20 &    100 &      2.60 &      0.22 &      0.87 &      5.97 &      5.89 &      9.64 &     12.81 &     64.06 &      1.02 &     48.55 \\
&    321 &      2.60 &    100 &      3.80 &      0.16 &      0.40 &      5.92 &      4.46 &      1.54 &      7.58 &     37.91 &      0.58 &    124.30 \\
&    385 &      2.50 &    100 &      3.30 &      0.09 &      0.45 &      8.90 &      6.86 &      0.89 &     11.28 &     56.41 &      0.79 &    303.80 \\
&    462 &      2.10 &    100 &      6.60 &      0.08 &      0.40 &     11.20 &      8.21 &      0.25 &     13.89 &     69.46 &      0.88 &    681.70 \\
&    555 &      1.50 &    100 &     46.50 &      0.16 &      0.27 &     11.41 &      7.47 &      0.03 &     13.64 &     68.19 &      0.80 &   1386 \\
&    666 &      1.30 &    100 &    164.00 &      0.14 &      0.24 &     13.35 &      8.44 &      0.0 &     15.80 &     78.99 &      0.85 &   2529 \\
&    799 &      1.10 &    100 &    816.00 &      0.12 &      0.21 &     14.47 &      8.58 &      0.0 &     16.82 &     84.11 &      0.84 &   4146 
\label{table:exp_space}
\end{tabular}
\end{table*}
\begin{table*}[!h]
\caption{Same as Table \ref{table:exp_space}, but for CMB ground-based experiments.}
\centering
\tiny
\begin{tabular}{l|llll|llllll|lll}
\hline
Experiment & Freq. & FWHM & f$_{sky}$ & $\sigma^P_{inst}$ & $\sigma_{inst}$ & $\sigma_{rad}$ & $\sigma_{IR}$ & $\sigma_{clust}$ & $\sigma_{CMB}$ & $\sigma_{tot}$ & S$_{lim}$ & SN$_{radio}$ & SN$_{IR}$ \\
 & GHz  & arcmin & \% & $\mu$K$_{CMB}$.arcmin &  mJy & mJy & mJy & mJy & mJy & mJy & mJy & Jy$^2$/sr & Jy$^2$/sr \\
\hline
C-BASS &      5 &     45.00 &    100 &   6000.00 &      9.32 &     87.02 &      - &      6.67 &      7.25 &     88.07 &    440.30 &     78.80 &      - \\
\hline
NEXT-BASS &      7 &     32.40 &    100 &    228.00 &      0.56 &     29.54 &      - &      2.70 &      6.57 &     30.39 &    151.90 &     24.84 &   -  \\
&      8 &     30.00 &    100 &    213.60 &      0.61 &     23.88 &      - &      2.19 &      6.71 &     24.91 &    124.60 &     18.33 &      - \\
&      9 &     27.60 &    100 &    204.60 &      0.69 &     20.20 &      - &      1.80 &      6.93 &     21.44 &    107.20 &     14.55 &      - \\
&     10 &     27.60 &    100 &    204.60 &      0.79 &     18.83 &      1.39 &      1.71 &      8.00 &     20.59 &    102.90 &     12.65 &      0.07 \\
&     11 &     25.20 &    100 &    195.60 &      0.93 &     16.24 &      1.23 &      1.42 &      8.55 &     18.47 &     92.37 &     10.58 &      0.06 \\
&     13 &     22.80 &    100 &    186.60 &      1.02 &     13.07 &      1.03 &      1.13 &      8.58 &     15.74 &     78.69 &      7.99 &      0.05 \\
&     14 &     22.80 &    100 &    196.20 &      1.36 &     13.17 &      1.02 &      1.09 &     10.92 &     17.23 &     86.13 &      8.12 &      0.05 \\
&     15 &     15.60 &    100 &     43.80 &      0.24 &      6.44 &      0.61 &      0.56 &      5.29 &      8.38 &     41.89 &      4.06 &      0.04 \\
&     17 &     13.20 &    100 &     38.40 &      0.22 &      4.59 &      0.47 &      0.40 &      4.62 &      6.54 &     32.70 &      2.89 &      0.03 \\
&     20 &     13.20 &    100 &     34.20 &      0.25 &      4.54 &      0.45 &      0.38 &      5.81 &      7.40 &     37.00 &      2.84 &      0.03 \\
&     22 &     10.80 &    100 &     39.00 &      0.29 &      3.15 &      0.34 &      0.26 &      4.56 &      5.57 &     27.85 &      2.05 &      0.02 \\
&     25 &     10.80 &    100 &     37.80 &      0.36 &      3.28 &      0.33 &      0.25 &      5.75 &      6.64 &     33.23 &      2.22 &      0.02 \\
&     28 &      8.40 &    100 &     36.00 &      0.33 &      2.02 &      0.23 &      0.15 &      3.87 &      4.38 &     21.91 &      1.39 &      0.02 \\
\hline
QUIJOTE &     11 &     55.20 &     16 &    300.00 &      2.76 &     95.94 &      5.09 &      7.07 &     55.10 &    111.00 &    555.00 &     43.52 &      0.12 \\
&     13 &     55.20 &     16 &    300.00 &      3.86 &     94.00 &      4.90 &      6.55 &     76.92 &    121.80 &    609.00 &     41.79 &      0.11 \\
&     17 &     36.00 &     16 &    300.00 &      4.29 &     28.24 &      1.73 &      2.26 &     46.12 &     54.32 &    271.60 &     17.87 &      0.07 \\
&     19 &     36.00 &     16 &    300.00 &      5.35 &     29.21 &      1.74 &      2.17 &     57.51 &     64.79 &    323.90 &     19.12 &      0.07 \\
&     30 &     22.20 &     16 &     60.00 &      1.61 &     14.29 &      0.95 &      0.85 &     40.76 &     43.24 &    216.20 &     10.01 &      0.04 \\
&     40 &     16.80 &     16 &     60.00 &      2.14 &      9.55 &      0.67 &      0.57 &     38.00 &     39.25 &    196.30 &      7.68 &      0.04 \\
\hline
AdvACTPOL &     90 &      2.20 &     50 &     11.00 &      0.22 &      0.13 &      0.12 &      0.07 &      0.48 &      0.56 &      2.80 &      0.08 &      0.07 \\
&    150 &      1.30 &     50 &      9.80 &      0.23 &      0.06 &      0.32 &      0.15 &      0.13 &      0.45 &      2.23 &      0.05 &      1.46 \\
&    230 &      0.90 &     50 &     35.40 &      0.69 &      0.06 &      0.79 &      0.36 &      0.04 &      1.11 &      5.53 &      0.11 &     18.39 \\
\hline
BICEP3+Keck[2018] & 95 &   24.00 &   1 &      2.10 &      0.50 &     29.58 &      2.37 &      4.97 &    399.20 &    400.30 &   2002 &     37.70 &      0.24 \\
&    150 &     30.00 &      1 &      2.70 &      1.45 &     44.93 &      8.65 &     27.86 &   1261 &   1262 &   6311 &     64.84 &      2.41 \\
\hline
BICEPArray[2023] & 30 &     76.00 &  1 &  5.60 &   0.52 &    182.30 &      7.65 &      6.94 &    550.40 &    579.90 &   2900 &     83.92 &      0.15 \\
&     40 &     57.00 &      1 &      6.20 &      0.75 &    142.60 &      5.96 &      5.56 &    739.50 &    753.20 &   3766 &     86.13 &      0.15 \\
&     95 &     24.00 &      1 &      1.00 &      0.24 &     29.58 &      2.37 &      4.97 &    399.20 &    400.30 &   2002 &     37.70 &      0.24 \\
&    150 &     15.00 &      1 &      1.00 &      0.28 &     13.77 &      4.28 &      9.56 &    249.30 &    249.90 &   1249 &     20.10 &      1.94 \\
&    220 &     11.00 &      1 &      4.40 &      1.05 &      7.31 &      9.27 &     18.41 &    149.60 &    151.20 &    755.90 &     10.61 &     17.06 \\
&    270 &      9.00 &      1 &      6.60 &      1.18 &      4.50 &     13.36 &     24.01 &     84.01 &     88.51 &    442.60 &      5.99 &     52.92 \\
\hline
CLASS &   38 &  90.00 & 70 &  39.00 &  6.72 &  158.70 &   6.61 &      6.04 &    740.30 &    757.20 &   3786 &     88.06 &      0.15 \\
& 93 &     40.00 &     70 &     10.00 &      3.83 &     73.16 &      4.68 &     11.50 &   1479 &   1481 &   7406 &     86.29 &      0.35 \\
&    148 &     24.00 &     70 &     15.00 &      6.34 &     32.51 &      6.89 &     19.31 &    711.40 &    712.40 &   3562 &     45.53 &      2.04 \\
&    217 &     18.00 &     70 &     43.00 &     16.77 &     19.05 &     15.10 &     41.49 &    453.10 &    455.90 &   2280 &     26.60 &     16.72 \\
\hline
SO-SAT &     27 &     91.00 &     10 &     49.50 &      4.43 &    136.50 &      5.92 &      5.84 &    372.30 &    396.60 &   1983 &     68.13 &      0.13 \\
&     39 &     63.00 &     10 &     29.70 &      3.77 &    171.40 &      7.12 &      6.39 &    817.60 &    835.40 &   4177 &     93.01 &      0.16 \\
&     93 &     30.00 &     10 &     3.70 &      1.05 &     41.25 &      2.88 &      6.47 &    675.70 &    677.00 &   3385 &     54.66 &      0.27 \\
&    145 &     17.00 &     10 &      4.70 &      1.37 &     17.49 &      4.47 &     10.63 &    317.90 &    318.60 &   1593 &     25.17 &      1.64 \\
&    225 &     11.00 &     10 &      8.90 &      2.12 &      7.27 &      9.89 &     19.74 &    149.30 &    151.20 &    755.80 &     10.49 &     19.40 \\
&    280 &      9.00 &     10 &     22.60 &      3.92 &      4.41 &     14.76 &     26.80 &     81.25 &     87.02 &    435.10 &      5.77 &     64.59 \\
\hline
SO-LAT &     27 &      7.40 &     40 &    100.40 &      0.73 &      1.49 &      0.19 &      0.12 &      2.50 &      3.00 &     15.02 &      0.97 &      0.02 \\
&     39 &      5.10 &     40 &     50.90 &      0.52 &      0.74 &      0.11 &      0.07 &      1.76 &      1.99 &      9.94 &      0.50 &      0.01 \\
&     93 &      2.20 &     40 &     11.30 &      0.24 &      0.13 &      0.13 &      0.08 &      0.50 &      0.59 &      2.96 &      0.09 &      0.09 \\
&    145 &      1.40 &     40 &     14.10 &      0.34 &      0.07 &      0.31 &      0.15 &      0.17 &      0.52 &      2.61 &      0.06 &      1.21 \\
&    225 &      1.00 &     40 &     31.10 &      0.67 &      0.07 &      0.83 &      0.38 &      0.06 &      1.14 &      5.68 &      0.11 &     16.40 \\
&    280 &      0.90 &     40 &     76.40 &      1.32 &      0.08 &      1.38 &      0.65 &      0.03 &      2.02 &     10.11 &      0.18 &     56.62 \\
\hline
SPT-3G &     95 &      1.60 &      6 &      6.00 &      0.10 &      0.06 &      0.10 &      0.05 &      0.16 &      0.22 &      1.12 &      0.03 &      0.09 \\
&    148 &      1.20 &      6 &      3.50 &      0.07 &      0.05 &      0.28 &      0.13 &      0.09 &      0.34 &      1.68 &      0.04 &      1.30 \\
&    223 &      1.00 &      6 &      6.00 &      0.13 &      0.06 &      0.79 &      0.37 &      0.06 &      0.89 &      4.45 &      0.08 &     15.05 \\
\hline
CMB-S4-SAT &   20 &     11.00 &     40 &      8.40 &      0.05 &      3.14 &      0.35 &      0.27 &      3.77 &      4.93 &     24.64 &      1.96 &      0.02 \\
&     30 &     72.80 &     40 &      3.50 &      0.31 &    181.70 &      7.62 &      6.92 &    548.80 &    578.20 &   2891 &     83.92 &      0.15 \\
&     40 &     72.80 &     40 &      4.50 &      0.69 &    197.60 &      8.18 &      7.20 &    958.30 &    978.50 &   4892 &     99.25 &      0.17 \\
&     85 &     25.50 &     40 &      0.90 &      0.19 &     31.25 &      2.10 &      3.94 &    387.70 &    388.90 &   1945 &     38.53 &      0.17 \\
&     95 &     25.50 &     40 &      0.80 &      0.20 &     32.54 &      2.53 &      5.41 &    463.10 &    464.30 &   2321 &     41.80 &      0.25 \\
&    145 &     22.70 &     40 &      1.20 &      0.47 &     29.54 &      6.15 &     16.73 &    613.70 &    614.60 &   3073 &     41.15 &      1.78 \\
&    155 &     22.70 &     40 &      1.30 &      0.54 &     29.76 &      7.41 &     20.64 &    654.00 &    655.00 &   3275 &     41.76 &      2.59 \\
&    220 &     13.00 &     40 &      3.50 &      0.99 &     10.25 &     11.07 &     24.80 &    220.80 &    222.70 &   1113 &     14.88 &     17.36 \\
&    270 &     13.00 &     40 &      6.00 &      1.56 &      9.44 &     19.58 &     46.53 &    202.90 &    209.30 &   1046 &     12.63 &     54.35 \\
\hline
CMB-S4-LAT &     30 &      7.40 &     40 &     30.80 &      0.28 &      1.54 &      0.19 &      0.12 &      3.07 &      3.45 &     17.24 &      1.03 &      0.02 \\
&     40 &      5.10 &     40 &     17.60 &      0.19 &      0.74 &      0.11 &      0.07 &      1.85 &      2.01 &     10.04 &      0.51 &      0.01 \\
&     95 &      2.20 &     40 &      2.90 &      0.06 &      0.13 &      0.14 &      0.08 &      0.52 &      0.56 &      2.82 &      0.08 &      0.10 \\
&    145 &      1.40 &     40 &      2.80 &      0.07 &      0.06 &      0.31 &      0.15 &      0.17 &      0.40 &      1.98 &      0.04 &      1.18 \\
&    220 &      1.00 &     40 &      9.80 &      0.21 &      0.06 &      0.76 &      0.36 &      0.06 &      0.87 &      4.37 &      0.09 &     13.97 \\
 &    270 &      0.90 &     40 &     23.60 &      0.42 &      0.07 &      1.23 &      0.58 &      0.03 &      1.42 &      7.12 &      0.13 &     44.55 
\label{table:exp_ground}
\end{tabular}
\end{table*}

\subsection{CIB power spectrum in polarisation \label{Polar_1h}}
The polarisation fraction $\Pi$ 
for a given intensity of dust emission I can be expressed in terms 
of the Stokes parameters Q and U as
\begin{eqnarray}
\Pi &=& \frac{\sqrt{(Q^2 + U^2)}}{\mathrm I},
\end{eqnarray}
where Q and U are related to the polarisation angle $\psi$, through
\begin{eqnarray}
\label{eqn:QU}
Q &=& \mathrm{I} \times  \Pi cos(2\psi) \\
U &=& \mathrm{-I} \times \Pi sin(2\psi).
\end{eqnarray}
Polarisation power spectra can be computed with 
the same formalism as we used to compute the 
CIB intensity power spectrum by substituting 
the galaxy luminosity $L_{(1+z)\nu}(M,z)$ for Q and U as
\begin{eqnarray}
L_{(1+z)\nu}(M,z) & \rightarrow & L^Q_{(1+z)\nu}(M,z) = L_{(1+z)\nu}(M,z) \Pi  cos(2\psi)\,\, \mathrm{(Q)} \\
L_{(1+z)\nu}(M,z) & \rightarrow & L^U_{(1+z)\nu}(M,z) = L_{(1+z)\nu}(M,z) \Pi  sin(2\psi)\,\, \mathrm{(U).}
\end{eqnarray} 
It is easy to see that if the polarisation among 
different sources is uncorrelated (as discussed in Sect.\,\ref{sect:intro}), the 
two-halo term cannot produce any polarisation 
power spectrum because computing it involves 
an average over the polarisation angle of all sources, which is zero.\\
The contribution from the one-halo term is slightly more complicated. The dark matter halos that contribute most to the CIB power spectra have a mass in the range $12.5<\mathrm{Log(M_H)}<13.5$, and the typical  number of satellite galaxies in this range is too small  (typically fewer than 5) to average the quantities $L^Q_{(1+z)\nu}(M,z)$ and $L^U_{(1+z)\nu}(M,z)$ to zero. 
As a result, when the one-halo contribution is computed, it is possible that terms proportional to 
\begin{eqnarray}
f_{\nu}^{\mathrm{sat}}(M,z)f_{\nu^\prime}^{\mathrm sat}(M,z) \Pi  ^2cos(2\psi)^2u(k,M,z)^2\,\,\mathrm{for\,\,Q}\\
f_{\nu}^{\mathrm{sat}}(M,z)f_{\nu^\prime}^{\mathrm sat}(M,z) \Pi  ^2sin(2\psi)^2u(k,M,z)^2\,\,\mathrm{for\,\,U}\,\end{eqnarray}
give a positive contribution to the polarisation power spectra. 
We did not consider here the terms proportional to $f^{\mathrm{sat}}f^{\mathrm cen}$ because it has been shown in simulations and observationally that the tidal field of a large central galaxy can torque its satellites such that the major axis of satellite galaxies points towards their hosts (see e.g. Fig.\,8 in \citealt{pereira08} or Fig. 6 in \citealt{Joachimi15}) and we therefore do not expect any polarised signal.
While accurate estimates of the amplitude of the  polarisation power spectrum would require numerical simulations, we here estimate the maximum contribution from the one-halo term, and we show that it is almost negligible with respect to the 
contribution from the shot noise (see Sect.\ref{compar_polar_Cell}). The maximum amplitude of 
polarisation can be obtained assuming the (unphysical) case where 
the polarisation angle $\psi$ of all sources is perfectly correlated 
and equal to zero (for Q) or $\mathrm{\pi/2}$ (for U). Assuming $\langle \Pi^{\mathrm{IR}} \rangle$  the mean fractional polarisation of all DSFG,  it is easy to see that the maximum amplitude of the  polarisation power spectra is simply $\langle \Pi^{\mathrm{IR}} \rangle ^2$  times the amplitude of the one-halo contribution to the CIB  intensity power spectrum, keeping only the term proportional to $f_{\nu}^{\mathrm{sat}}(M,z)^2$.  Thus, the EE of BB CIB power spectra are computed following:
\begin{equation}
C_\ell^{EE} = C_\ell^{BB} = \frac{1}{2} \times P_{\mathrm 1h}[\propto f^{\mathrm{sat}}(M,z)^2] \times \langle \Pi^{\mathrm{IR}} \rangle ^2 \,.
\end{equation}
Maximising the contribution of the one-halo term is supported by the evidence of strong clustering of dusty star-forming galaxy on sub-arcmin scales \citep{Chen2016} as well as the observed abundance of proto-cluster cores on such scales \citep{negrello2017}. Deriving the polarised CIB power spectrum by simply scaling the total (two- and one-halo) CIB power spectrum in temperature using a fractional polarisation (as done in \citealt{Curto2013} or \citealt{Trombetti2018}) obviously overestimates its contribution.

 \section{Confusion noise for future polarised experiments \label{sec:conf_noise}}
 Using our models for radio and DSFG number counts and for the CIB anisotropies, we can now compute the confusion noise and the point-source flux limit (Sect.\,\ref{Sect:conf}) for any CMB experiments, given their characteristics (Sect\,\ref{sect:CMB_expe} and Sect\,\ref{sect:CMB_conversions}). We describe our method and its validation in Sect\,\ref{CF_validation}, and we  discuss the contributions of the different components (instrument noise, radio, DSFG, CMB) to the point-source sensitivity limit in Sect.\,\ref{sect:contrib}. Section\,\ref{sect:spica} is dedicated to our predictions of confusion noise (in intensity and polarisation) for SPICA B-BOP.
   
 \subsection{Future CMB experiments \label{sect:CMB_expe}}
We considered all future CMB experiments, either already selected, funded, or in advanced discussion.  Their name, frequency, angular resolution, sky coverage, and instrument noise (in intensity) are given in Table \,\ref{table:exp_space} for balloon-borne and space-based experiments and in Table\,\ref{table:exp_ground} for ground-based experiments.  We also considered \textit{Planck} for reference and for cross-checks of our computations. The characteristics of each experiment were extracted from\\
- \cite{Planck_over_2018} for \textit{Planck},  \\
- \cite{CBASS} for C-BASS, \\
- \cite{LP2014} for QUIJOTE,  \\
- \cite{calabrese14} for AdvACTPOL,   \\
- \cite{Hui2018} for BICEP+keck and BICEPArray,  \\ 
- \cite{CLASS} for CLASS,  \\
- \cite{Errard2016} for PIPER, \\
- \cite{SO} for Simons Observatory, \\
- \cite{SPIDER}  for SPIDER, \\
- \cite{PICO} for PICO,\\
- \cite{CMB-S4} for CMB-S4,\\
- \cite{jaz2019a} for NEXT-BASS,  \\
- The following presentation for SPT-3G:\\
https://indico.fnal.gov/event/20244/session/6/contribution/69/\\
material/slides/0.pdf, \\
- The following presentation for LiteBIRD: \\
https://agenda.infn.it/event/15448/contributions/95798/\\
attachments/65895/80698/sugai\_public.pdf,\\
- The following presentation for IDS:\\
http://research.iac.es/congreso/cmbforegrounds18/media/talks/\\
day2/IDS\_v1.pdf.

\subsection{Unit conversions and bandpass corrections \label{sect:CMB_conversions}}
In the mm wavelength domain, two different units are often used. While for studies of Galactic emission or extragalactic  sources, the unit is Jansky ($Jy$), $K_{CMB}$ is the natural unit for CMB. Transforming $Jy$ into $K_{CMB}$ is not only a unit conversion, but also requires a colour correction (to account for the different spectral energy distribution that is implicitly assumed in the two units). This transformation is detailed in Appendix\,\ref{CC_UC}.
The conversion factors that are given in Tables\,\ref{table:Cell_space} and \,\ref{table:Cell_ground} assume a square bandpass, with a $\delta \nu$ and a central frequency $\nu$ given in the tables. Colour corrections are not computed for each experiment as it requires precisely knowing the bandpasses (e.g. for \textit{Planck}, assuming a square bandpass rather than the true bandpass leads to error in the colour corrections that are of the same order as the correction). Consequently, all the numbers given in the tables in $Jy$ are given for the true spectra (but $\sigma_{inst}$ and $\sigma_{CMB}$ , which are given for the convention $\nu I \nu=$constant, use the square bandpasses).

For current experiments with known bandpass, accurate unit conversions are given in Appendix\,\ref{CC_UC}. For current experiments, a comparison of foreground levels (CIB and SZ especially) also necessitates their extrapolation between nearby frequencies of different experiments. To this end, useful conversion factors are given in Appendix\,\ref{CC_UC}.

\subsection{Confusion noise and flux limit \label{Sect:conf}}  

As we showed in Eq.\,\ref{eq_polar}, we chose to use a flux cut in total intensity rather than in polarised intensity mainly for two reasons: i) we assumed that sources are removed or masked from polarisation maps using total intensity data, for which we could have a high-resolution survey complete to some level in total intensity, as opposed to the equivalent in polarised intensity \cite[e.g.][]{bat11, Datta2019},  and ii) source number counts in polarisation are very scarce, and more polarisation data are required to constrain $dN/dP$. By contrast, thanks to the numerous data in intensity obtained in the past decade, accurate modelling is available for number counts in intensity. 
Consequently, we computed the confusion noise and flux limit in intensity for each CMB experiment listed in Sect.\,\ref{sect:CMB_expe}.

\subsubsection{Method and validation \label{CF_validation}}
The confusion noise{\footnote{We only considered the confusion noise due to extragalactic sources because in the high Galactic latitude cosmological fields, the cirrus confusion noise is negligible, contributes very little to the total noise \citep{dole2003}, or can be mitigated using component separation methods.}} is usually defined as fluctuations of the background sky brightness below which sources cannot be detected individually.  These fluctuations are caused by intrinsically discrete extragalactic sources.  In the far-IR, sub-mm, and mm, the confusion noise is an important part of the
total noise budget because of the limited size of the telescopes compared to the wavelength. The confusion noise is even often greater than the instrument noise and therefore severely limits the survey depth \citep[e.g.][]{lagache2003, dole2003, negrello2004,Nguyen2010}.\\

When the flux of a point source is measured, the root mean square (rms) fluctuations due to extragalactic point sources are the sum of three components:
\begin{equation}
\label{eq:sigma}
\sigma_{conf}^2 = \sigma^2_{SNrad} +  \sigma^2_{SNir} +  \sigma^2_{Clus}  \,, 
\end{equation}
where $\sigma_{SNrad}$,  $\sigma_{SNir}$, and $\sigma_{Clus}$  are the rms fluctuations associated with the radio shot noise, dusty galaxy shot noise, and dusty galaxy clustering, respectively (we recall that clustering from radio sources is neglected, see Sect.\,\ref{sect:intro}). They are related to the power spectrum $P_{k}$ following
\begin{equation}
\sigma^2_i = \int 2 \pi k P_k^i T_{k} W_{k} dk \,,
\label{eq:sigma_Pk}
\end{equation}
where $W_{k}$ is the power spectrum of the beam (we assume Gaussian beams), and $i$ stands for $SNrad$, $SNir$, and $Clus$, respectively. $T_{k}$ is the transfer function linked to the flux measurement of the sources. We assumed that fluxes are measured using aperture photometry,  
\begin{equation}
f(r)= h_1\prod \left ( \frac{r}{2R_1} \right ) - h_2\prod \left (  \frac{r}{2R_2} \right ) \, ,
\end{equation}
where $\prod$ is the rectangular function, and $R_1$ and $R_2$ are the radii of the two circular apertures (with $R_2 > R_1$) and
\begin{equation}
h_{1}= \frac{R_{2}^2}{R_{2}^2-R_{1}^2}
\end{equation} 
\begin{equation}
h_{2}= \frac{R_{1}^2}{R_{2}^2-R_{1}^2} \,.
\end{equation} 
The Fourier transform of $f(r)$ is
\begin{equation}
F(k)= \pi R_{1}^2 \frac{2J_{1}(2 \pi kR_{1})}{2 \pi kR_{1}}h_{1} -
 \pi R_{2}^2 \frac{2J_{1}(2 \pi kR_{2})}{2 \pi kR_{2}}h_{2} \,,
\end{equation}
which gives the following power spectrum for our aperture photometry filter:
\begin{equation}
T_k=\left( \frac{\pi R_{1}^2 R_{2}^2}{R_{2}^2-R_{1}^2}  \right)^2 
\left[\frac{2J_{1}(2 \pi kR_{1})}{2 \pi kR_{1}}-
\frac{2J_{1}(2 \pi kR_{2})}{2 \pi kR_{2}} \right]^2
.\end{equation}

The confusion noise can be determined using two criteria, the so-called photometric and source density criteria \citep{dole2003,lagache2003}. The photometric case is derived from the fluctuations of the signal due to the sources below the detection threshold $S_{lim}$ in the beam.  The source density case is derived from a completeness limit and evaluates the density of the sources detected above the detection threshold $S_{lim}$, such that only a small fraction of sources is missed because they cannot be separated from their nearest neighbour.
The choice of the criterion depends on the shape of the source counts and the solid angle of the beam \citep{dole2003}. The transition between the two is at about 200\,$\mu$m, depending on telescope diameters \citep{lagache2003}. In this paper, we therefore use the photometric criterion. \\

The photometric criterion is related to the quality of the photometry
of detected sources, the flux measured near S$_{lim}$ being severely affected by fainter sources in the beam. It is defined by the implicit equation,
\begin{equation}
\label{eq:Slim}
S_{lim} = q_{phot} \times \sigma_{tot}(S_{lim}) \,,
\end{equation}
where q$_{phot}$ measures the photometric accuracy (we assume q$_{phot}$=5{\footnote{We chose a standard signal-to-noise ratio S/N=5$\sigma$, which is usually sufficient to obtain a reliability close from 100\% (e.g. $>$95\% at S/N = 5 in \citealt{planck_PCCS2}). It is extremely difficult to assess the reliability of a survey as a function of S/N before actual data are available because it is sensitive to many unknown parameters (non-Gaussian noise and systematics, non-Gaussian foregrounds, exact statistics of the sources, and choice of source extraction method). In addition, the exact threshold associated with a given reliability can also vary with regions in case of heterogeneous depth and/or foreground contamination, as for \textit{Planck}.}}), and S$_{lim}$ is the confusion limit. $\sigma_{tot}$ is defined as
\begin{equation}
\sigma_{tot} = \sqrt{F^2 \times [\sigma_{conf}^2 + \sigma^2_{CMB}] + \sigma^2_{inst} } \,, 
\end{equation}
where $\sigma_{conf}^2$ is given in Eq.\,\ref{eq:sigma} and $\sigma_{inst}$ is the instrument noise per beam (given in Tables\,\ref{table:exp_space} and \ref{table:exp_ground}). We also added the noise introduced by CMB fluctuations, $\sigma_{CMB}$, which is given by Eq.\,\ref{eq:sigma_Pk}, where we replaced $P_k$ by the power spectrum of the CMB. $F$ is a correction factor that accounts for the flux lost by the aperture photometry procedure (which does not cover the entire beam size). With our choice of $R_{1}$ and $R_{2}$ (see below), and assuming Gaussian beams, F$\simeq$3 for all experiments considered here. \\

In the range of confusion limits of CMB experiments, only  $P_{k}^{SNrad}$ and $P_{k}^{SNir}$ depend on $S_{lim}$. They are derived following
\begin{equation}
P_{k} = \int_0^{S_{lim}} S^2 \frac{dN}{dS} dS\,,
\end{equation}
where dN/dS are the number counts given by the models described in Sect.\,\ref{mod:rad} and Sect.\,\ref{mod:ir} for radio and dusty galaxies, respectively.\\

Confusion noises and flux limits are given in Tables\,\ref{table:exp_space} and \,\ref{table:exp_ground}. They were obtained using $R_1$=FWHM/2 and $R_2 =2 \times R_1$. \\

We confirmed that our confusion noises agree very well with those measured by \textit{ISO}/ISOPHOT,  \textit{Herschel}/SPIRE, and \textit{Planck}. For SPIRE, we obtain $\sigma_{conf}$=6.4, 6.6, and 5.3 mJy/beam, while \cite{Nguyen2010} measured 5.8$\pm$0.3, 6.3$\pm$0.4, and 6.8$\pm$0.4 mJy/beam at 250, 350 and 500 $\mu$m, respectively. For \textit{Planck}, we compared our flux limit to the flux cuts given in the PCCS2 source catalogue for  90\% completeness (in the extragalactic zone) in Table\,\ref{sigma_conf_planck}. This comparison is indicative as the  90\% completeness flux limit is not strictly equivalent to the confusion noise\footnotemark[\value{footnote}]. The overall agreement is better than $\sim$2$\sigma$. However, our flux cut is systematically below the PCCS2 flux limit for the highest frequencies (217, 353, 545, and 857\,GHz). We verified that this underestimate can be easily explained by the cirrus contamination, which may be quite high in the extragalactic zone (covering $|b|>30^{\mathrm o}$) and which is ignored in the present paper. 
Finally, we also verified our results for SPT by substituting $\sigma_{inst}$ from SPT-3G in the SPT-SZ survey. 
Considering $\sigma^{SPT-SZ}_{inst}$ = 2, 1.2, and 4\,mJy, we obtain S$_{lim}$ = 11, 7.1 and 20.5\,mJy, at 95, 150, and 220\,GHz, respectively, which agrees very well with  \cite{Mocanu2013} (see their Table 3, for 95\% completeness limit).The very good agreement with previous far-IR, sub-mm, and mm experiments gives us confidence in our computations.
 
\begin{table}
\begin{center}
\begin{tabular}{c|c|c|c}
\hline
Frequency & PCCS2 & This paper  & N \\ 
GHz & mJy & mJy \\ \hline
30 & 426$\pm$87 & 541  & + 1.3 \\
44 & 676$\pm$134 & 761 & + 0.6 \\
70 & 489$\pm$101 & 330 & - 1.6 \\
100 & 269$\pm$55 & 278 & + 0.2\\
143 & 177$\pm$35 & 207 & - 0.9\\
217 & 152$\pm$29 & 105 & - 1.6 \\
353 & 304$\pm$55 & 190 & - 2.1 \\
545 & 555$\pm$105 & 330 & - 2.1\\
857 & 791$\pm$168 & 569 & - 1.3\\ \hline
\end{tabular}
\end{center}
\caption{Flux limits for \textit{Planck} frequencies from the PCCS2 source catalogue \citep{planck_PCCS2} for 90\% completeness in the extragalactic zone and using our model. The last column gives the N$\sigma$ difference between the two estimates (considering only the uncertainty on the flux limit given for the PCCS2).}
\label{sigma_conf_planck}
\end{table}

\subsubsection{Contributions to the point-source sensitivity \label{sect:contrib}}

Ground-based experiments have a maximum frequency of 280\,GHz. The contribution of the different components to the point-source sensitivity mostly depends on the frequency and size of the telescope apertures.

The smallest telescopes, with sizes $<$1m (BICEP, CLASS, SO-SAT, and CMB-S4-SAT) or the low-frequency telescopes (C-BASS, NEXT-BASS, and QUIJOTE, with $\nu<40$\,GHz) have quite poor angular resolutions. The contribution of radio sources dominates up to $\sim$10-15GHz, then the confusion noise from the CMB becomes dominant. If we can remove the CMB, the CIB clustering dominates the noise budget at the higher frequencies ($\nu>$200\,GHz). Instrument noise is always much lower than the astrophysical components.

As expected, a telescope with a larger aperture returns lower flux limits because the confusion noise is much lower (and the instrument noise is generally lower as well). For larger aperture telescopes (AdvACTPOL, SO-LAT, SPT-3G, and CMB-S4-LAT), the instrument noise is at the same order of magnitude as confusion noises. For $\nu>$145\,GHz, the dominant contribution to the  $\sigma_{tot}$ comes from the shot noise of DSFG.\\

In space, telescopes have smaller apertures in general and instrument noise is always negligible compared to confusion noise. Confusion from the CMB always dominates, except at the highest frequencies ($\nu \gtrsim$300\,GHz). Except for the CMB, galaxy clustering above $\sim$150-200\,GHz contributes much. PIPER, SPIDER, and LiteBIRD have large S$_{lim}$ ($>$1Jy) that will consequently lead to a large contamination to the CMB-B mode measurements.

\subsubsection{The case of B-POP \label{sect:spica}}

We also considered the SPICA B-POP polarised experiment, which is at shorter wavelength. B-POP will provide 100-350\,$\mu$m images of linearly polarised dust emission with an angular resolution, signal-to-noise ratio, and dynamic ranges comparable to those achieved by Herschel images of the cold ISM in total intensity. The angular resolution of B-BOP at 200\,$\mu$m will also be a factor $\sim$30 better than Planck polarisation data.

At these wavelengths and with this high angular resolution, only the shot noise of DSFG contributes to the confusion noise ($\sigma_{conf}$). Flux limits are about 0.4, 19.6, and 35.3\,mJy at 100, 200, and 350\,$\mu$m, respectively (see Table\,\ref{tab:B-BOP}). This is sightly above the SPIRE/Herschel 350\,$\mu$m flux limit due to the smaller telescope aperture.
For one pointing (2.5'$\times$2.5'), confusion noise levels are reached in 9.9, 0.02, and 0.02\,seconds at 100, 200, and 350\,$\mu$m, respectively\footnote{These values were computed using the \citet{Andre2019} sensitivity forecasts (see their Table\,1). They correspond to the time needed to reach $\sigma_{inst}$ = $\sigma_{conf}$.}. For a 1 Sq. Deg. survey, they are reached in 1.6\,hours, 9.7\,seconds, and 12.1\,seconds. This shows that the 200 and 350\,$\mu$m maps, even on large areas, will be severely limited in depth by extragalactic confusion.\\

In polarisation, after masking all the sources detected in intensity, up to $S_{lim}$, the r.m.s of polarised intensity due to confusion is
\begin{equation}
\sigma^P_{conf} = \sqrt{(\sigma^Q_{conf})^2 + (\sigma^U_{conf})^2} = \sigma_{conf} \times \langle \Pi^{\mathrm{IR}} \rangle \,.
\end{equation}
Assuming a fractional polarisation for DSFG $\langle \Pi^{\mathrm{IR}} \rangle$=1.4\% (see Sect.\,\ref{DSFG_polar}) and $\sigma^Q_{conf} = \sigma^U_{conf}$, we obtain a confusion noise in polarisation $\sigma^{Q,U}_{conf}$= 0.79,  38.6,  70.3\,$\mu$Jy  after masking all the sources detected in intensity at 100, 200, and 350\,$\mu$m, respectively. These $\sigma^{Q,U}_{conf}$ levels are reached in 57\,hours, 5.8\,minutes, and 7.0\,minutes for a single pointing, and 33\,737, 57, and 69\,hours for a 1 Sq. Deg. survey, at 100, 200, and 350\,$\mu$m, respectively. In polarisation, confusion is therefore not expected to be reached at 100\,$\mu$m, but could be reached for the deepest integrations at longer wavelengths. Confusion from galaxies could ultimately limit the sensitivity of the high-latitude polarimetric deep surveys of the interstellar medium of our Galaxy at 200 and 350\,$\mu$m.

\begin{table}
\begin{center}
\begin{tabular}{c|c|c|c|c}
\hline
$\lambda$  & FWHM & $\sigma_{conf}$ &  S$_{lim}$& SN$_{IR}$\\
$\mathrm{\mu}$m & arcsec & mJy & mJy & Jy$^2$/sr \\ \hline
100 &  9 & 8.0$\times$10$^{-2}$ & 0.40 & 6.4 \\
200 &  18 &  3.9 & 19.6 &  3.9$\times$10$^{3}$ \\
350 &  32 &  7.1 & 35.3 & 4.1$\times$10$^{3}$ \\
\hline
\end{tabular}
\end{center}
\caption{Confusion noise, flux limit, and DSFG shot noise level for the SPICA B-POP experiment.}
\label{tab:B-BOP}
\end{table}

\begin{table*}[!h]
\caption{C$_{\ell}^{BB}$ of the extragalactic foreground components for space-based and balloon-borne experiments: radio galaxies, dusty galaxies (IR) , and CIB one-halo (completely negligible for $\nu \le 90$\,GHz and thus not computed). They are given in Jy$^2$/sr. The unit conversion factor is also given (C = MJy sr$^{-1}$[$\nu I_{\nu}=$~constant] K$_{\mathrm CMB}^{-1}$). The power spectra in Jy$^2$/sr have to be divided by C$^2$ to obtain power spectra in $\mu$K$^2_{\mathrm CMB}$.}
\centering
\tiny
\begin{tabular}{l|lll|ccccc}
\hline
Experiment & $\nu$ & $\delta\nu$ & $C$ & C$_{\ell}^{BB}$ Radio &  C$_{\ell}^{BB}$ IR &  C$_{\ell}^{BB}$ CIB ($\ell$=80) &  C$_{\ell}^{BB}$ CIB ($\ell$=1000) &  C$_{\ell}^{BB}$ CIB ($\ell$=4000)  \\
 & GHz & \% & & Jy$^2$/sr & Jy$^2$/sr &  Jy$^2$/sr & Jy$^2$/sr & Jy$^2$/sr  \\
\hline
PLANCK &     30 &     30 &     26.81 &     8.918$\, 10^{-3}$ &     6.542$\, 10^{-6}$ &     -- &    -- &    -- \\
&     44 &     30 &     56.17 &     9.882$\, 10^{-3}$ &     6.766$\, 10^{-6}$ &     -- &     -- &     -- \\
&     70 &     30 &    131.85 &     3.537$\, 10^{-3}$ &     5.973$\, 10^{-6}$ &     -- &     -- &     -- \\
&    100 &     30 &    237.01 &     2.515$\, 10^{-3}$ &     1.840$\, 10^{-5}$ &     4.517$\, 10^{-6}$ &     4.367$\, 10^{-6}$ &     4.032$\, 10^{-6}$ \\
&    143 &     30 &    377.14 &     1.583$\, 10^{-3}$ &     1.267$\, 10^{-4}$ &     4.387$\, 10^{-5}$ &     4.212$\, 10^{-5}$ &     3.835$\, 10^{-5}$ \\
&    217 &     30 &    480.18 &     7.009$\, 10^{-4}$ &     1.465$\, 10^{-3}$ &     4.986$\, 10^{-4}$ &     4.724$\, 10^{-4}$ &     4.189$\, 10^{-4}$ \\
&    353 &     30 &    294.65 &     9.675$\, 10^{-4}$ &     2.052$\, 10^{-2}$ &     6.724$\, 10^{-3}$ &     6.164$\, 10^{-3}$ &     5.125$\, 10^{-3}$ \\
\hline
IDS &    150 &     30 &    395.55 &     1.581$\, 10^{-3}$ &     1.684$\, 10^{-4}$ &     5.998$\, 10^{-5}$ &     5.751$\, 10^{-5}$ &     5.223$\, 10^{-5}$ \\
&    180 &     30 &    454.58 &     1.027$\, 10^{-3}$ &     4.920$\, 10^{-4}$ &     1.671$\, 10^{-4}$ &     1.595$\, 10^{-4}$ &     1.435$\, 10^{-4}$ \\
&    220 &     30 &    480.08 &     4.473$\, 10^{-4}$ &     1.569$\, 10^{-3}$ &     5.321$\, 10^{-4}$ &     5.040$\, 10^{-4}$ &     4.466$\, 10^{-4}$ \\
&    250 &     30 &    463.95 &     2.754$\, 10^{-4}$ &     3.209$\, 10^{-3}$ &     1.106$\, 10^{-3}$ &     1.041$\, 10^{-3}$ &     9.097$\, 10^{-4}$ \\
&    280 &     30 &    426.14 &     5.747$\, 10^{-4}$ &     6.116$\, 10^{-3}$ &     2.008$\, 10^{-3}$ &     1.877$\, 10^{-3}$ &     1.620$\, 10^{-3}$ \\
&    320 &     30 &    356.86 &     4.049$\, 10^{-4}$ &     1.223$\, 10^{-2}$ &     4.052$\, 10^{-3}$ &     3.750$\, 10^{-3}$ &     3.174$\, 10^{-3}$ \\
&    360 &     30 &    281.64 &     3.683$\, 10^{-4}$ &     2.186$\, 10^{-2}$ &     7.369$\, 10^{-3}$ &     6.745$\, 10^{-3}$ &     5.589$\, 10^{-3}$ \\
\hline
PIPER &    200 &     30 &    474.77 &     1.412$\, 10^{-2}$ &     1.060$\, 10^{-3}$ &     3.087$\, 10^{-4}$ &     2.935$\, 10^{-4}$ &     2.622$\, 10^{-4}$ \\
&    270 &     30 &    440.61 &     6.468$\, 10^{-3}$ &     5.397$\, 10^{-3}$ &     1.677$\, 10^{-3}$ &     1.570$\, 10^{-3}$ &     1.361$\, 10^{-3}$ \\
&    350 &     16 &    301.91 &     4.355$\, 10^{-3}$ &     2.058$\, 10^{-2}$ &     6.448$\, 10^{-3}$ &     5.916$\, 10^{-3}$ &     4.925$\, 10^{-3}$ \\
&    600 &     10 &     31.88 &     5.911$\, 10^{-3}$ &     2.083$\, 10^{-1}$ &     7.684$\, 10^{-2}$ &     6.487$\, 10^{-2}$ &     4.589$\, 10^{-2}$ \\
\hline
SPIDER &     94 &     24 &    216.11 &     3.652$\, 10^{-2}$ &     3.789$\, 10^{-5}$ &     1.595$\, 10^{-6}$ &     1.554$\, 10^{-6}$ &     1.457$\, 10^{-6}$ \\
&    150 &     24 &    396.64 &     2.542$\, 10^{-2}$ &     2.359$\, 10^{-4}$ &     5.998$\, 10^{-5}$ &     5.751$\, 10^{-5}$ &     5.223$\, 10^{-5}$ \\
\hline
LiteBIRD &     40 &     30 &     46.82 &     3.891$\, 10^{-2}$ &     1.667$\, 10^{-5}$ &     -- &     -- &     -- \\
&     50 &     30 &     71.49 &     3.855$\, 10^{-2}$ &     1.731$\, 10^{-5}$ &     -- &     -- &     -- \\
&     60 &     23 &    100.42 &     3.617$\, 10^{-2}$ &     1.796$\, 10^{-5}$ &     -- &    -- &     -- \\
&     68 &     23 &    125.69 &     3.094$\, 10^{-2}$ &     1.765$\, 10^{-5}$ &    -- &     -- &     -- \\
&     78 &     23 &    159.42 &     2.798$\, 10^{-2}$ &     1.993$\, 10^{-5}$ &    -- &   -- &     -- \\
&     89 &     23 &    198.26 &     2.494$\, 10^{-2}$ &     2.476$\, 10^{-5}$ &     -- &    -- &     -- \\
&    100 &     23 &    237.76 &     2.263$\, 10^{-2}$ &     3.390$\, 10^{-5}$ &     4.517$\, 10^{-6}$ &     4.367$\, 10^{-6}$ &     4.032$\, 10^{-6}$ \\
&    119 &     30 &    303.30 &     1.929$\, 10^{-2}$ &     6.655$\, 10^{-5}$ &     1.377$\, 10^{-5}$ &     1.327$\, 10^{-5}$ &     1.219$\, 10^{-5}$ \\
&    140 &     30 &    368.76 &     1.724$\, 10^{-2}$ &     1.465$\, 10^{-4}$ &     3.696$\, 10^{-5}$ &     3.553$\, 10^{-5}$ &     3.240$\, 10^{-5}$ \\
&    166 &     30 &    431.24 &     1.964$\, 10^{-2}$ &     3.864$\, 10^{-4}$ &     1.082$\, 10^{-4}$ &     1.035$\, 10^{-4}$ &     9.349$\, 10^{-5}$ \\
&    195 &     30 &    471.16 &     1.656$\, 10^{-2}$ &     9.365$\, 10^{-4}$ &     2.733$\, 10^{-4}$ &     2.600$\, 10^{-4}$ &     2.325$\, 10^{-4}$ \\
&    235 &     30 &    475.27 &     1.322$\, 10^{-2}$ &     2.629$\, 10^{-3}$ &     7.830$\, 10^{-4}$ &     7.390$\, 10^{-4}$ &     6.504$\, 10^{-4}$ \\
&    280 &     30 &    426.14 &     5.453$\, 10^{-3}$ &     6.534$\, 10^{-3}$ &     2.008$\, 10^{-3}$ &     1.877$\, 10^{-3}$ &     1.620$\, 10^{-3}$ \\
&    337 &     30 &    324.81 &     3.446$\, 10^{-3}$ &     1.687$\, 10^{-2}$ &     5.305$\, 10^{-3}$ &     4.886$\, 10^{-3}$ &     4.098$\, 10^{-3}$ \\
&    402 &     23 &    209.33 &     2.715$\, 10^{-3}$ &     3.910$\, 10^{-2}$ &     1.269$\, 10^{-2}$ &     1.147$\, 10^{-2}$ &     9.275$\, 10^{-3}$ \\
\hline
PICO &     21 &     25 &     13.33 &     9.236$\, 10^{-3}$ &     7.227$\, 10^{-6}$ &     -- &     -- &     -- \\
&     25 &     25 &     18.80 &     7.127$\, 10^{-3}$ &     6.041$\, 10^{-6}$ &     -- &     -- &     -- \\
&     30 &     25 &     26.88 &     6.550$\, 10^{-3}$ &     5.586$\, 10^{-6}$ &     -- &     -- &     -- \\
&     36 &     25 &     38.31 &     5.421$\, 10^{-3}$ &     4.964$\, 10^{-6}$ &     -- &     -- &     -- \\
&     43 &     25 &     53.89 &     5.817$\, 10^{-3}$ &     5.160$\, 10^{-6}$ &     -- &     -- &     -- \\
&     52 &     25 &     77.10 &     4.826$\, 10^{-3}$ &     4.953$\, 10^{-6}$ &     -- &     -- &     -- \\
&     62 &     25 &    106.48 &     2.782$\, 10^{-3}$ &     4.479$\, 10^{-6}$ &     -- &     -- &     -- \\
&     75 &     25 &    148.98 &     2.354$\, 10^{-3}$ &     6.148$\, 10^{-6}$ &     -- &     -- &     -- \\
&     90 &     25 &    201.67 &     2.136$\, 10^{-3}$ &     1.114$\, 10^{-5}$ &     -- &     -- &     -- \\
&    108 &     25 &    266.03 &     1.599$\, 10^{-3}$ &     2.575$\, 10^{-5}$ &     8.413$\, 10^{-6}$ &     8.117$\, 10^{-6}$ &     7.465$\, 10^{-6}$ \\
&    129 &     25 &    336.61 &     1.558$\, 10^{-3}$ &     6.925$\, 10^{-5}$ &     2.447$\, 10^{-5}$ &     2.354$\, 10^{-5}$ &     2.152$\, 10^{-5}$ \\
&    155 &     25 &    408.65 &     1.080$\, 10^{-3}$ &     2.015$\, 10^{-4}$ &     7.149$\, 10^{-5}$ &     6.851$\, 10^{-5}$ &     6.214$\, 10^{-5}$ \\
&    186 &     25 &    463.34 &     4.249$\, 10^{-4}$ &     5.855$\, 10^{-4}$ &     2.096$\, 10^{-4}$ &     1.997$\, 10^{-4}$ &     1.791$\, 10^{-4}$ \\
&    223 &     25 &    480.81 &     2.666$\, 10^{-4}$ &     1.673$\, 10^{-3}$ &     5.823$\, 10^{-4}$ &     5.510$\, 10^{-4}$ &     4.874$\, 10^{-4}$ \\
&    268 &     25 &    444.34 &     4.014$\, 10^{-4}$ &     4.758$\, 10^{-3}$ &     1.611$\, 10^{-3}$ &     1.509$\, 10^{-3}$ &     1.309$\, 10^{-3}$ \\
&    321 &     25 &    355.82 &     2.258$\, 10^{-4}$ &     1.218$\, 10^{-2}$ &     4.126$\, 10^{-3}$ &     3.817$\, 10^{-3}$ &     3.229$\, 10^{-3}$ \\
&    385 &     25 &    237.49 &     3.078$\, 10^{-4}$ &     2.977$\, 10^{-2}$ &     1.031$\, 10^{-2}$ &     9.368$\, 10^{-3}$ &     7.650$\, 10^{-3}$ \\
&    462 &     25 &    126.74 &     3.463$\, 10^{-4}$ &     6.681$\, 10^{-2}$ &     2.442$\, 10^{-2}$ &     2.166$\, 10^{-2}$ &     1.688$\, 10^{-2}$ \\
&    555 &     25 &     51.31 &     3.142$\, 10^{-4}$ &     1.358$\, 10^{-1}$ &     5.530$\, 10^{-2}$ &     4.746$\, 10^{-2}$ &     3.470$\, 10^{-2}$ \\
&    666 &     25 &     15.07 &     3.317$\, 10^{-4}$ &     2.478$\, 10^{-1}$ &     1.177$\, 10^{-1}$ &     9.688$\, 10^{-2}$ &     6.511$\, 10^{-2}$ \\
&    799 &     25 &      3.00 &     3.298$\, 10^{-4}$ &     4.063$\, 10^{-1}$ &     2.357$\, 10^{-1}$ &     1.839$\, 10^{-1}$ &     1.109$\, 10^{-1}$ 
\label{table:Cell_space}
\end{tabular}
\end{table*}
\begin{table*}[!htbp]
\vspace{-.15cm}
\caption{Same as Table \ref{table:Cell_space}, but for CMB ground-based experiments. }
\centering
\tiny
\begin{tabular}{l|lll|ccccc}
\hline
Experiment & $\nu$ & $\delta\nu$ & $C$ & C$_{\ell}^{BB}$ Radio &  C$_{\ell}^{BB}$ IR &  C$_{\ell}^{BB}$ CIB ($\ell$=80) &  C$_{\ell}^{BB}$ CIB ($\ell$=1000) &  C$_{\ell}^{BB}$ CIB ($\ell$=4000)  \\
 & GHz & \% & & Jy$^2$/sr & Jy$^2$/sr &  Jy$^2$/sr & Jy$^2$/sr & Jy$^2$/sr  \\
\hline
C-BASS &      5 &     20 &      0.77 &     3.089$\, 10^{-2}$ &     --&     --&     --&     --\\
\hline
NEXT-BASS &      7 &     20 &      1.67 &     9.737$\, 10^{-3}$ &     --&     --&     --&     --\\
&      8 &     20 &      2.11 &     7.185$\, 10^{-3}$ &     --&     --&     --&     --\\
&      9 &     20 &      2.70 &     5.704$\, 10^{-3}$ &     --&     --&     --&     --\\
&     10 &     20 &      3.12 &     4.959$\, 10^{-3}$ &     6.716$\, 10^{-6}$ &     --&     --&     --\\
&     11 &     20 &      4.18 &     4.147$\, 10^{-3}$ &     5.949$\, 10^{-6}$ &     --&     --&     --\\
&     13 &     20 &      5.31 &     3.133$\, 10^{-3}$ &     4.899$\, 10^{-6}$ &     --&     --&     --\\
&     14 &     20 &      6.76 &     3.183$\, 10^{-3}$ &     4.779$\, 10^{-6}$ &     --&     --&     --\\
&     15 &     20 &      7.69 &     1.592$\, 10^{-3}$ &     3.519$\, 10^{-6}$ &     --&     --&     --\\
&     17 &     20 &      9.73 &     1.134$\, 10^{-3}$ &     2.925$\, 10^{-6}$ &     --&     --&     --\\
&     20 &     20 &     12.24 &     1.112$\, 10^{-3}$ &     2.687$\, 10^{-6}$ &     --&     --&     --\\
&     22 &     20 &     15.30 &     8.020$\, 10^{-4}$ &     2.282$\, 10^{-6}$ &     --&     --&     --\\
&     25 &     20 &     19.28 &     8.706$\, 10^{-4}$ &     2.234$\, 10^{-6}$ &     --&     --&     --\\
 &     28 &     20 &     24.19 &     5.437$\, 10^{-4}$ &     1.799$\, 10^{-6}$ &     --&     --&     --\\
\hline
QUIJOTE &     11 &     18 &      3.70 &     1.706$\, 10^{-2}$ &     1.203$\, 10^{-5}$ &     --&     --&     --\\
&     13 &     15 &      5.16 &     1.638$\, 10^{-2}$ &     1.114$\, 10^{-5}$ &     --&     --&     --\\
&     17 &     12 &      8.80 &     7.005$\, 10^{-3}$ &     6.590$\, 10^{-6}$ &     --&     --&     --\\
&     19 &     11 &     10.98 &     7.495$\, 10^{-3}$ &     6.649$\, 10^{-6}$ &     --&     --&     --\\
&     30 &     27 &     26.85 &     3.924$\, 10^{-3}$ &     4.318$\, 10^{-6}$ &     --&     --&     --\\
&     40 &     24 &     46.95 &     3.011$\, 10^{-3}$ &     3.723$\, 10^{-6}$ &     --&     --&     --\\
\hline
AdvACTPOL &     90 &     30 &    201.20 &     3.272$\, 10^{-5}$ &     7.137$\, 10^{-6}$ &     --&     --&     --\\
 &    150 &     30 &    395.55 &     1.956$\, 10^{-5}$ &     1.427$\, 10^{-4}$ &     5.998$\, 10^{-5}$ &     5.751$\, 10^{-5}$ &     5.223$\, 10^{-5}$ \\
&    230 &     30 &    477.65 &     4.163$\, 10^{-5}$ &     1.802$\, 10^{-3}$ &     6.994$\, 10^{-4}$ &     6.607$\, 10^{-4}$ &     5.825$\, 10^{-4}$ \\
\hline
BICEP3+Keck[2018] &     95 &     30 &    219.11 &     1.478$\, 10^{-2}$ &     2.375$\, 10^{-5}$ &     2.082$\, 10^{-6}$ &     2.023$\, 10^{-6}$ &     1.886$\, 10^{-6}$ \\
&    150 &     30 &    395.55 &     2.547$\, 10^{-2}$ &     2.359$\, 10^{-4}$ &     5.998$\, 10^{-5}$ &     5.751$\, 10^{-5}$ &     5.223$\, 10^{-5}$ \\
\hline
BICEPArray[2023] &     30 &     30 &     26.81 &     3.290$\, 10^{-2}$ &     1.448$\, 10^{-5}$ &     --&     --&     --\\
&     40 &     30 &     46.82 &     3.376$\, 10^{-2}$ &     1.474$\, 10^{-5}$ &     --&     --&     --\\
&     95 &     30 &    219.11 &     1.478$\, 10^{-2}$ &     2.375$\, 10^{-5}$ &     2.082$\, 10^{-6}$ &     2.023$\, 10^{-6}$ &     1.886$\, 10^{-6}$ \\
&    150 &     30 &    395.55 &     7.879$\, 10^{-3}$ &     1.898$\, 10^{-4}$ &     5.998$\, 10^{-5}$ &     5.751$\, 10^{-5}$ &     5.223$\, 10^{-5}$ \\
&    220 &     30 &    480.08 &     4.159$\, 10^{-3}$ &     1.672$\, 10^{-3}$ &     5.321$\, 10^{-4}$ &     5.040$\, 10^{-4}$ &     4.466$\, 10^{-4}$ \\
&    270 &     30 &    440.61 &     2.349$\, 10^{-3}$ &     5.186$\, 10^{-3}$ &     1.677$\, 10^{-3}$ &     1.570$\, 10^{-3}$ &     1.361$\, 10^{-3}$ \\
\hline
CLASS &     38 &     30 &     42.42 &     3.452$\, 10^{-2}$ &     1.497$\, 10^{-5}$ &     --&     --&     --\\
&     93 &     30 &    211.94 &     3.383$\, 10^{-2}$ &     3.465$\, 10^{-5}$ &     1.108$\, 10^{-6}$ &     1.085$\, 10^{-6}$ &     1.028$\, 10^{-6}$ \\
&    148 &     30 &    390.46 &     1.785$\, 10^{-2}$ &     2.002$\, 10^{-4}$ &     5.538$\, 10^{-5}$ &     5.311$\, 10^{-5}$ &     4.826$\, 10^{-5}$ \\
&    217 &     30 &    480.18 &     1.043$\, 10^{-2}$ &     1.639$\, 10^{-3}$ &     4.986$\, 10^{-4}$ &     4.724$\, 10^{-4}$ &     4.189$\, 10^{-4}$ \\
\hline
SO-SAT &     27 &     30 &     21.81 &     2.671$\, 10^{-2}$ &     1.257$\, 10^{-5}$ &     --&     --&     --\\
&     39 &     30 &     44.60 &     3.646$\, 10^{-2}$ &     1.570$\, 10^{-5}$ &     --&     --&     --\\
&     93 &     30 &    211.94 &     2.143$\, 10^{-2}$ &     2.615$\, 10^{-5}$ &     1.108$\, 10^{-6}$ &     1.085$\, 10^{-6}$ &     1.028$\, 10^{-6}$ \\
&    145 &     30 &    382.56 &     9.867$\, 10^{-3}$ &     1.610$\, 10^{-4}$ &     4.847$\, 10^{-5}$ &     4.651$\, 10^{-5}$ &     4.231$\, 10^{-5}$ \\
&    225 &     30 &    479.26 &     4.112$\, 10^{-3}$ &     1.901$\, 10^{-3}$ &     6.157$\, 10^{-4}$ &     5.823$\, 10^{-4}$ &     5.145$\, 10^{-4}$ \\
 &    280 &     30 &    426.14 &     2.262$\, 10^{-3}$ &     6.330$\, 10^{-3}$ &     2.008$\, 10^{-3}$ &     1.877$\, 10^{-3}$ &     1.620$\, 10^{-3}$ \\
\hline
SO-LAT &     27 &     30 &     21.81 &     3.791$\, 10^{-4}$ &     1.614$\, 10^{-6}$ &     --&     --&     --\\
&     39 &     30 &     44.60 &     1.968$\, 10^{-4}$ &     1.186$\, 10^{-6}$ &     --&     --&     --\\
&     93 &     30 &    211.94 &     3.467$\, 10^{-5}$ &     8.623$\, 10^{-6}$ &     1.108$\, 10^{-6}$ &     1.085$\, 10^{-6}$ &     1.028$\, 10^{-6}$ \\
&    145 &     30 &    382.56 &     2.374$\, 10^{-5}$ &     1.184$\, 10^{-4}$ &     4.847$\, 10^{-5}$ &     4.651$\, 10^{-5}$ &     4.231$\, 10^{-5}$ \\
&    225 &     30 &    479.26 &     4.308$\, 10^{-5}$ &     1.607$\, 10^{-3}$ &     6.157$\, 10^{-4}$ &     5.823$\, 10^{-4}$ &     5.145$\, 10^{-4}$ \\
&    280 &     30 &    426.14 &     6.880$\, 10^{-5}$ &     5.549$\, 10^{-3}$ &     2.008$\, 10^{-3}$ &     1.877$\, 10^{-3}$ &     1.620$\, 10^{-3}$ \\
\hline
SPT-3G &     95 &     27 &    219.42 &     1.175$\, 10^{-5}$ &     9.260$\, 10^{-6}$ &     2.082$\, 10^{-6}$ &     2.023$\, 10^{-6}$ &     1.886$\, 10^{-6}$ \\
&    148 &     26 &    391.20 &     1.453$\, 10^{-5}$ &     1.277$\, 10^{-4}$ &     5.538$\, 10^{-5}$ &     5.311$\, 10^{-5}$ &     4.826$\, 10^{-5}$ \\
&    223 &     23 &    481.20 &     3.326$\, 10^{-5}$ &     1.4745$\, 10^{-3}$ &     5.823$\, 10^{-4}$ &     5.510$\, 10^{-4}$ &     4.874$\, 10^{-4}$ \\
\hline
CMB-S4-SAT &     20 &     25 &     12.10 &     7.679$\, 10^{-4}$ &     2.352$\, 10^{-6}$ &     --&     --&     --\\
 &     30 &     30 &     26.81 &     3.290$\, 10^{-2}$ &     1.448$\, 10^{-5}$ &     --&     --&     --\\
&     40 &     30 &     46.82 &     3.891$\, 10^{-2}$ &     1.667$\, 10^{-5}$ &     --&     --&     --\\
&     85 &     24 &    183.92 &     1.510$\, 10^{-2}$ &     1.699$\, 10^{-5}$ &     --&     --&     --\\
&     95 &     24 &    219.71 &     1.639$\, 10^{-2}$ &     2.477$\, 10^{-5}$ &     2.082$\, 10^{-6}$ &     2.023$\, 10^{-6}$ &     1.886$\, 10^{-6}$ \\
&    145 &     22 &    383.91 &     1.613$\, 10^{-2}$ &     1.749$\, 10^{-4}$ &     4.847$\, 10^{-5}$ &     4.651$\, 10^{-5}$ &     4.231$\, 10^{-5}$ \\
&    155 &     22 &    409.13 &     1.637$\, 10^{-2}$ &     2.539$\, 10^{-4}$ &     7.149$\, 10^{-5}$ &     6.851$\, 10^{-5}$ &     6.214$\, 10^{-5}$ \\
&    220 &     22 &    481.78 &     5.833$\, 10^{-3}$ &     1.701$\, 10^{-3}$ &     5.321$\, 10^{-4}$ &     5.040$\, 10^{-4}$ &     4.466$\, 10^{-4}$ \\
&    270 &     18 &    442.72 &     4.951$\, 10^{-3}$ &     5.326$\, 10^{-3}$ &     1.677$\, 10^{-3}$ &     1.570$\, 10^{-3}$ &     1.361$\, 10^{-3}$ \\
\hline
CMB-S4-LAT &     30 &     30.0 &     26.81 &     4.057$\, 10^{-4}$ &     1.587$\, 10^{-6}$ &     --&     --&     --\\
&     40 &     30 &     46.82 &     1.983$\, 10^{-4}$ &     1.189$\, 10^{-6}$ &     --&     --&     --\\
&     95 &     30 &    219.11 &     3.261$\, 10^{-5}$ &     9.703$\, 10^{-6}$ &     2.082$\, 10^{-6}$ &     2.023$\, 10^{-6}$ &     1.886$\, 10^{-6}$ \\
&    145 &     30 &    382.56 &     1.755$\, 10^{-5}$ &     1.156$\, 10^{-4}$ &     4.847$\, 10^{-5}$ &     4.651$\, 10^{-5}$ &     4.231$\, 10^{-5}$ \\
&    220 &     30 &    480.08 &     3.350$\, 10^{-5}$ &     1.369$\, 10^{-3}$ &     5.321$\, 10^{-4}$ &     5.040$\, 10^{-4}$ &     4.466$\, 10^{-4}$ \\
 &    270 &     30 &    440.61 &     4.971$\, 10^{-5}$ &     4.366$\, 10^{-3}$ &     1.677$\, 10^{-3}$ &     1.570$\, 10^{-3}$ &     1.361$\, 10^{-3}$ 
\label{table:Cell_ground}
\end{tabular}
\end{table*}

\section{Contamination of the CMB B-modes}
\label{contrib_Bmode}

In order to provide reliable predictions of the radio source and DSFG contamination to CMB anisotropy polarisation measurements, we have to assume a fractional polarisation for each population of galaxies. For radio sources, at the frequencies where the contamination of the B-modes is minimum (i.e. $\sim$90-300\,GHz), there are still few polarisation measurements and very scarce polarisation fraction measurements for the different types of radio sources (see Sect.\,\ref{radio_polar}).  Thus, we used a constant $\langle \Pi^{\mathrm{rad}} \rangle$=2.8\%, in agreement with the recent \textit{Planck}, SPT, and ACT measurements and radio source follow-ups from 90 to 220\,GHz. For DSFG, the situation is even worse and polarisation properties are almost completely unexplored. As discussed in Sect.\,\ref{DSFG_polar}, we adopted $\langle \Pi^{\mathrm{IR}} \rangle$=1.4\%. As all our BB power spectra are proportional to the square of the fractional polarisation, it is very easy to obtain polarised power spectra for other choices of fractional polarisation:
\begin{eqnarray}
C_{\ell}^{\mathrm{BB}, Radio} (\Pi^{rad})= C_{\ell}^{\mathrm{BB}, Radio} \left( \frac{\Pi^{rad}}{0.028} \right) ^2 \,, \\
C_{\ell}^{\mathrm{BB}, CIB} (\Pi^{IR})= C_{\ell}^{\mathrm{BB}, CIB} \left( \frac{\Pi^{IR}}{0.014} \right) ^2\,, \\
C_{\ell}^{\mathrm{BB}, IR} (\Pi^{IR})= C_{\ell}^{\mathrm{BB}, IR} \left( \frac{\Pi^{IR}}{0.014} \right) ^2\,.
\end{eqnarray}

\subsection{Polarised power spectra of the extragalactic components \label{compar_polar_Cell}}
We list in Tables\,\ref{table:Cell_space} and \ref{table:Cell_ground} the level of BB power spectra for radio (C$_{\ell}^{rad}$) and DSFG (C$_{\ell}^{IR}$) shot noise, and the clustering (C$_{\ell}^{CIB}$) for three multipoles ($\ell$=80, 1000, and 4000).\\
We first compare in Fig.~\ref{fig:IR_1h} the relative level of DSFG shot noise and clustering power spectra at $\ell$=80. We recall that the clustering power spectra are an upper limit as we estimated the maximum contribution of the one-halo term (see Sect.\ref{Polar_1h}). The ratio C$_{\ell}^{IR}$/ C$_{\ell}^{CIB}$ is mostly constant, and between 2 and 3 for $120<\nu<700$\,GHz. At lower frequencies, it is much higher (from 4 to 30) and thus C$_{\ell}^{CIB}$ can be neglected. Consequently, we did not compute the clustering power spectra for frequencies $\nu \le$90\,GHz. The ratio increases very slowly with $\ell$, by up to $\sim$30\% at $\ell$=4000 and $\nu<$400\,GHz.\\

We then compare in Fig.\,\ref{fig:IR1h_rad} the level of the radio power spectra and DSFG+clustering power spectra as a function of frequency. As expected, the general trend is an increase in  $\Delta = \frac{C_{\ell}^{IR} + C_{\ell}^{CIB}}{C_{\ell}^{Rad}}$ with frequency, roughly proportional to $\nu^7$ for 80$<\nu<$400\,GHz. We can distinguish three families of points, depending on the telescope size, with $\Delta$ varying by a factor $\sim$250:

\begin{figure}
\hspace{-0.5cm}
\includegraphics[width=9.cm]{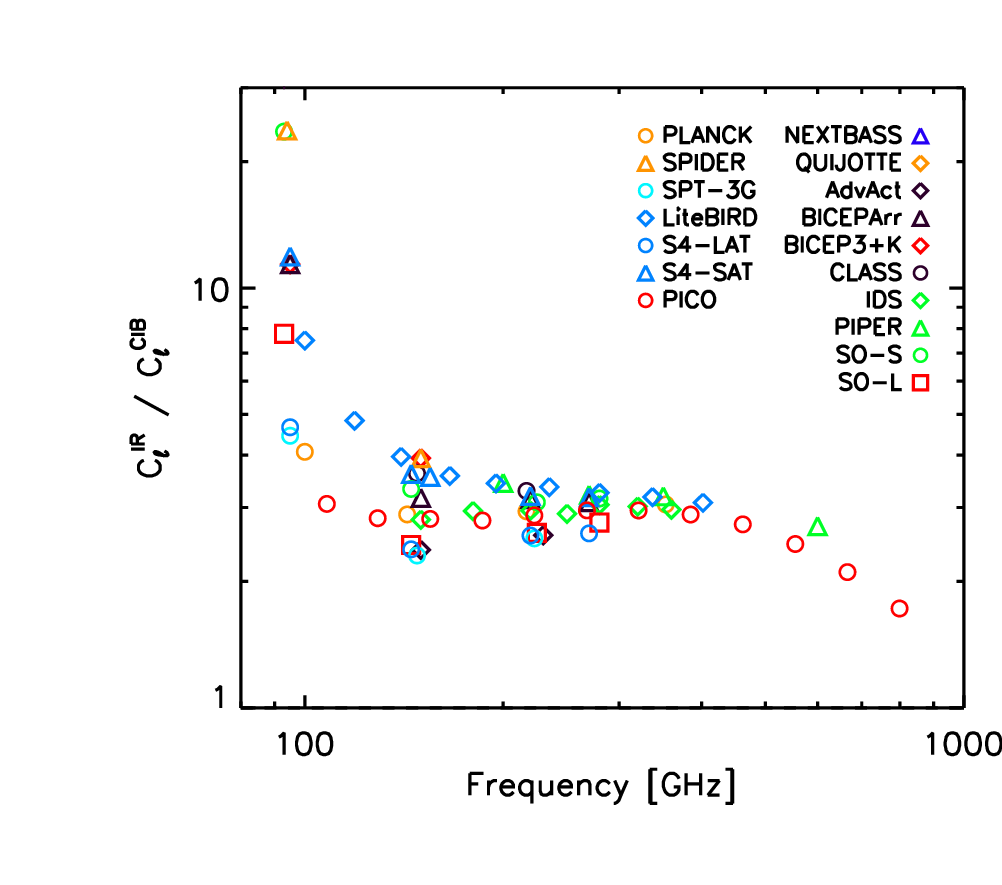}
\caption{\label{fig:IR_1h} Ratio of shot noise and clustering (one-halo CIB anisotropies) for dusty galaxies at $\ell$=80 for all CMB experiments ($\ell$=80  corresponds to the recombination B-peak).} 
\end{figure}

\begin{figure}
\hspace{-0.5cm}
\includegraphics[width=9.cm]{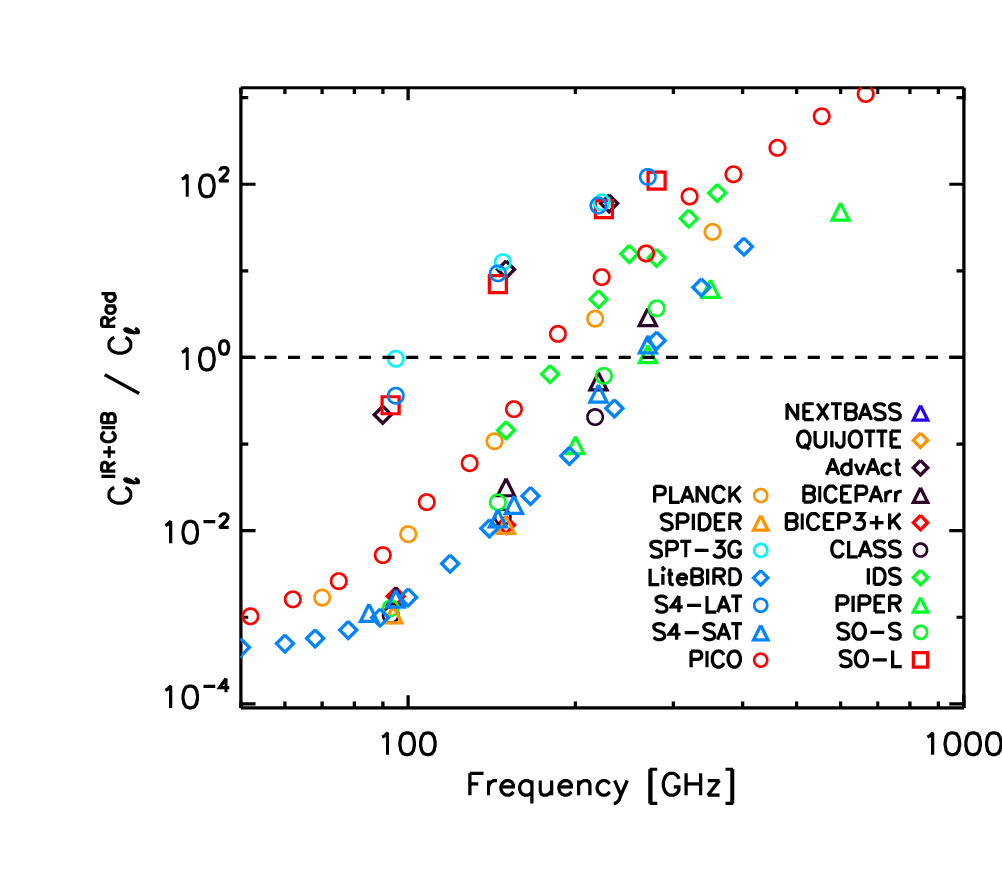}
\caption{\label{fig:IR1h_rad} Ratio between the BB power spectra of [IR shot noise + clustering] and radio shot noise, at $\ell$=80 for all CMB experiments. } 
\end{figure}

\begin{itemize}
\item For the large-aperture telescopes ($\ge$ 6m, i.e. SPT-3G, S4-LAT, SO-LAT, AdvActPol), $\Delta \simeq 100 \times \left ( \frac{\nu}{220\,[GHz]} \right)^7$.
\item For the medium-aperture telescopes ($\sim$1.5m, i.e. Planck, IDS, PICO), $\Delta \simeq 4 \left ( \frac{\nu}{220\,[GHz]} \right)^7$.
\item For the small-aperture telescopes ($\le$0.6m, i.e. LiteBIRD, SPIDER, CLASS, SO-SAT, S4-SAT, BICEP), $\Delta \simeq 0.4 \times \left ( \frac{\nu}{220\,[GHz]} \right)^7$.
\end{itemize}
Thus, the DSFG power spectra level is higher than that of radio galaxies at a frequency that decreases with telescope size: $\sim$247, 180, and 114\,GHz, from small to large apertures. These results do not depend on the multipole (as  C$_{\ell}^{IR}$/ C$_{\ell}^{CIB}$ varies weakly with $\ell$). 
\begin{figure}
\begin{center}
\includegraphics[width=8.5cm]{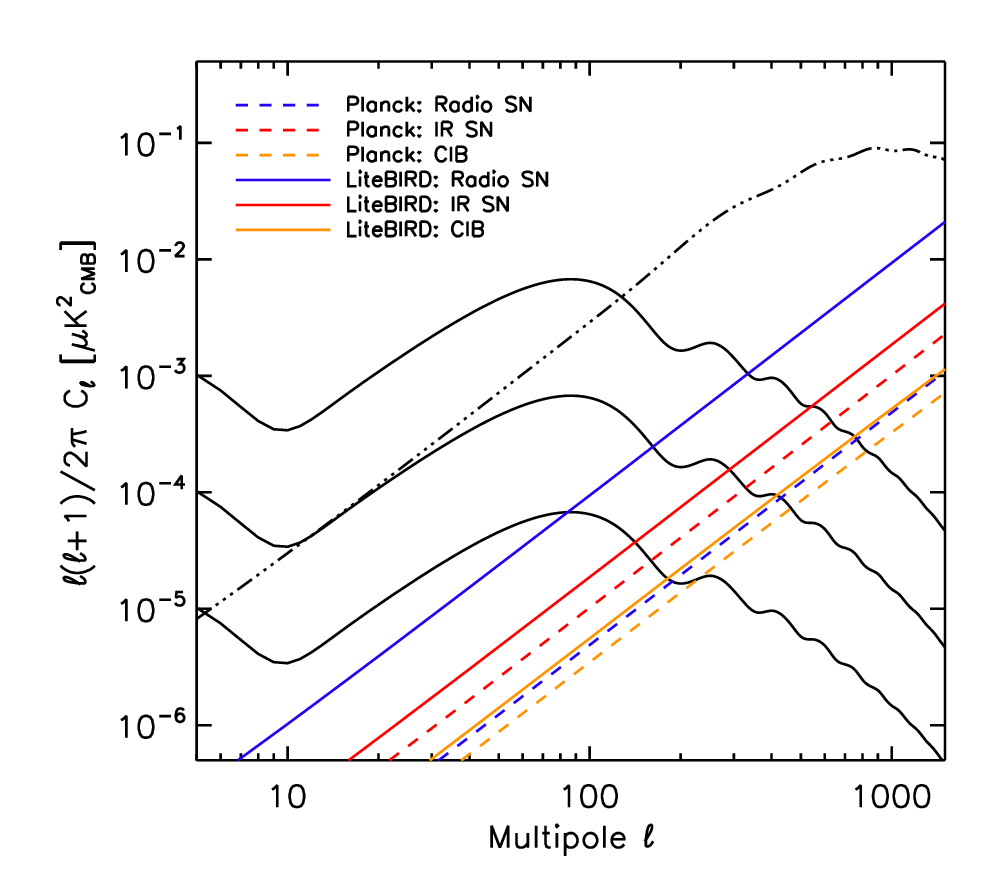}
\end{center}
\caption{\label{fig:LP100} Extragalactic foreground power spectra for \textit{Planck} (coloured dashed lines) and LiteBIRD (coloured continuous lines) at 217 and 235 \,GHz, respectively. The three continuous black lines are the primordial CMB B-mode power spectrum for $r=0.1, 0.01,\text{and } 0.001$ from top to bottom. The dash-three-dots line is the lensing B-mode.} 
\end{figure}

\begin{figure}
\begin{center}
\includegraphics[width=8.5cm]{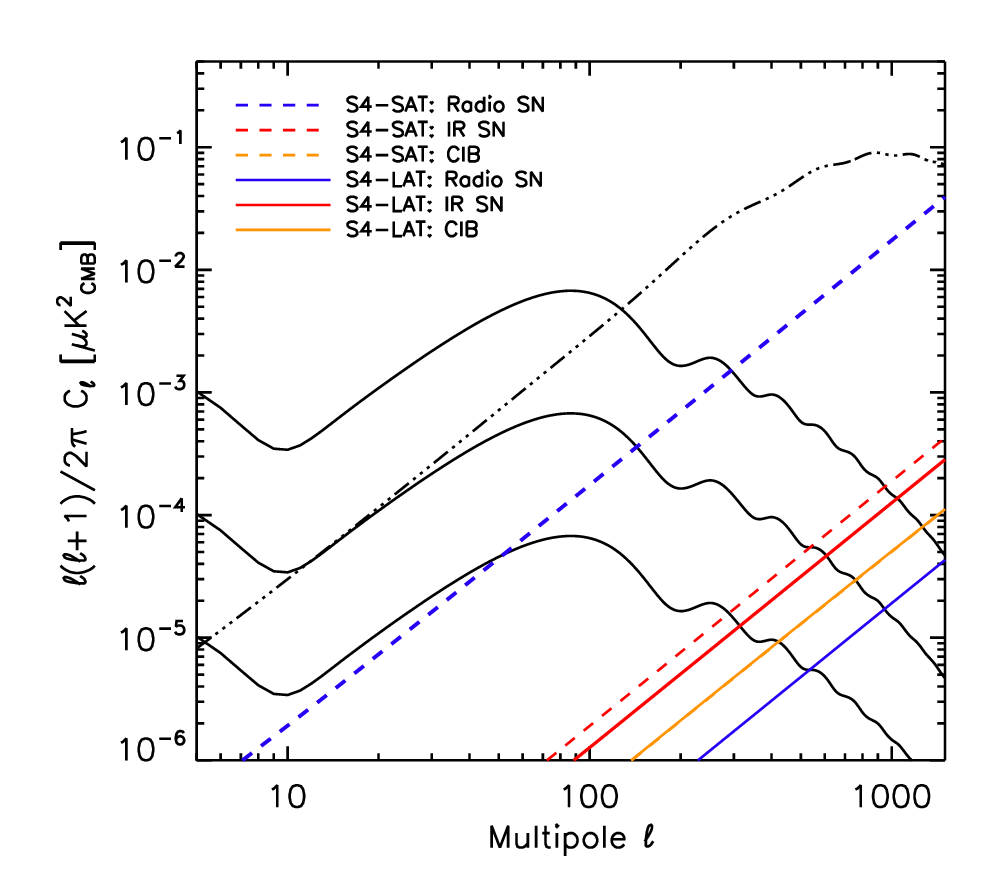}
\end{center}
\caption{\label{fig:S4} Extragalactic foreground power spectra for S4-SAT (coloured dashed lines) and S4-LAT (coloured continuous lines) at 145\,GHz. The three continuous black lines are the primordial CMB B-mode power spectrum for $r=0.1, 0.01,\text{and } 0.001$ from top to bottom. The dash-three-dots line is the lensing B-mode. As the two experiments are at the same frequency, the two C$_{\ell}^{CIB}$ curves are confounded.} 
\end{figure}

\begin{figure*}
\begin{center}
\includegraphics[width=15cm]{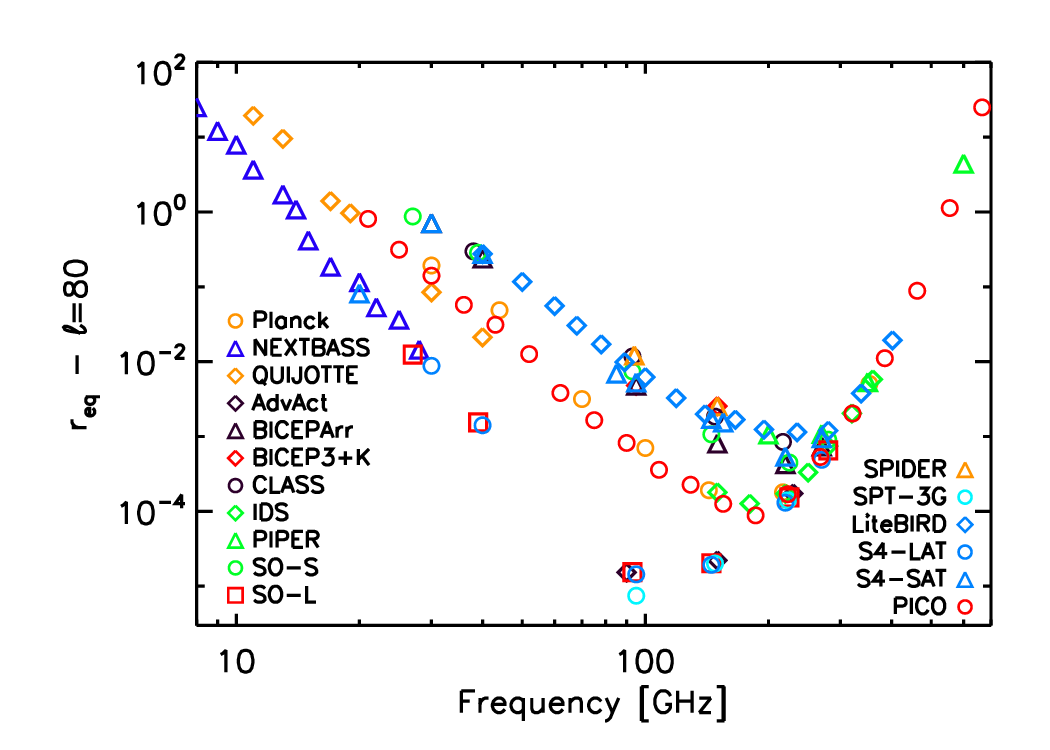}
\end{center}
\caption{\label{fig:req80} Equivalent tensor-to-scalar ratio ($\mathrm r_{eq}$) of the sum of the extragalactic foregrounds at the recombination B-peak, $\ell$=80, for the different CMB experiments ($\mathrm r_{eq}$ is computed for each individual frequency).} 
\end{figure*}

\subsection{Comparison with the CMB B-modes \label{compar_CMB_Cell}}

We first illustrate the contaminations of extragalactic components to the CMB B-mode power spectrum at two frequencies, $\sim$220\,GHz (Fig.\,\ref{fig:LP100})  and 145\,GHz (Fig.\,\ref{fig:S4}). At each frequency, we plot the power spectra for two different aperture telescopes to illustrate the turnover between radio/DSFG dominant contaminations. The CMB B-mode power spectrum was calculated for the Planck 2018 cosmology (using TT, TE, EE+lowE+lensing+BAO and a pivot scale for $r$ of 0.002 Mpc$^{-1}$, \citealt{Planck2018_r}). \\

We compare in Fig.\,\ref{fig:LP100} \textit{Planck} at 217\,GHz with LiteBIRD at 235\,GHz. While the contamination by radio galaxies is twice lower than by DSFG for \textit{Planck} , the power spectrum of radio galaxies is five times larger than that of DSFG for LiteBIRD (even if the frequency of 235\,GHz is higher). It is at the same level of the $r=0.01$ ($r=0.001$) B-mode power spectrum for $\ell$=160 ($\ell$= 83). For \textit{Planck}, the total contamination is negligible compared to the last 95\% CL upper limit $r_{0.002}<0.056$ \citep{planck2018_cosmo}. 
In Fig.\,\ref{fig:S4} we show the level of the extragalactic components for the ground-based S4-SAT and S4-LAT experiments. Contamination by radio sources dominates for S4-SAT at a level of $r$=1.7$\times$10$^{-3}$ at $\ell$=80. For S4-LAT, the dominant contamination comes from DSFG shot noise, at a level of $r$=1.2$\times$10$^{-5}$ at $\ell$=80.\\

We finally compute the equivalent tensor-to-scalar ratio ($r_{eq}$) of the total extragalactic contamination (radio galaxy shot noise, DSFG shot noise, and clustering) for each individual frequency at given multipoles.  We show in Fig.\,\ref{fig:req80} the variation in $r_{eq}$ as a function of  frequencies at the recombination B-peak, $\ell$=80. Minimum $r_{eq}$ is reached for 90$\lesssim \nu \lesssim$300\,GHz depending on the experiment.
Similarly to Fig.\,\ref{fig:IR1h_rad} (and see Sect\,\ref{compar_polar_Cell}), we can distinguish three cases according to the telescope aperture size:
\begin{itemize}
\item Large-aperture telescopes. The minimum contamination is at the level of $r_{eq}$=7.4$\times$10$^{-6}$ for SPT-3G at 95\,GHz. For SO-LAT, AdvACT, and S4-LAT, $r_{eq}$ is about 1.5 and 2$\times$10$^{-5}$ at 90-93 and 145-150\,GHz, respectively. These levels are well below the targeted $\sigma _\mathrm{r}$ of these experiments (by a factor of $\gtrsim$20-400).
\item Medium-aperture telescopes. The minimum contamination is at the level of $r_{eq}\simeq$10$^{-4}$ and is reached at $\nu\simeq$200\,GHz. While this is $\sim$40 times higher than $\sigma _\mathrm{r}$ for IDS alone, it is at the same level as $\sigma _\mathrm{r}$ for PICO \citep{Hanany_2019}.
\item Small-aperture telescopes. The contamination reaches a level of $4.3-5.4\times$10$^{-4}$ for S4-SAT, SO-SAT, and BICEPArray at $\sim$220\,GHz. It increases to 8.5$\times$10$^{-4}$ for CLASS at 217\,GHz, and 1.1$\times$10$^{-3}$ for LiteBIRD at 235 and 280\,GHz and PIPER at 200 and 270\,GHz. Finally, it is about 2.5$\times$10$^{-3}$ for SPIDER at 150\,GHz. 
The level of contamination (of 4 to $8\times$10$^{-4}$ from 150 to 270\,GHz) is below the targeted $\sigma _\mathrm{r}$ for the Bicep/Keck experiment, for which they project $0.002<\sigma _\mathrm{r}<0.006$ by the end of the planned BICEP Array program,  assuming current modelling of polarised Galactic foregrounds and depending on the level of delensing that can be achieved with higher angular resolution maps from the South Pole Telescope \citep{Hui2018}. For LiteBIRD, the contamination reaches the 68\% confidence level uncertainty, that is $\sigma _\mathrm{r}<10^{-3}$ (this $\sigma _\mathrm{r}$ includes statistical, instrumental systematic, and Galactic foreground uncertainties, \citealt{Matsumura2016}).
\end{itemize}

\begin{figure*}
\begin{center}
\includegraphics[width=15cm]{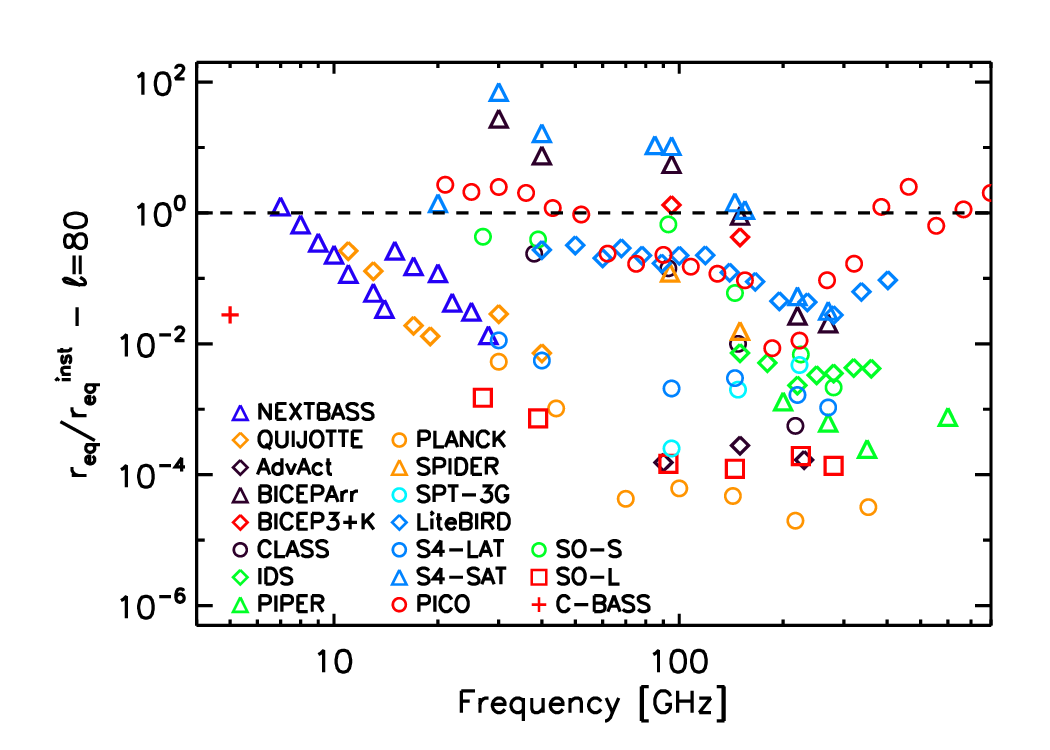}
\end{center}
\caption{\label{fig:reqinst} Ratio of equivalent $r$ of the extragalactic foregrounds ($r_{eq}$) and instrument noise ($r_{eq}^{inst}$), at $\ell$=80.} 
\end{figure*}

\begin{figure}
\begin{center}
\includegraphics[width=9cm]{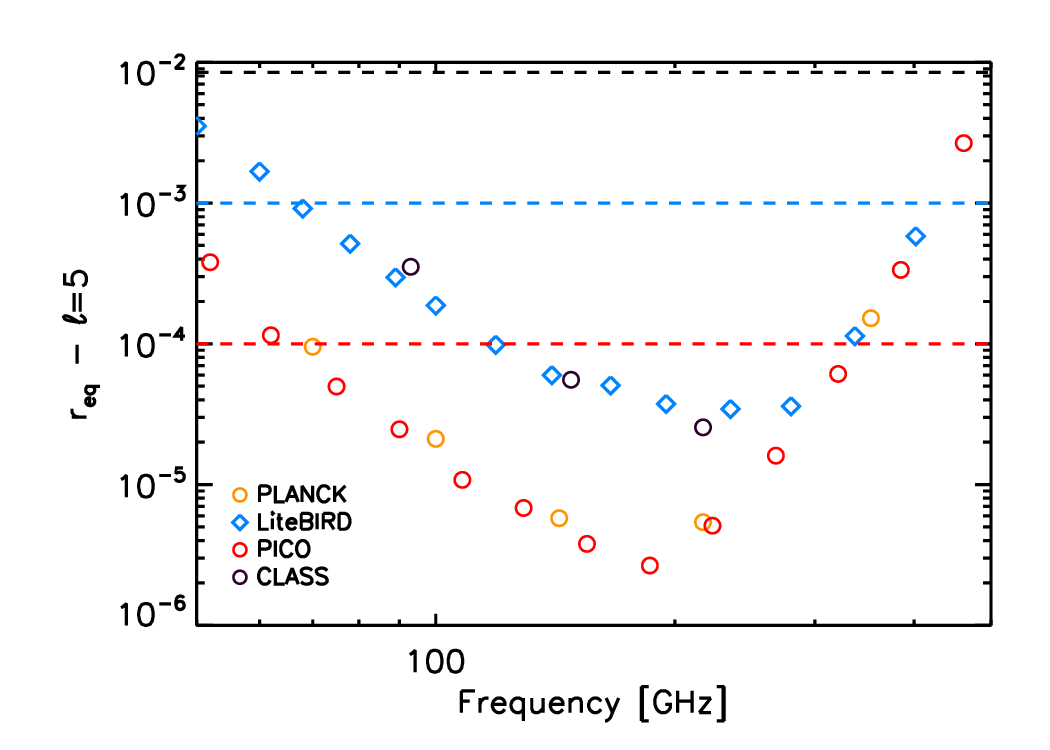}
\end{center}
\caption{\label{fig:req5} Equivalent tensor-to-scalar ratio ($\mathrm r_{eq}$) of the sum of the extragalactic foregrounds at $\ell$=5, corresponding to the reionisation B-bump. We also show the $\sigma _\mathrm{r}$ for LiteBIRD, PICO, and CLASS (dashed lines). For \textit{Planck}, the current 1$\sigma$ upper limit is $r <0.028$ and is thus not visible in the figure.}
\end{figure}

This comparison between $r_{eq}$ and $\sigma _\mathrm{r}$ was made considering each frequency for $r_{eq}$ independently, while $\sigma _\mathrm{r}$ is usually estimated for each experiment by combining the whole set of available bands and under specific assumptions (e.g. taking systematic effects or foregrounds residual impacts into account). Multi-frequency component separations should be able to decrease the level of extragalactic foreground contamination.\\

To offer a complementary view, rather than comparing $r_{eq}$ with $\sigma _\mathrm{r}$, we could compare $r_{eq}$  with the equivalent instrument noise $r_{eq}^{inst}$ computed independently at each frequency. We calculate $r_{eq}^{inst}$ following
\begin{equation}
r_{eq}^{inst}= \left( \frac{\sigma^P_{inst}}{\frac{180}{\ell} \times 60} \right)^2 \times \frac{1}{D^{BB}_\ell(r=1, \ell)}
,\end{equation}
where $\sigma^P_{inst}$ is the instrument noise in polarisation (given in Tables\,\ref{table:exp_space} and \ref{table:exp_ground}). We show in Fig.\,\ref{fig:reqinst} the ratio of $r_{eq}$ and $r_{eq}^{inst}$  for all frequencies and experiments. A contamination of at least 10\%\ ($r_{eq}$/$r_{eq}^{inst} \ge 0.1$) for $70\le \nu \le 250$\,GHz is reached for BICEP at 95 and 150\,GHz,  CLASS, SO-SAT, and SPIDER at 93\,GHz, LiteBIRD from 78 to 140\,GHz, S4-SAT from 85 to 155\,GHz, and PICO from 75 to 129\,GHz. Combining higher and lower frequencies to decrease the Galactic foreground residuals may also add more contamination from extragalactic sources (because of their different mean polarised SEDs and because they are not correlated from high to low frequencies). For example, for PICO, $0.9 \le r_{eq}$/$r_{eq}^{inst} \le 2.7$ for $21\le \nu \le 52$\,GHz and for S4-SAT, it is $>10$ for $\nu$=30-95\,GHz.
\\

The scale dependency of extragalactic foregrounds compared to the CMB makes the ratio of the primordial CMB signal over foregrounds more favourable at  larger scale, in particular at the reionisation B-bump ($\ell$=5). Only nearly full-sky ($f_{sky}\ge$70\%) experiments can provide some measurements at such low multipoles. The $r$ equivalent in this case is very small (2.7$\times10^{-6}$ for PICO at 186\,GHz, 3.7-3.4$\times10^{-5}$ for LiteBIRD at 195-235\,GHz,  and 2.5$\times10^{-5}$ for CLASS at 217\,GHz; see Fig.\,\ref{fig:req5}). They are much smaller than the targeted limits on the primordial $r$ for PICO and LiteBIRD, and $\sigma _\mathrm{r}=8.5\times10^{-3}$ for CLASS \citep[including diffuse Galactic thermal dust and synchrotron foregrounds,][]{Watts2015}. For \textit{Planck}, the level of contamination by polarised extragalactic sources is much lower than the current B-mode upper limit \citep{Planck2018_r}.\\

Finally, we consider the ratio of the extragalactic foreground and CMB lensing BB power spectra (at $\ell$=1000). This ratio is $\sim$120 times higher than the equivalent tensor-to-sclar ratio $r_{eq}$ at $\ell$=80. It extends from $\sim\,10^{-3}$ for large-aperture experiments to $\sim\,10^{-1}$ for small-aperture experiments. As is already known, ground-based large-aperture telescopes will provide the ability to delens the maps from future satellite CMB  missions, such as LiteBIRD \citep[e.g.][]{Namikawa2014}.

\section{Conclusion \label{conclu}}
We have computed the expected level of polarised fluctuations from the shot noise of radio galaxies and DSFG and from the CIB clustering using current or updated models.  Using these models, we predicted the point-source detection limits (confusion noises, in intensity) for future CMB space-based or balloon-borne experiments (IDS, PIPER, SPIDER, LiteBIRD, and PICO) and ground-based experiments (C-BASS, NEXT-BASS, QUIJOTE, AdvACTPOL, BICEP3+Keck, BICEPArray, CLASS, SO, SPT3G, and S4). 
These limits were computed by taking  the instrument noise, the three extragalactic foregrounds, and the CMB into account. The models, as well as the point-source detection flux limits, were validated using most recent measurements on number counts, CIB power spectra, confusion noises, and shot noise levels.  As expected, we found that the confusion noise levels are mostly driven by the telescope-aperture sizes and frequency.  \\

Assuming a constant polarisation fraction consistent with current observational results for the radio sources of  $\langle \Pi^{\mathrm{rad}} \rangle$=2.8\%, and assuming for the dusty source $\langle \Pi^{\mathrm{IR}} \rangle$=1.4\%, we then predicted the shot noises and CIB one-halo clustering B-mode power spectra. We compared the amplitude of the different extragalactic foregrounds as a function of frequency and telescope-aperture size. We found that CIB clustering is almost negligible. The relative levels of radio and DSFG shot noises are mainly driven by the telescope sizes, which can be classified into three categories: large-aperture ($\ge$ 6m, i.e. SPT-3G, S4-LAT, SO-LAT, AdvActPol), medium-aperture ($\sim$1.5m, i.e. Planck, IDS, PICO), and small-aperture ($\le$0.6m, i.e. LiteBIRD, SPIDER, CLASS, SO-SAT, S4-SAT, BICEP-Keck) telescopes. While we have an equal contribution between radio shot noise and DSFG shot noises (+ clustering) at $\nu \simeq$ 120\,GHz for large-aperture telescopes, it reaches $\nu \simeq$280\,GHz  for small-aperture telescopes, which are thus dominated by the radio shot noise at the frequencies dedicated to the CMB measurement.
\cite{gon05} showed that the contribution of radio source clustering to the temperature angular power spectrum is small and can be neglected if sources are not subtracted down to very faint flux limits, $S\ll10$\,mJy. However, future ground-based experiments such as S4-LAT will be able to reach flux limits of the order of 2-3\,mJy. At these levels, the clustering of radio sources might not be negligible for $\ell<30$ compared to the shot-noise level \citep{gon05}. 

We also predict the confusion noise for SPICA B-BOP and showed that confusion could ultimately limit the sensitivity of deep polarised surveys at 200 and 350\,$\mu$m (with the confusion noise in polarisation reached in 57 and 69\,hours for a 1 square degree field at 200 and 350\,$\mu$m, respectively).\\

Finally, we computed the equivalent tensor-to-scalar ratio ($r_{eq}$) of the total extragalactic contamination (radio galaxy shot noise, DSFG shot noise, and clustering) for given multipoles. At the reionisation B-bump ($\ell$=5), the extragalactic contamination will not limit the measurements.  At the recombination B-peak ($\ell$=80), the contamination for large-aperture telescope experiments is much below the targeted primordial $r$, but this is not the case for some of the small- and medium- aperture telescopes. For example for the LiteBIRD and PICO space experiments, the contamination is at the level of the 68\% confidence level uncertainty on the primordial $r$ (not considering a multi-frequency component separation that should globally decrease $r_{eq}$). On the other side of the multipole range, extragalactic components represent 10-20\% of the CMB lensing BB power spectrum at $\ell$=1000 for LiteBIRD.  Moreover, a similar slope is observed between the extragalactic components and the CMB lensing BB power spectrum up to $\ell$=200 and between the extragalactic components and the primordial B-mode power spectrum for 15$\lesssim \ell \lesssim$50, leading to degeneracies in any model fitting. Removing this extragalactic contamination from the data is thus mandatory for some of the small- and medium-aperture telescope experiments. \\

Foreground mitigation was studied for the Galactic components. We showed that it requires a multi-frequency coverage (but see \citealt{Philcox2018} for a method based on anisotropy statistics, or \citealt{Aylor2019} for the use of neural network). It will be difficult to apply this multi-frequency approach to extragalactic foregrounds, as the three extragalactic components are degenerated (i.e. same power spectra at the multipole of interest) and the sum of the three does not have a well-defined frequency dependency. Moreover, even if more precise polarised source counts for radio galaxies will be obtained in the near future, the variation in radio shot noise with flux limit (changing the flux cut by 30\% affects the shot noise by 30\%, see Table\,\ref{SN_radio_Planck}), together with the variability of radio sources, may prevent us from using more accurate modelling to precisely predict the shot-noise level.

Polarised Galactic foregrounds are dominated by dust and  synchrotron  emissions with spatial variation of their SEDs. 
Using a parametric maximum-likelihood approach, \cite{Errard2016} found that combinations from ground- and space-based and balloon-borne experiments can significantly improve component separation performance, delensing, and cosmological constraints over individual datasets.  In particular, they reported that a combination of post-2020 ground- and space-based experiments could achieve constraints such as $\sigma_\mathrm{r} \sim$1.3$\times$10$^{-4}$ after component separation and iterative delensing. However, such results \citep[see also e.g.][]{Stompor2016}  are often derived ignoring complexities in the Galactic foreground emission due to synchrotron and dust, and neglecting potential other contaminants such as anomalous microwave emission and extragalactic foregrounds.  Moreover, they adopted component separation methods that essentially assume a model that matches the simulated foregrounds under study well. \citet{Remazeilles2016} tested some of these assumptions explicitly and reported biases in the derived value of $r$ of more than 1$\sigma$ by  neglecting the curvature of the synchrotron emission law, for instance. Given their levels for some of mid- and small-aperture telescopes, extragalactic foregrounds have clearly to be considered in the component separation methods dedicated to the extraction of the CMB B-modes. For this purpose, our detailed computation of flux limits and shot-noise levels will allow including these foregrounds precisely into the input sky models.  

\begin{acknowledgements}
GL warmly thank F. Boulanger for insightful discussions on the CIB polarisation, J. Delabrouille and J. Gonzales-Nuevo  for discussions on  the transfer function linked to the source flux measurements, G. De Zotti for discussions around the results presented in this paper, and N. Ponthieu and M. Tristram for their help in manipulating polarised quantities. PS acknowledges hospitality from LAM, where part of this work was completed. GL, MB and LM acknowledge support from the OCEVU Labex (ANR-11-LABX-0060) and the A*MIDEX project (ANR-11-IDEX-0001-02) funded by the "Investissements d'Avenir" French government program managed by the ANR. GL has received funding from the European Research Council (ERC) under the European Union's Horizon 2020 research and innovation programme (grant agreement No 788212). We acknowledge financial support from the "Programme National de Cosmologie and Galaxies" (PNCG) funded by CNRS/INSU-IN2P3-INP, CEA and CNES, France.
\end{acknowledgements}

\bibliographystyle{aa} 

\bibliography{astrobib}


\appendix

\section{\label{CC_UC} Colour corrections and unit conversions for \textit{Planck}/HFI, ACT, SPT, and \textit{Herschel}/SPIRE }

\subsection{Colour corrections} 
Following the IRAS convention, the spectral intensity data $I_{\nu}$ are often  expressed at fixed nominal
frequencies, assuming the source spectrum is $\nu I_{\nu}=$~constant (i.e. constant
intensity per logarithmic frequency interval, labelled ``ref'' hereafter). The colour-correction factor $\cal C$ is defined such that
\begin{equation}
\label{cc_2}I_{\nu_0}^{\mathrm{act}} = \frac{I_{\nu_0}^{\mathrm{ref}}}{\cal C} \, ,
\end{equation}
where $I_{\nu_0}^{\mathrm{act}}$ is the actual specific intensity of the sky at frequency $\nu_0$, $I_{\nu_0}^{\mathrm{ref}}$ is the corresponding value given with the IRAS \citep{neugebauer1984} or DIRBE \citep{silverberg1993} convention{\footnote{The DIRBE and {{\it IRAS} data products give $I_{\nu_0}(\nu I_{\nu}=constant).$}} , and $\nu_0$ is the frequency corresponding to the nominal wavelength of the band. With these definitions,
\begin{equation}
\label{cc_1}{\cal C} = \frac{\int (I_{\nu}/I_{\nu_0})^{\mathrm{act}}{ R_{\nu} d \nu}}{\int (\nu_0/\nu)  R_{\nu} d \nu} \, ,
\end{equation}
where $(I_{\nu}/I_{\nu_0})^{\mathrm{act}}$ is the actual specific intensity of the sky (SED) normalised to the 
intensity at frequency $\nu_0$, and $R_{\nu}$ is the spectral response. 

\subsection{Colour corrections for CIB and IR shot-noise} 
We give here colour corrections that are useful for joined CIB analyses in HFI, ACT, SPT, and \textit{Herschel}/SPIRE. To have an idea of the errors linked to the SED used to compute ${\cal C}$, we used two different CIB SEDs,
\begin{itemize}
\item from \cite{gispert2000} fit of FIRAS measurements 
\item from \cite{bethermin2012_model} empirical model of galaxy evolution.
\end{itemize}
We recommend using the CIB from \cite{bethermin2012_model} as it comes from a unified model based on our current understanding of the evolution of main-sequence and starburst galaxies.  It reproduces all recent measurements of galaxy counts from the mid-IR to the radio, including counts per redshift slice. It is probably more accurate than the FIRAS measurements. 
Colour corrections ${\cal C}$ are given in Table~\ref{tab:cc}. We can use the same colour corrections  for the star-forming galaxy shot noise and the clustered power spectrum (as the SEDs are very similar).
\begin{table}
\begin{center}
\begin{tabular}{l|c|cc} 
\hline
Experiment & Frequency & ${\cal C}^{model}$ & ${\cal C}^{measure}$ \\
& [GHz] & \\ \hline
& 100 & 1.0759 & 1.0824 \\
& 143 & 1.0171 & 1.0124\\
\textit{Planck}/HFI & 217 & 1.1190 & 1.1076\\
&  353 & 1.0973 & 1.0941\\
& 545 &  1.0677 & 1.0675\\
& 857 & 0.9948 & 0.9939\\\hline
IRAS & 3000 & 0.9605 & 0.9446\\\hline
 & 148 & 1.0720 & 1.0719\\
ACT & 218 & 1.0422 & 1.0384\\
& 277 & 1.0227 & 1.0217\\\hline
& 150 & 1.1411 & 1.1350\\
SPT & 220 & 1.0059 & 1.0046\\
& 95 & 1.1386 & 1.1525\\\hline
&  1200 &  0.9880 & 0.9808\\
\textit{Herschel}/SPIRE &  857 & 0.9887 & 0.9875\\
(extended RSRF) &  600 & 0.9739 & 0.9763\\\hline
& 1200 & 1.0053 & 0.9945\\
\textit{Herschel}/SPIRE & 857  & 1.0193 & 1.0187 \\
(point-source RSRF) &  600 &  1.0469 & 1.0503 \\ \hline
\end{tabular}\\
\caption{\label{tab:cc} Colour corrections ${\cal C}$ (Eq.\,\ref{cc_1}) for dusty star-forming galaxies are given for two different CIB spectral energy distributions ("model" refers to the model of \cite{bethermin2012_model}, while "measure" refers to the  \cite{gispert2000} fit of FIRAS measurements). For SPIRE, we give the colour corrections for the two spectral responses (extended or point-source RSRF). 
}
\end{center}
\end{table}

\subsection{Colour corrections for radio shot noise} 
For the radio galaxy shot noise SED we can use a power law $S_\nu \propto \nu^\alpha$, with $\alpha=-0.5/-0.6$. This is the average spectral index for radio sources that mainly contribute to the shot-noise power spectrum. With this SED, we find that the colour corrections are all lower than 0.7\% for $20\le \nu \le 857$\,GHz. We can thus  neglect them.

\subsection{Unit conversions (tSZ, K$_{\mathrm CMB}$, MJy sr$^{-1}$)}
In unit conversion, data are presented in a different unit, but remain consistent with a given SED (e.g. MJy sr$^{-1}$ can be expressed as an equivalent brightness in K). With colour correction, data are expressed with respect to a different assumed SED at the same reference frequency. Changing from K$_{\mathrm CMB}$ to MJy sr$^{-1}$ with a different spectral index involves both a unit conversion and a colour correction. We give some unit conversions for SPT, ACT, and HFI in Tables~\ref{unit_conv} and \ref{unit_conv_2}. Spectral responses are the official 2013 released ones for \textit{Planck}/HFI. For ACT and SPT, they have been provided by the teams. For SPT, we use the SPT-SZ bandpasses.

\begin{table}
\begin{center}
\begin{tabular}{l|c|c} 
\hline
Experiment  & Frequency  & MJy sr$^{-1}$[$\nu I_{\nu}=$~constant] K$_{\mathrm CMB}^{-1}$\\ \hline & 857 & 2.288 \\ & 545 &  57.980 \\  & 353 & 287.228 \\ \textit{Planck}/HFI & 217 & 483.485 \\& 143 & 371.658 \\& 100 &  244.059\\& 70 & 133.69\\& 44 & 56.82\\& 30 & 24.33 \\ \hline 
& 148 & 401.936 \\ACT  & 218 & 485.311 \\ & 277 & 431.584\\ \hline   & 95 &  234.042 \\SPT & 150 & 413.540 \\ & 220 & 477.017 \\ \hline & 1200 & 3.0568$\times$10$^{-2}$\\ 
\textit{Herschel}/SPIRE & 857 &  2.124 \\  (extended RSRF)  & 600 & 41.275 \\ \hline 
\end{tabular}\\
\caption{\label{unit_conv} MJy sr$^{-1}$[$\nu I_{\nu}=$~constant] to K$_{\mathrm CMB}$ unit conversion. To convert an intensity in K$_{\mathrm CMB}$ into an equivalent specific intensity MJy sr$^{-1}$, the original intensity has to be multiplied by the factors given in the table.}
\end{center}
\end{table}

\begin{table}
\begin{center}
\begin{tabular}{l|c|c} 
\hline
Experiment  & Frequency  & $y_{SZ}$ K$_{\mathrm CMB}^{-1}$\\ \hline 
& 857 & 0.0383 \\ & 545 &  0.0692 \\ \textit{Planck}/HFI   & 353 & 0.1611 \\ & 217 & 5.142\\& 143 & -0.3594 \\& 100 &  -0.2482\\ \hline 
& 148 & -0.390 \\ACT  & 218 & 9.16$^{\ast}$ \\ & 277 & 0.379\\ \hline
& 95 &  -0.243 \\SPT & 150 &  -0.416 \\ & 220 & 9.44 \\ \hline
& 1200 & 0.0240 \\ \textit{Herschel}/SPIRE & 857 &  0.0365 \\  (extended RSRF) & 600 & 0.0646\\ \hline 
\end{tabular}\\
\end{center}
\vspace{-0.2cm}
{\tiny $^{\ast}$This number varies by about 10$\%$ w.r.t. to the boundaries of the bandpass taken in the integrals.}
\caption{\label{unit_conv_2} $y_{SZ}$ to K$_{\mathrm CMB}$ unit conversion. To convert an intensity in K$_{\mathrm CMB}$ to $y_{SZ}$, the original intensity has to be multiplied by the factors given in the table.}
\end{table}

\subsection{Converting CIB power spectra between HFI, ACT, SPT, and SPIRE}
The purpose here is to convert the measurement through one bandpass into a measurement as it would be obtained through another bandpass (often close in frequency, e.g. HFI at 143\,GHz versus SPT at 150\,GHz). This means that we wish to find $K$ such that 
\begin{equation} I^{ref}_{\nu_{0_1}} = K\ I^{ref}_{\nu_{0_2}}\, . 
\label{K_eq}
\end{equation}

For clarity, we write $I_1$ and $I_2$ the fiducial monochromatic flux densities from spectral response 1 and 2 (with the convention $\nu I_{\nu}=constant$) at their respective reference frequencies $\nu_1$ and $\nu_2$. 
Combining Eq.~\ref{cc_2} and \ref{cc_1} gives
\begin{equation}
I^{ref}_{\nu_0} = \cfrac{1}{\nu_0} \times \cfrac{\int \text I_\nu^{act}\, {R}_\nu\,d\nu}{\int \text R_\nu / \nu \,d\nu} \,.
\end{equation}
It then follows\begin{equation}K = \cfrac{I^{ref}_{1}}{I^{ref}_{2}} = \cfrac{\nu_2}{\nu_1} \times \cfrac{\int R_2/\nu \,d\nu}{\int R_1/ \nu \,d\nu} \times \cfrac{\int R_1 I_\nu^{act}\,d\nu}{\int R_2 I_\nu^{act} \,d\nu}
\label{K_factor_eq}
,\end{equation}

\noindent where $R_1$ and $R_2$ are the normalised spectral responses 1 and 2, respectively. Values for $K$ for HFI, ACT, and SPT are given in Table~\ref{K_factors}. For HFI 545 and 857 GHz and \textit{Herschel}/SPIRE 500 and 350\,$\mathrm \mu$m channels, $K(545\,\mathrm GHz, 500\,\mathrm \mu m)$= 0.899808 and $K(857\,\mathrm GHz, 350\,\mathrm \mu m)$=1.00685.\\
We can note that $K(143,148)$ and $K(143,150)$ are $<$1 because the HFI 143\,GHz bandpass is sensitive to lower frequencies than ACT 148\,GHz and SPT 150\,GHz. \\

\underline{Example of the use of K factors:} the HFI-alone likelihood gives the best $C_{\ell}$ CIB amplitude at 143\,GHz in $\mathrm \mu K_{CMB}$. To convert it for ACT at 148\,GHz into $\mathrm \mu K_{CMB}$ follows
\begin{equation}
C_{\ell}^{148} = C_{\ell}^{143} 
 \times 371.658^2 \times \frac{1}{0.8500^2} \times \frac{1}{401.936^2}
.\end{equation}

\begin{table}
\begin{center}
\rotatebox{90}{\begin{tabular}{ll|cccccc|ccc|ccc}  
\hline
& & & & & \textit{Planck}/HFI  & & & & ACT & & & SPT & \\  & & 857 & 545 & 353 & 217 & 143 & 100 & 148 & 218 & 277 & 95 & 150 & 220 \\\hline & 857 & 1 &  1.989  & 5.712 & 24.155 &  97.650 & 269.51 & 83.00 & 25.50 & 12.43 & 294.03 & 75.08 &  25.62 \\  & 545 &  \dots &  1 & 2.872 & 12.15 & 49.10 & 135.52 & 41.74 & 12.82 & 6.25 & 147.85 &  37.75 &  12.88   \\ HFI  & 353 & \dots & \dots &  1 & 4.229 & 17.10 & 47.18 & 14.53 & 4.465 & 2.176 & 51.48 & 13.14 & 4.485 \\ & 217 & \dots & \dots & \dots & 1 &  4.043 & 11.16 & 3.436 & 1.056 & 0.5146 & 12.17 & 3.108 & 1.061 \\ & 143 & \dots & \dots & \dots & \dots & 1 & 2.760 & 0.8500 & 0.2612 & 0.1273 & 3.011 & 0.7688 &  0.2624 \\ & 100 & \dots  & \dots  & \dots & \dots & \dots & 1 & 0.3090 & 0.09463 & 0.046121 & 1.090 & 0.2786 & 0.09506 \\ \hline & 148 & \dots & \dots & \dots & \dots & \dots & \dots & 1 &  0.3073 &  0.1498 & 3.542 & 0.9045 & 0.3087 \\ ACT  & 218 & \dots & \dots & \dots & \dots & \dots & \dots & \dots  & 1 & 0.4874 & 11.53 & 2.944 &  1.005 \\ & 277 & \dots & \dots & \dots & \dots & \dots & \dots & \dots & \dots & 1 & 23.66 & 6.040 &  2.061 \\ \hline  & 95 & \dots & \dots & \dots & \dots & \dots & \dots & \dots & \dots & \dots & 1 & 0.2553 &  0.08713 \\ SPT & 150 & \dots & \dots &\dots & \dots & \dots & \dots & \dots & \dots & \dots & \dots & 1 &  0.3413 \\ & 220 & \dots & \dots & \dots  & \dots & \dots  & \dots & \dots & \dots & \dots & \dots & \dots & 1   \\ \hline 
\end{tabular}}
\caption{\label{K_factors} Factors to convert the CIB intensity (in Jy/sr with the convention $\nu \mathrm I_{\nu}$=constant) into the HFI, ACT, and SPT bandpasses (see Eqs.~\ref{K_eq} and \ref{K_factor_eq}). $\nu_1$ and $\nu_2$ are given in the first column and first line, respectively (e.g. $K(\nu_1, \nu_2) = K(857, 545) = 1.989$. The factors were computed using the \cite{bethermin2012_model} CIB SED. Some factors can be deduced from combinations of others. We give all of them for convenience.}
\end{center}\end{table}

\end{document}